\documentclass{ar2e}
\usepackage{caption}
\usepackage[dvips]{epsfig}
\usepackage[dvips]{color}     
\usepackage{ARAstroBib}
\usepackage{lscape}
\usepackage{amsmath}    
\begin{document}

\newcommand{\aap}    {Astron.Astrophys.}
\newcommand{\aaps}   {Astron.Astrophys.Suppl.}
\newcommand{\aj}     {Astron.J.}
\newcommand{\AJ}     {Astron.J.}
\newcommand{\apj}    {Astrophys.J.}
\newcommand{\ApJ}    {Astrophys.J.}
\newcommand{\apjl}   {Astrophys.J.Lett.}
\newcommand{\ApJL}   {Astrophys.J.Lett.}
\newcommand{\apjs}   {Astrophys.J.Suppl.}
\newcommand{\ApJS}   {Astrophys.J.Suppl.}
\newcommand{\araa}   {Annu.Rev.Astron.Astrophys.}
\newcommand{\arnps}  {Annu.Rev.Nucl.Part.Sci.}

\newcommand{\jcap}    {J.Cosmology and Astroparticle Physics}
\newcommand{\memsai}  {Mem.Soc.Astron.Ital.}
\newcommand{\msais}   {Mem.Soc.Astron.Ital.Suppl.}
\newcommand{\mnras}   {Mon.Not.R.Astron.Soc.}
\newcommand{\MNRAS}   {Mon.Not.R.Astron.Soc.}
\newcommand{\nat}     {Nature}
\newcommand{\nar}     {New.Astron.Rev.}
\newcommand{\pasj}    {Publ.Astron.Soc.Jpn.}
\newcommand{\pasa}    {Publ.Astron.Soc.Aust.}
\newcommand{\solphys} {Solar Phys.}

\newcommand{\my}   {metal-poor}
\newcommand{\lya}  {Lyman-$\alpha$}
\newcommand{\bd}   {BD$+44^{\circ}\,493$}
\newcommand{\cd}   {CD$-38^{\circ}\,245$}
\newcommand{\hen}  {HE~0107$-$5240}
\newcommand{\hea}  {HE~1327$-$2326}
\newcommand{\hej}  {HE~0557$-$4840}
\newcommand{\smk}  {SM~0313$-$6708}
\newcommand{\smklong} {SMSS~J031300.36$-$670839.3}
\newcommand{\sdc}  {SD~1029+1729}
\newcommand{\sdssc} {SDSS~J102915+172927}
\newcommand{\sda}  {SD~0018$-$0939}
\newcommand{\sdssa} {SDSS~J001820$-$093939}
\newcommand{\kms}  {\rm km\,s$^{-1}$}
\newcommand{\feh}  {[Fe/H]} 
\newcommand{\teff} {$T_{\rm eff}$} 
\newcommand{\logg} {log\,$g$} 
\newcommand{\loggf} {log\,$gf$} 
\newcommand{\mel}   {Mel\'{e}ndez et al.}

\def\gsim{\;\lower.6ex\hbox{$\sim$}\kern-8.75pt\raise.65ex\hbox{$>$}\;}
\def\lsim{\;\lower.6ex\hbox{$\sim$}\kern-8.75pt\raise.65ex\hbox{$<$}\;}

\title{NEAR-FIELD COSMOLOGY WITH EXTREMELY METAL-POOR STARS}

\author{Anna Frebel \affiliation{Department of Physics and Kavli
    Institute for Astrophysics and Space Research, Massachusetts
    Institute of Technology, 77 Massachusetts Avenue, Cambridge, MA
    02139, USA; email: afrebel@mit.edu} John E. Norris
  \affiliation{Research School of Astronomy \& Astrophysics, The
    Australian National University, Mount Stromlo Observatory, Cotter
    Road, Weston, ACT 2611, Australia; email: jen@mso.anu.edu.au}}

\markboth{FREBEL \& NORRIS}{NEAR-FIELD COSMOLOGY WITH EXTREMELY METAL-POOR STARS}


\begin{abstract}

The oldest, most metal-poor stars in the Galactic halo and satellite
dwarf galaxies present an opportunity to explore the chemical and
physical conditions of the earliest star forming environments in the
Universe.  We review the fields of stellar archaeology and dwarf
galaxy archaeology by examining the chemical abundance measurements
of various elements in extremely metal-poor stars. Focus on the
carbon-rich and carbon-normal halo star populations illustrates how
these provide insight into the Population\,III star progenitors
responsible for the first metal enrichment events.  We extend the
discussion to near-field cosmology, which is concerned with the
formation of the first stars and galaxies and how metal-poor stars can
be used to constrain these processes. Complementary abundance
measurements in high-redshift gas clouds further help to establish the
early chemical evolution of the Universe.  The data appear consistent
with the existence of two distinct channels of star formation at the
earliest times.

\end{abstract}

\newpage
\maketitle

\section{INTRODUCTION}\label{sec:intro}

\subsection{The Stellar Path to the Early Universe}\label{sec:path}

The paradigm that stars in our Galaxy can tell us about conditions
that existed during the first few billion years after the Big Bang, as
set down by \citet{fbh02} and described by them as ``near-field''
cosmology, is an idea that formed slowly, over some decades.  During
the last century, the study of stars provided a major tool for
exploring the nature of the Milky Way Galaxy (``the'' Galaxy), its
structure and size, and its origin and history.  It also became clear
that we live in just one of many galaxies in the Universe \citep[and
  references therein]{sandage86}.

In the absence of accurate means of directly determining the ages of
all but a few individual oldest stars, it has been assumed that the
best available proxy for age, however imperfect, is a star's chemical
abundance profile.  In particular, the underlying assumption is that
the most metal-poor stars (where as usual we refer to all elements
heavier than lithium) are most likely to be the oldest stars that
exist today.  The second premise, based on theoretical arguments
\citep{brommARAA}, is that stars and galaxies began to form at
redshifts $z \sim 20 - 30$, some 100 -- 200\,Myr after the Big Bang.
Against this background, the aim of the present review is to use stars
and their chemical abundances, in particular stars with
$\mbox{[Fe/H]}<-3.0$, to constrain conditions that existed during the
first $\sim$ 500\,Myr.

The secret to the viability of this near-field cosmology lies in the
mass dependent lifetimes of stars.  While high mass stars soon die as
supernovae that can be observed over large distances, their low-mass
counterparts (of mass M) live for some $\sim 10\,(M/M_{\odot})^{-3}$
Gyr and have witnessed eras long gone.  Stellar populations thus
contain detailed information about the past of their host systems,
connecting the present state of a galaxy to its history of formation
and evolution.  This fortuitous relationship can be used to study the
early Universe and the beginning of star and galaxy formation with
long-lived stars.  Since, however, the approach requires detailed
observations of individual stars, it is only feasible for the
unraveling of the detailed histories of the Milky Way and its dwarf
galaxy satellites.

The key to characterizing individual stars, and indeed entire stellar
populations, is their chemical composition, kinematics, and age (where
possible).  Composition is of particular importance: it yields
information concerning a star's formation era, since in their
atmospheres stars preserve information on the chemical and physical
conditions of their birth gas clouds.  Overall, the amount of elements
heavier than lithium in a star reflects the extent of chemical
enrichment within its natal cloud.

During the past four decades, extensive study has been devoted to the
search for extremely metal-poor stars within the halo of the Galaxy,
in order to piece together early chemical evolution soon after the Big
Bang (see \citealp{beers&christlieb05}).  Specifically, these stars
provide an exceptionally versatile means for studying a large variety
of open questions regarding the nature and evolution of the early
Universe. Consequently, this field is often referred to as ``stellar
archaeology'' since these low-metallicity, low-mass ($M\le$
0.8\,M$_{\odot}$) stars reveal unique observational clues about the
formation of the very first stars and their supernova explosions, the
onset of cosmic chemical evolution, the physics of nucleosynthesis,
early metal- and gas-mixing processes, and even early (proto) galaxy
formation and the assembly processes of larger galaxies. The latter
approach has been termed ``dwarf galaxy archaeology'' because entire
metal-poor dwarf galaxies are now being used to study star and galaxy
formation processes in the early Universe within the actual dwarf
galaxy environment, just as individual stars are used for stellar
archaeology (\citealt{bovill09}, \citealt{frebel12}).

From a more technical point of view, stars are also the easily
accessible \textit{local} equivalent of the high-redshift Universe,
offering a complementary approach to the study of photon-starved,
high-redshift observations (far-field cosmology) of, for example,
damped Ly$\alpha$ (DLA) and sub-DLA systems (\citealp{pettini11},
\citealp{becker12}).  By providing detailed observational data on the
era of the first stars and galaxies, stellar and dwarf galaxy
archaeology have become increasingly attractive for comparison with
theoretical predictions about early Universe physics and galaxy
assembly processes \citep{bromm_araa11}. In the field of near-field
cosmology, the necessary observational and theoretical ingredients are
now being effectively combined for comprehensive studies of how metal
enrichment drove the evolution of the early Universe, and the role
that extremely metal-poor stars and dwarf galaxies played in galactic
halo formation (e.g., \citealp{cooke14}, \citealp{ritter14}).

\subsection{Exploring the Earliest Times}\label{sec:earliest}

Within the metal-poor star discipline, theoretical and observational
works alike have benefited enormously from the discovery, since the
turn of the century, of seven individual halo stars in the Galactic
halo with abundances in the range $\mbox{[Fe/H]}\sim-7.3$ to $-4.5$.
All but one of them have a very large overabundance of carbon relative
to iron.  During the same period, it has become clear that the Milky
Way's dwarf galaxy satellites, in particular the ultra-faint systems,
contain a surprisingly large fraction of stars with
$\mbox{[Fe/H]}<-3.0$, including carbon-rich (C-rich) stars similar to
those found in the Galactic halo.

Spectroscopic abundance measurements of these and other metal-poor
halo stars have provided critical missing information in the following
areas: the chemical enrichment of the Universe at the earliest times;
a more complete characterization of the chemical nature and frequency
of carbon enhancement of stars with $\mbox{[Fe/H]}<-3.0$; the
relationship between the Galaxy's halo and its dwarf satellites; the
evolution of lithium in the very early Universe; and r-process
nucleosynthesis and the various sources of neutron-capture elements.
Comparison of observations with the results of cosmological
simulations is leading to a more complete characterization of the
chemical nature of the first galaxies, with the aim of understanding
their relationship to the surviving dwarfs, as well as the ``building
blocks'' of the Milky Way's halo.

These observational and theoretical endeavors are leading to a deeper
understanding of the origin of the elements, the nature of the first
stars and galaxies in the Universe, and the chemical enrichment of the
Milky Way.  The aim of the present review is to examine the progress
that has been made in these areas over the past decade.

\subsection{Background Material}\label{sec:background}

\begin{figure}[!htbp]
\begin{center}
\includegraphics[clip=true,width=1.0\textwidth,angle=0, bbllx=14, bblly=34,
    bburx=1039, bbury=802]{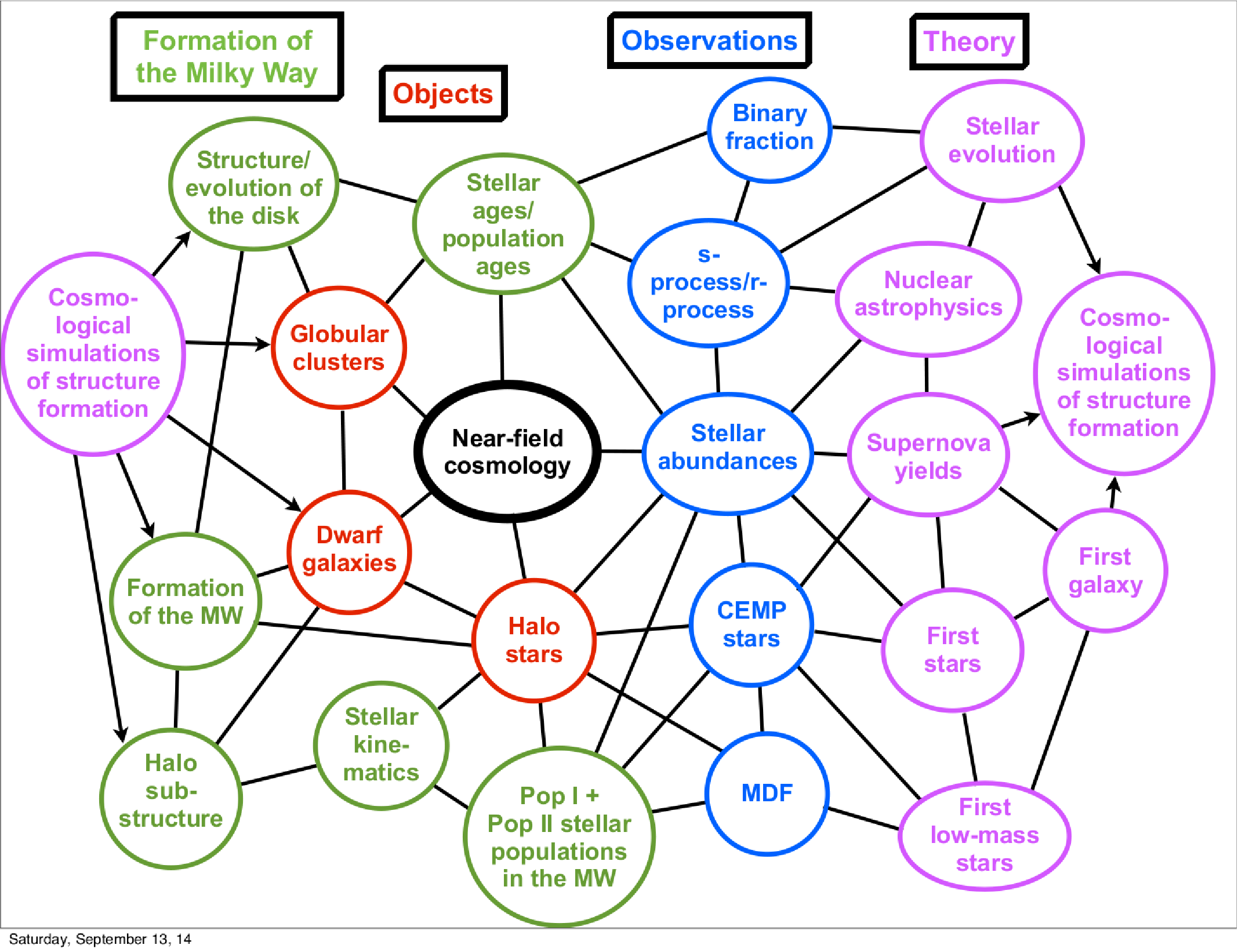}
  \caption{\label{fig:cartoon}\scriptsize Schematic overview of the
    many scientific topics related to near-field cosmology. Example
    connections illustrate how closely connected the many different
    areas are.}
\end{center}
\end{figure}

Interest in the varied results of metal-poor star studies has
dramatically increased in recent years, due to their relevance for
other subfields of astronomy.  These include areas such as galaxy
formation, Galactic dynamics, supernovae, nucleosynthesis, and nuclear
astrophysics. In Figure~\ref{fig:cartoon}, we give a schematic
overview of the substantial connections between the different fields
of study, and how they connect to near-field cosmology and its main
technique, stellar abundance determinations.

Given space constraints, the present review is limited principally to
recent core near-field cosmology results and their implications.  To
address this restriction, we refer the reader to the following broad
range of other relevant works that cover many of the related areas
highlighted in Figure~\ref{fig:cartoon}.  These address the following
topics:

\begin{itemize}

\item

The first stars and galaxies and associated environments for star
formation --- \citet{brommARAA}, \citet{bromm_araa11}, \citet{galli13},
and \citet{loeb13}.

\item

The context of the early chemical enrichment of the Universe, and how
one might use metal-poor stars to explore back in time to the Big Bang
-- \citet{pagel97}, \citet{frebel10},
\citet{karlsson13}, and \citet{fn13}.

\item

The determination of the chemical abundances of
  stars, the important abundance patterns, and the reliability of the
  results -- \citet{wheeleretal89}, \citet{asplund05}, \citet{gray05},
  \citet{sneden_araa}, and \citet{asplund09}.

\item

The search for and discovery of metal-poor stars -- \citet{beers&christlieb05}. 

\item

The role of abundances in the stellar population paradigm, and the
manner in which they inform our understanding of galactic chemical
enrichment -- \citet{sandage86}, \citet{gilmoreetal89},
\citet{mcwilliam97}, \citet{fbh02}, and \citet{ivezic12}. 

\item

Progress in understanding how supernovae produce the chemical elements
-- \citet{woosley_weaver_1995}, \citet{woosley02},
\citet{kobayashietal06}, \citet{heger10}, and \citet{nomoto13}.

\item

The discovery, observation, and interpretation of the Milky Way's
dwarf galaxy satellites -- \citet{mateo98}, \citet{tolstoy_araa},
\citet{willman10}, and \citet{belokurov13}.

\end{itemize}

These reviews are, of course, not one-dimensional, and in many cases a given
review discusses matters in several of the above topics.
 
\subsection{Terms and Assumptions}\label{sec:terms}

\subsubsection{Stellar archaeology, dwarf galaxy archaeology, and near-field cosmology}

Throughout this review the term ``extremely metal-poor
  star'' refers to stars having $\mbox{[Fe/H]}<-3.0$
  \citep{beers&christlieb05}.  For the discussion that follows, we
  begin with definition of terms, together with their significance in
  the study of the early Universe using extremely metal-poor stars.

{\bf Stellar Archaeology}: The study of the astrophysical sites and
conditions of nucleosynthesis and the major physical processes that
drove early star formation, by using stellar chemical abundance
analyses of old stars.  This rests on the abundance determination of
many elements of metal-poor halo stars throughout the periodic table
to enable the detailed documentation of the earliest chemical
enrichment events, and their subsequent interpretation.

{\bf Dwarf Galaxy Archaeology}: The investigation of galaxy formation
on small scales and the associated early metal mixing processes. By
comparing abundances of their most metal-poor stars, particularly
those in ultra-faint galaxies, with those of equivalent
Galactic halo stars, the (beginning of) cosmic chemical evolution
within a galaxy can be tested, providing insight into the relationship
between the dwarf galaxies and the ``building blocks'' of the Galactic
halo, and whether these systems are the survivors of the first
galaxies.

{\bf Near-Field Cosmology}: The interpretation of the results obtained
from stellar and dwarf galaxy archaeology to gain insight into the
physical conditions at the earliest times and the assembly history of
the Galactic halo. The coupling of low-metallicity stellar abundances
with results from cosmological simulations enables studies ranging
from the nature of the first stars all the way to the formation
mechanisms of the metal-poor halo of the Milky Way.

\subsubsection{Abundance definitions}\label{sec:defs}
We recall the following basic definitions.

The abundance of element A, $\epsilon$(A), is presented
logarithmically, relative to that of hydrogen (H), in terms of N$_{\rm
  A}$ and N$_{\rm H}$, the numbers of atoms of A and H.
\begin{equation*}
      \log_{10}{\epsilon} {\rm{(A)}} = \log_{10}(N_{\rm A}/N_{\rm H}) + 12
\end{equation*}

\noindent (For lithium, an alternative and more frequently used
abundance nomenclature is A(Li), where by definition
A(Li) = log$_{10}{\epsilon}$(Li). For hydrogen, by definition,
log$_{10}{\epsilon}$(H) = 12.)

For stellar abundances in the literature, results are generally
presented relative to their values in the Sun, using the so-called
``bracket notation'',
\begin{equation*}
{\rm{[A/H]}} = \log_{10}(N_{\rm A}/N_{\rm H})_\star - \log_{10}(N_{\rm A}/N_{\rm H})_\odot 
\end{equation*}
\noindent and for two elements A and B, one then has
\begin{equation*}
{\rm {[A/B]}} = \log_{10}(N_{\rm A}/N_{\rm B})_\star - \log_{10}(N_{\rm A}/N_{\rm B})_\odot 
\end{equation*}
\noindent 

In the case of the Fe metallicity, [Fe/H] = $\log_{10}(N_{\rm
  Fe}/N_{\rm H})_\star - \log_{10}(N_{\rm Fe}/N_{\rm H})_\odot$. For
example, $\mbox{[Fe/H]}=-4.0$ corresponds to an iron abundance 1/10000
that of the Sun.  With this notation, one needs the abundance not only
of the star being analyzed, but also of the Sun (see, e.g.,
\citealp{asplund09} and \citealp{caffausun11} for details of this
aspect of the problem).

In stars, the most commonly measured element is iron, given that it is
one of the most abundant and the most readily measurable of the
elements in stellar spectra.  From a semantic point of view and for
the discussion that follows, we note that the term ``metal-poor'' is
not always synonymous with ``Fe-poor''.  For stars with
$\mbox{[Fe/H]}>-4.0$, ``metal-poor'' in most cases well describes the
amount of both Fe and metals in the star, but for those with
$\mbox{[Fe/H]}<-4.0$, the assumed equality appears to generally break
down given the common and huge amounts of carbon and other elements
with respect to iron.  Needless to say, there are exceptions above and
below $\mbox{[Fe/H]}=-4.0$.  We shall refer to stars with
$\mbox{[Fe/H]}<-4.5$ as the most iron-poor stars and use ``most
metal-poor stars'' when referring generically to stars of the lowest
metallicity, in the absence of {\it a priori} knowledge of the details
of their heavy-element abundance distributions.

\subsubsection{Model Atmosphere Assumptions}\label{sec:assumptions}

Most chemical abundance determinations of stars are based on
one-dimensional (1D) model stellar atmosphere analyses that assume
hydrostatic equilibrium, flux constancy, Local Thermodynamic
Equilibrium (LTE), and treat convection in terms of a rudimentary
mixing length theory.  To first order, the basic atmospheric
parameters that define the model are effective temperature ({\teff}),
surface gravity ({\logg}), and chemical composition ([M/H], where M
refers to ``metals''). We refer the reader to \citet{gray05} and
\citet{gustafsson08} for the concepts associated with stellar
atmosphere modeling.  For details of the more realistic and rigorous
three dimensional (3D) model atmosphere and non-LTE (hereafter NLTE)
formalisms, rather than the 1D/LTE approximations, see
\citet{asplund05}, and references therein; examples of recent
developments in these areas include \citet{lind12b}, \citet{ludwig12},
\citet{tremblay13} and \citet{magic13}.  We have discussed the corresponding abundance
differences that result from these two approaches previously
\citep[see their Figure~11]{fn13}, to which we refer the reader for
details. In brief, differences between LTE and NLTE results, on the
one hand, and between 1D and 3D analyses, on the other, are of order
$\sim0.5$\,dex for some atomic species.  3D corrections for the
hydrides of C, N, and O are $\sim0.5$ -- 1.0\,dex, in the sense that
1D abundances are higher. We note also that LTE/NLTE and 1D/3D
differences, determined independently, are not always necessarily in
the same sense; and, thus, to some extent, the criticisms that apply
to 1D/LTE results remain if only one of the two improvements has been
made, as opposed to a complete 3D/NLTE analysis. In general the
results of the more rigorous, but very computationally challenging
3D/NLTE formalism are always to be preferred, if available.  In what
follows, however, unless otherwise stated, we shall present abundances
from the more widely used 1D/LTE procedures, which have produced
results for a much larger sample of stars.  The rationale for this is
that it is unwise to mix the results of the 1D/LTE and 3D/NLTE
formalisms when investigating abundance trends and objects with
apparently anomalous compositions.  On the other hand, one must bear
in mind the problems that could occur if 1D/LTE abundances, rather
than the more rigorous 3D/NLTE values, are compared with those of, for
example, stellar evolution computations, other theoretical
predictions, and non-stellar (e.g., Damped Ly$\alpha$) systems.

\subsection{Overview}

We begin in Section~\ref{sec:milkyway} with a discussion of the basic
properties of the Galaxy, its components, and the different types of
stars and objects that are being studied to explore its nature and
evolution.  This sets the scene for stellar archaeology in
Section~\ref{sec:archaeology} and the role of metal-poor stars, which
are at the heart of near-field cosmology.  Here we trace the search
for the most metal-poor stars and the discovery of a small number that
have only 10$^{-5}$ -- 10$^{-7}$ the amount of iron one finds in the
Sun.  In stark contrast, the ratio of carbon to iron in these objects
is enormous and $\sim$\,10 -- 10$^{+5}$ times the solar ratio.  We
also discuss the metallicity distribution function for stars with
$\mbox{[Fe/H]}<-3.0$ and the behavior of other elements that place
strong constraints of the nature of the first stars.  Here we reach
the crux of the matter: below $\mbox{[Fe/H]}\sim-4.0$ the chemical
abundance patterns are fundamentally different from those of stars
above this limit. Section~\ref{sec:dwarfgal} addresses the archaeology
of the Milky Way dwarf satellite galaxies and a comparison of their
properties with those of Galactic halo population.  In
Section~\ref{sec:nearfield}, with the archaeology complete, we move to
near-field cosmology.  Here we trace the development of the
theoretical interpretation of the manner in which the earliest stars
and galaxies formed within the framework of the Lambda Cold Dark
Matter ($\Lambda$CDM) paradigm, and seek to infer how the first stars
and the earliest populations formed in the Universe, in light of the
chemical abundances of the most metal-poor stars.  In
Section~\ref{sec:nearfar} we compare what we have learned from
near-field cosmology with the results of far-field endeavors.
Fonally, in Section~\ref{sec:future} we conclude with expectations for
the future.

\newpage

\section{THE MILKY WAY}\label{sec:milkyway}
  
\subsection{Galactic Populations}\label{sec:pops}

In a masterful review of the structure and evolution of the Milky Way,
\citet{sandage86} described the population concept as one of the
``grand unifying themes in science''.  At its  center is
the concept that stellar populations may be well described in terms of
their spatial, kinematic, chemical abundance, and age distributions
that distinguish them one from another, and which provide the basic
information necessary and essential for an understanding of their
origin and evolution.  In what follows, we shall
concentrate mainly on chemical abundances, given our
premise that these are the principal population
characteristics that will be used in our discussion of near-field
cosmology.  A current review of the structure and
properties of the Galaxy is given by Bland-Hawthorn in this
volume, to which we refer the interested reader.

\subsection{Baryonic Components}\label{sec:baryonic}

\subsubsection{DISK, HALO, AND BULGE} 

The Milky Way consists of several directly observable components --
most notably the disk, the halo, and the bulge.  The disk can be well
described in terms of a ``thin'' disk together with a more vertically
extended ``thick'' disk (\citealp{yoshii82}, \citealp{gilmore83}) and
an even more extended ``metal-weak thick-disk'' sub-component
(\citealp{morrisonetal90}, \citealp{chiba&beers00},
\citealp{beers14}).  Young, metal-rich, Population\,I stars similar to
the Sun are primarily located in the thin disk, with an average
metallicity of $\mbox{[Fe/H]}\sim-0.2$; for the thick disk the mean
value is $\mbox{[Fe/H]}\sim-0.6$; and the metal-weak thick disk is
even more metal-poor, with $-2.5 \lsim \mbox{[Fe/H]} \lsim -1.0$.  The
stellar halo has a spheroidal distribution that envelops the disk and
bulge and reaches out to some 150\,kpc.  It contains older,
Population\,II, stars which are generally more metal-poor: average
halo metallicities for inner and outer halo sub-components are [Fe/H]
$<-1.6$ and $-2.2$, respectively (Carollo et al. 2007, 2010), with a
distribution that stretches down to at least $\mbox{[Fe/H]}= -7.3$
\citep{keller14}. Given the extent of the halo, the outer regions
beyond a distance of 30\,kpc are not well explored in terms of
high-resolution spectroscopic abundance analyses, given that the stars
are faint, with $V \gsim 16$.
\nocite{carollo07, carollo10}

Spatial, kinematic, and abundance distributions of halo stars indicate
that the Milky Way's halo may not be a single monolithic component,
but contains substructure resulting from accretion events,
superimposed on a dichotomy of inner and outer components
(\citealp{hartwick87}, \citealp{norris89}, Carollo et al. 2007, 2010,
2014, \citealp{dejong10}, \citealp{deason11}, \citealp{beers12}, and
references therein).  We refer the reader to \citet{schoenrich11,
  schoenrich14} for an alternative viewpoint on the dual nature of the
Galactic halo based on SEGUE/SDSS data.  The details of the
relationship between the substructure and the ``inner'' and ``outer''
components remains to be clarified.  In general, the halo is well
described in terms on an inner component that may have formed {\it in
  situ} during the evolution of the Milky Way, together with a more
diffuse outer one that originated from past accretion and tidal
disruption of dwarf galaxies (\citealp{zolotov09, zolotov10},
\citealp{font11}, \citealt{mccarthy12}, \citealt{tissera14}).  The
metal-poor component of the Galactic bulge is, observationally,
largely unexplored territory.  While its oldest stars formed coevally
with the early assembly phases of the Milky Way (e.g.,
\citealp{brook}, \citealp{tumlinson10}), several younger populations
are also present, making it challenging to efficiently isolate
individual members of the first population.  Additionally, crowding
effects and the large amount of dust extinction towards the bulge
complicate this endeavor.  Accordingly, only limited progress has been
made in identifying its most metal-poor and oldest components.  We
refer the reader to the works of \citet{garcia13}, \citet{ness13}, and
\citet{howes14}, and references therein, for important recent advances
in this area.  \nocite{carollo07, carollo12, carollo14}

Galactic globular clusters have been studied for many decades, and
were the first probes used to map the structure and extent of the
Milky Way (see \citealp{shapley30}).  They comprise two sub-systems,
one associated with the bulge and thick disk, the other with the halo
\citep{zinn85}.  The latter is the more metal-poor group, with mean
and minimum [Fe/H] $\sim-1.6$ and $\sim-2.3$, respectively.
Individual clusters possess extremely complicated chemical abundance
patterns, which preclude any clear understanding of their formation
and evolution.  Poorly understood problems include the following: some
systems have sub-populations with spectacularly large helium
abundance, $Y \sim 0.35$ -- 0.40 (e.g., \citealp{king12}); all display
chemical abundance signatures of the lighter elements (correlations
and anti-correlations among C, N, O, Na, Mg, and Al) that differ
fundamentally from the relationships found among field halo stars
(e.g., \citealp{gratton04}); and detailed spectroscopic observations
have revealed iron and heavy neutron-capture abundance spreads in an
increasing number of clusters (e.g., \citealp[and references
  therein]{yong14}).

\subsubsection{SATELLITE DWARF GALAXIES} 

The Milky Way is surrounded by a host of satellite dwarf galaxies that
orbit in its outer halo (see \citealp{mateo98},
\citealp{tolstoy_araa}, \citealp{belokurov13}). The most massive of
them are the Large and Small Magellanic Clouds which are gaseous
irregular systems with ongoing star formation. Other less massive
dwarf irregular (dIrr) galaxies are complemented by gas-poor dwarf
spheroidal (dSph and ultra-faint) galaxies. In the context of the
present review, the dSph and the even less-luminous ultra-faint
systems are of prime importance, since their old stellar population(s)
encode their early history of star formation and chemical enrichment.
While dSph have been studied extensively in their own right, it has
become clear in recent years that our understanding of the formation
of the Milky Way is closely connected to their nature and history, and
the role dwarf galaxies may play as ``building blocks'' of larger
galaxies.

\subsection{The Dark Halo and Putative Population\,III}\label{sec:darkpops}

Associated with the Galaxy's luminous material is a massive halo of
dark matter, extending to $\sim$300\,kpc from the Galactic center, and
having a mass in the range 1.0 -- 2.4$\times$10$^{12}$\,M$_{\odot}$
(at the 90\% confidence level) \citep[and references
  therein]{boylan13}.  This topic lies outside the terms of the
present discussion, except insofar as the formation of structure
within the Universe was determined by the properties of the dark
matter, and its entrainment of baryonic material.

In addition to Population\,I and II stars, which are traditionally
observed to study the nature of the disk and halo of the Milky Way,
there is also a putative ``Population\,III'' -- a metal-free first
population that lit up the Universe 100 --
200\,Myr after the Big Bang. Early theoretical models of first star
formation favored a top-heavy mass function for this
population, which renders them unobservable today, given their
corresponding short lifetimes of a few tens of Myr.  More recent work,
however, suggests this population may have contained stars of
significantly lower mass \citep{stacy14}.  If such Population\,III
stars formed, in particular objects having masses less than
1\,M$_{\odot}$, they would have sufficiently long lifetimes to be
still observable today.  In that case, we would expect them to be
eventually found in the Galaxy's halo and/or bulge.

\subsection{The Galaxy in the Cosmological Context}

Fundamental impetus to an understanding of galaxy formation was
provided by \citet{white&rees78}, who proposed their Cold Dark Matter
(CDM) hierarchical clustering paradigm, in which ``The entire
luminosity content of galaxies ... results from the cooling and
fragmentation of residual gas within the transient potential wells
provided by the dark matter.''  At the same time, the observational
community was addressing two essentially different paradigms for the
formation of the halo of the Milky Way.  On the one hand, the
monolithic collapse model of \citet[hereafter ELS]{els} had predicted
a very rapid collapse phase (of a few 10$^8$\,yr), and a dependence of
kinematics on abundance together with a radial abundance gradient for
halo material.  On the other, the fragment accretion model of
\citet[hereafter SZ]{sz78} proposed a longer formation period of a few
10$^9$\,yr, no dependence of kinematics on abundance, and no radial
abundance gradient.  We refer the reader to our previous review of
this confrontation \citep{fn13}.  Suffice it here to say, neither
fully explains the observations. Rather, as stated there ``it seems
likely the answer will be found within the hierarchical $\Lambda$CDM
paradigm ... The work of \citet{zolotov09}, for example, while
supporting the SZ paradigm of halo formation, also produces a dual
halo configuration of ``{\it in situ}'' and ``{\it accreted}''
components, not unlike those envisaged in the ELS and SZ observational
paradigms.  Remarkably, these paradigms were first established on
essentially observational grounds only. They are now being explained
in terms of a theoretical framework based on tracing the dark matter
evolution from initial density fluctuations early in the Universe''.
The {\it in situ} and {\it accreted} components can readily be
identified with the inner and outer halo components 
(Carollo et al. 2007, 2010) 
which is also supported by
simulations such as those of \citet{tissera14}.
\nocite{carollo07, carollo10}

Models of cosmic structure formation and galaxy evolution based on
cosmological initial conditions and following the gravitational
evolution of collapsing dark matter halos were first undertaken by
\citet{moore} and \citet{klypin99}. These works showed that galaxy
evolution is an ongoing assembly process. The emerging central halo
(``the galaxy'') accretes gas from large scale filaments of the cosmic
web as well as smaller halos (``dwarf galaxies'') that orbit it. In this way, the
galaxy grows over billions of years.  Recently, hydrodynamical
cosmological models have emerged which are capable of modeling the
evolution of not only dark matter but also gravitationally
entrained baryons as part of the formation of disk galaxies
(\citealp{agertz11}, \citealp{guedes11}, \citealp{vogelsberger14}).

Indeed, we can see the accretion process in the Milky Way.  It is
still ongoing, and the many stellar streams and overdensities in the
halo are evidence of it (e.g., \citealp{pila14}).  The most
prominent example of the phenomenon is the Sagittarius dwarf galaxy
\citep{ibataetal95}, which is in the process of being disrupted,
leaving a stellar stream wrapped more than once around the Galaxy.

\begin{figure}[!htb]
\begin{center}
  \includegraphics[width=1.0\textwidth,angle=0]{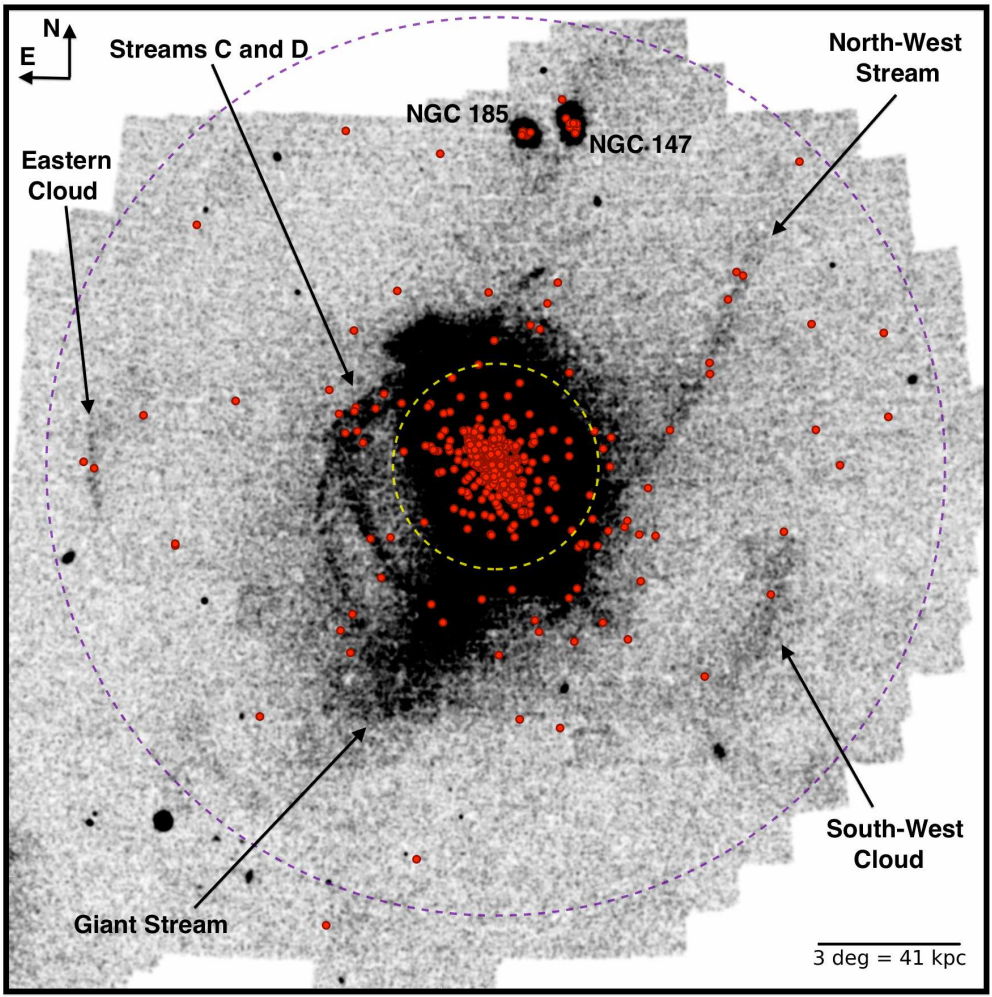}
  \caption{\label{fig:andromeda} {\scriptsize A spatial density map of
      metal-poor stars ($\mbox{[Fe/H]}\lsim-1.4$) in the halo of the
      Andromeda galaxy obtained by the PAndAS team \citep{pandas09}.
      The red circles represent globular clusters, and their
      association with stellar streams is clear.  The two circles in
      the figure have radii of 30 and 130\,kpc. In this map the
      central spiral structure is lost in the lack of contrast of the
      exposure.  Credit: A. D. Mackey and the PAndAS collaboration.}}
\end{center}
\end{figure}

In this context, the oldest stars found in the Galactic halo are the
best tracers of the earliest accretion events.  They formed as part of
the earliest generations of stars in their respective host systems
which means that their current elemental abundances reflect those
first local chemical enrichment events.  This scenario is broadly
supported by stellar age measurements that have shown individual
metal-poor halo stars with overabundances of heavy r-process
neutron-capture elements to be 13 -- 14\,Gyr old (e.g.,
\citealp{Hilletal:2002}, \citealp{he1523}). Similar ages have been
obtained for globular clusters \citep{marinfranch09} as well as
several ultra-faint dwarf galaxies (\citealp{brown12},
\citealp{brown13}).  These ages are commensurate with the
independently determined age of the Universe of 13.8\,Gyr by
\citet[the Planck Collaboration]{planck13}.

Being situated inside the Milky Way makes it difficult for us to
appreciate these accretion events directly.  That said, we are ideally
placed to examine the effect in our nearest spiral galaxy neighbor,
Andromeda. Figure~\ref{fig:andromeda} shows an extremely deep
Pan-Andromeda Archaeological Survey (PAndAS) \citep{pandas09} spatial
density map of metal-poor stars ($\mbox{[Fe/H]}\lsim -1.4$) in the
halo of Andromeda, reaching to a projected galactocentric radius of up
to 150\,kpc.  The remnants of numerous accretion events are clearly
seen in Andromeda's halo.  Globular clusters are cataloged as well,
and are represented here by superimposed red circles.  It is evident
that in many cases the clusters are associated with the halo streams
and thus accreted with the captured galaxies.  \citet{gilbert14}
provides new support for such a history and that Andromeda has an
inner and outer halo (not unlike that proposed for the Milky Way),
and rather similar to at least one of the simulation results of
\citet{tissera14}.

\newpage

\section{STELLAR ARCHAEOLOGY}\label{sec:archaeology}

Stellar archaeology seeks to understand the manner in which stellar
populations formed and have evolved.  The principal parameters
available for the endeavor, as noted in Section~\ref{sec:pops}, are
their spatial distribution, kinematics, chemical abundances, and ages.
In the cosmological context, we seek to understand the earliest times,
and are interested in stars that formed within the first billion years
after the Big Bang singularity.  Clearly, age is of fundamental
importance.  The basic problem that confronts us, however, is that we
are unable to obtain accurate ages for all but a very small minority
of individual stars from this era.  We must look for proxies that
provide related and complementary information.

\subsection{The Role of Chemical Abundances}\label{sec:abundance_role}

A basic premise is that some minutes after the singularity, at the era
of decoupling of radiation and matter, the only chemical elements in
the Universe were hydrogen, helium and lithium.  Within the framework
of Standard Big Bang Nucleosynthesis (SBBN), constrained by the
results of the Wilkinson Microwave Anisotropy Probe (WMAP), the
predicted relative mass densities of these elements are 0.75, 0.25,
2.3$\times$10$^{-9}$, respectively \citep{WMAP}. All other elements
(with the exceptions of beryllium and boron) have been produced
subsequently in stars and supernova explosions, a process that
continues to the present day.

The assumption that follows is that the oldest stars are those that
have the lowest total heavy element abundance, Z.  In the absence of
an understanding of the nature and evolution of the first stars,
however, and in particular our lack of insight into their
nucleosynthetic yields of individual elements, this is a potentially
fraught assumption, to which we shall return in Sections~\ref{sec:mdf}
and ~\ref{sec:twopops}.  The assumption that we shall make is that the
most Fe-poor stars are the best candidates we have to take us to the
earliest stellar generations and closest to the Big Bang.  This is the
working hypothesis that drives stellar archaeology.

In what follows, we shall confine our attention to the two most
accessible observational areas.  The first is the halo of the Milky
Way Galaxy, while the second is the Galaxy's dwarf satellite galaxies
(its dwarf spheroidal and the relatively recently discovered
ultra-faint galaxies). We limit the present discussion to results of
high-resolution, high $S/N$ spectroscopic chemical abundance analyses
in these two areas and almost exclusively to objects that have
$\mbox{[Fe/H]}<-3.0$, in an effort to take us closer in time to the
Big Bang than would stars of higher Fe abundances.  We note that this
restriction excludes the Galaxy's globular clusters, and its
metal-weak thick-disk and bulge stars from the discussion.  While the
bulge is believed to be the site of some of the very first star
formation, its admixture with later stellar generations at the
Galaxy's center has to this point precluded discovery of more than a
few stars with $\mbox{[Fe/H]}<-3.0$.

\subsection{The Search for Extremely Metal-poor Stars in the Galactic Halo}\label{sec:search}

\noindent In the middle of the twentieth century, the concept of
``metal-poor'' stars did not exist.  To quote \citet{sandage86},
``There had grown up a general opinion, shared by nearly all
spectroscopists, that there was a single universal curve of all the
elements, and that the Sun and all the stars shared ... precisely
... identical ratios of any element A to hydrogen''.  Against this
background, \citet{chamberlain51j} demonstrated that the ``A-type
subdwarfs'' HD~19445 and HD~140283 have abundances [Fe/H]
($\mbox{[Ca/H]}$) $= -0.8$ ($-1.4$) and $-1.0$ ($-1.6$), respectively,
thereby establishing a profound paradigm shift, which opened the way
to the discovery of stars having progressively lower and lower
chemical abundances, culminating in the recent discovery by
\citet{keller14} of the halo giant {\smklong} (hereafter {\smk}), with
$\mbox{[Fe/H]}<-7.3$ and $\mbox{[Ca/H]} = -7.2$ -- i.e., with Fe
undetected at the 10$^{-7}$ level and Ca at some 10$^{-7}$ times
relative to those of the Sun.

The decades long searches have shown that metal-poor stars are rare:
in the solar neighborhood, below $\mbox{[Fe/H]} = -3.5$, we expect to
find of order only one such star among 100000 field stars.  Several
techniques have therefore been used to improve the odds of finding
these objects in the Galactic halo, based on different criteria.  We
shall give only a brief description of these efforts here, and refer
the reader to the comprehensive discussions of
\citet{beers&christlieb05}, \citet{ivezic12}, and \citet{fn13} for more
detail.

Important contributions that provide ``candidate'' metal-poor stars
suitable for further observation and analysis include the following.

\subsubsection {Informed serendipity}  Some bright, extremely metal-poor stars have
been found by what can best be described as informed serendipity.
Perhaps the two best known of these are {\cd} ($\mbox{[Fe/H]}=-4.0$,
$V = 12.8$, \citealt{bessell&norris84}) and {\bd} ($\mbox{[Fe/H]} =
-3.7$, $V = 9.1$, \citealt{ito2009}), which were first recognized
following their inclusion in A-star and radial-velocity surveys,
respectively.

\subsubsection {High-proper-motion surveys}  A large fraction of halo 
stars has high proper motion relative to those of the Galactic disk.
The first star with $\mbox{[Fe/H]}<-3.0$ (G64-12;
\citealt{carney&peterson81}) was discovered as the result of its high
proper motion.  The surveys of \citet{ryan&norris91a} and
\citet{carneyetal96} have utilized this technique, and their samples
each comprise a few hundred halo main-sequence dwarfs with
$\mbox{[Fe/H]}<-1.0$.  Taken together these authors obtained some
$\sim$10 stars having $\mbox{[Fe/H]}<-3.0$.  It is important to note
that this technique produces a sample that does not have any explicit
abundance bias.  That said, the reader should bear in mind that the
sample may be an admixture of subpopulations having different origins.
Another example of the role of kinematic selection is provided by the
current ESA Gaia Mission, which will revolutionize our view of the
Milky Way (see \citealp{perryman01}).

\subsubsection {Schmidt Objective-prism surveys} These permit one to obtain large
numbers of low-resolution spectra (resolving power $R$ ($=
\lambda/\Delta\lambda$) $\sim400$) simultaneously, over several square
degrees.  Examination of the strength of the Ca\,II\,K line at
3933.6\,{\AA} with respect to that of nearby hydrogen lines, or the
color of the star, leads to a first estimate of metal weakness.
Candidate metal-poor stars are then observed at intermediate
resolution ($R\sim2000$) to obtain a first estimate of metallicity,
based again on the Ca II\,K line. The techniques are described in
detail by \citet{beers&christlieb05}, who also document important
surveys of this type that have obtained these first abundance
estimates for some tens of thousands of stars brighter than
$B\sim16.5$ with $\mbox{[Fe/H]}<-1.0$.  The first major surveys of
this type for metal-poor stars were those of \citet{bond70} and
\citet{bidelman73}.  The most comprehensive to date have been the HK
survey \citep{bps92} and the Hamburg/ESO Survey (HES) \citep[see also
  \citealp{frebel_bmps} and \citealp{placco11}]{christliebetal08}.

Until now, this has been the most efficient way to find metal-poor
stars, and hence the essential source of objects with
$\mbox{[Fe/H]}<-3.0$ for which high-resolution, high $S/N$ chemical
abundance analyses are currently available.  By its nature, the method
is strongly biased towards stars of lowest abundance.  Current
experience suggests that samples are relatively complete for stars
with $\mbox{[Fe/H]}{\lsim} -3.0$ \citep{schoerck}.

\subsubsection {Spectroscopic surveys}  The Sloan Digital Sky Survey (SDSS) and its
subsequent SEGUE surveys have obtained spectra with resolving power $R
\sim2000$, and have proved to be a prolific source of metal-poor
stars.  The Tenth SDSS Data Release may be found at
http://www.sdss3.org/dr10/.  Currently, the LAMOST survey
\citep{deng12} is underway to provide metal-poor candidates in the
northern hemisphere.

\subsubsection {Photometric surveys}  Photometric systems have the potential to
provide an alternative, low-resolution, method for discovering
metal-poor candidates.  In the Johnson UBV system, for example, the
ultraviolet excess, $\delta$(U-B) (driven by the sensitivity of the U
magnitude to line blocking by the plethora of metal lines in the U
bandpass) was useful, in the infancy of this subject, for isolating
and obtaining metallicity estimates for metal-poor stars
(\citealt{roman54}, \citealt{wallerstein60}).

A survey currently underway to discover the most metal-poor stars in
the southern hemisphere by using a tailor-made intermediate band
filter system is the SkyMapper Southern Sky Survey
\citep{kelleretal07, kellbess07}. The filter set was designed
specifically for the determination of the atmospheric parameters of
metal-poor stars.  During the commissioning period, the most iron-poor
star currently known, {\smk}, with $\mbox{[Fe/H]} <-7.3$, was
discovered \citep{keller14}.

\subsection{High-Resolution, High $S/N$ Follow-Up Spectroscopic Abundance Analyses}\label{sec:hires}

After identifying promising candidate metal-poor stars, the next step
is to obtain high-resolution ($R \gsim 30000$), high $S/N$ optical
spectra to provide the data for the determination of basic stellar
information such as accurate chemical abundances, isotopic ratios (of
only very few elements), and in some cases stellar ages.  The required
high-resolution, high $S/N$ requirements are best achieved with 6 --
10\,m telescope/´echelle spectrograph combinations -- currently
HET/HRS, Keck/HIRES, Magellan/MIKE, Subaru/ HDS, and VLT/UVES.  In
using these facilities, some investigators take short (``snapshot'')
exposures to enable them to choose the most interesting metal-poor
stars in their sample, before embarking on long exposures (several
hours) to obtain the high $S/N$ necessary for a detailed analysis.
Two examples of this are \citet{barklem05} and \citet{aoki13}. Most
stellar abundances are then determined from equivalent width (EW)
measurements of spectral absorption lines, which are obtained with an
observational uncertainty $\sigma$(EW) that varies as
FWHM$^{0.5}/(S/N)$, where FWHM is the full width at half maximum of
the line \citep{cayrel2004}.  As the resolving power of the
spectrograph and the $S/N$ of the spectrum increase, the line
measurement uncertainties decrease allowing for the detection of
weaker features. This is of particular importance when observing the
extremely weak-lined most metal-poor stars.

As noted in Section~\ref{sec:abundance_role}, we shall confine our
attention in the present work primarily to stars having
$\mbox{[Fe/H]}<-3.0$.  The following investigations contain abundances
for a significant number of stars ($\gsim15$) having [Fe/H] less than
this limit: \citet{McWilliametal}, \citet{ryan96}, \citet{cayrel2004},
\citet{lai08}, \citet{hollek11}, \citet{bonifacio12a},
\citet{yong13a}, \citet{cohen13}, \citet{roederer14}.  For a more
comprehensive introduction to the literature for surveys and analyses
of metal-poor stars of all abundances, we refer the reader to
\citet[Section~1]{roederer14}.  Two comprehensive and useful
compilations of abundances for metal-poor stars, based on
high-resolution analyses, are those of
\citet[http://saga.sci.hokudai.ac.jp]{suda08}
and\\\citet[http://www.metalpoorstars.com]{frebel10}, the latter of
which we shall use in what follows.

\subsection{The Most Metal-poor Stars}\label{sec:mmps}

\subsubsection{Metallicity as a Function of Epoch of Discovery}  
\begin{figure}[!htbp]
\begin{center}
\includegraphics[width=1.0\textwidth,angle=0]{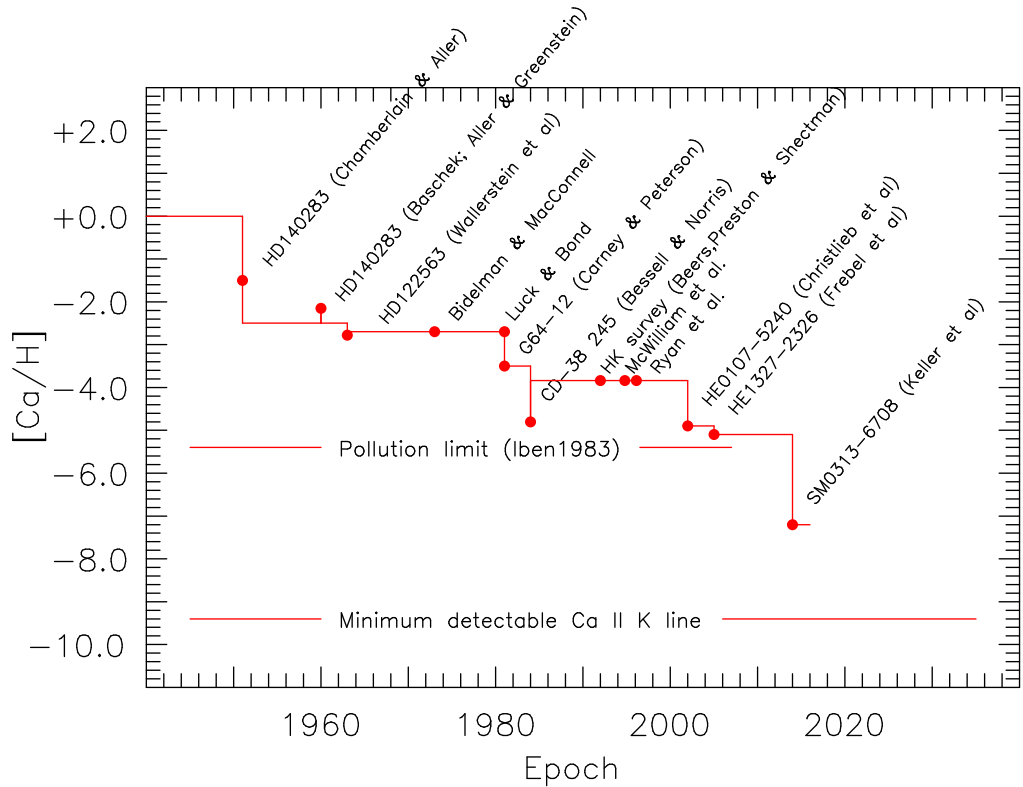}

  \caption{\label{fig:cah_epo}\scriptsize The calcium abundance,
    [Ca/H], of the most Fe-poor star then known as a function of epoch
    of discovery.  (We use [Ca/H] since calcium is still observable at
    lowest metallicities, when iron is no longer detectable.) The
    circles denote the abundance determined by the authors identified
    in the figure, while the horizontal connecting lines refer,
    approximately, to currently accepted values.  The two lower
    limiting lines are discussed in the text. }

\end{center}
\end{figure}

The history of the 
discovery of the most metal-poor stars in the halo is shown in
Figure~\ref{fig:cah_epo}, which presents the abundance of calcium,
[Ca/H], of the most calcium-poor star then known, as a function of
epoch.  Also shown, as reality checks, are two estimates of predicted
lower limits below which one might not expect to find any star -- for
physical or technical reasons.  The higher of the two limits is the
prediction by \citet{iben83} of $\mbox{[Fe/H]}=-5.7$ (assuming
$\mbox{[Ca/Fe]} = 0.3$) as the putative surface value a zero
heavy-element-abundance star would acquire by accretion of the
Galactic interstellar material over the past 10\,Gyr, assuming a
time-averaged accreted abundance of $\mbox{[Fe/H]}=-0.6$.  Insofar as
the Iben calculation is a ``back-of-the-envelope'' calculation, one
should not be surprised that a star has been observed below this
limit.  The lower line, on the other hand, is the more restrictive
limit of \citet{fn13} which is the calcium abundance one might
determine for a very low upper limit of the observed Ca\,II\,K line
strength of 20\,m{\AA} in a putative metal-poor red giant having
{\teff} = 4500\,K and {\logg} = 0.5.  (The Ca\,II\,K line is the
strongest atomic feature in the spectra of metal-poor stars (see
Figure~\ref{fig:spectra_araa} below), and its strength is
intrinsically greater in red giants than in near-main-sequence
dwarfs.)  We make two further points based on the figure.  First,
steady progress has been made over the past six decades to find stars
with lower and lower metallicities (the one to two decade periods
without progress notwithstanding).  Second, we have most likely come
to the point where further progress to lower abundance will be very
slow.  The spectrum of the most metal-poor star currently known,
{\smk} \citep{keller14} has $\mbox{[Ca/H]} = -7.2$, and no Fe lines
have been observed.  The only other elements so far detected are H, C,
O, Mg, and Si; with the strength of the Ca\,II\,K line being a mere
90\, m{\AA}. For comparison, the strength of this line in {\cd} (which
has [Fe/H] = --4.0) is 1485\,m{\AA}.
 

\subsubsection{The Seven Most Fe-poor Stars}
A very challenging aspect of the study of the most metal-poor stars is
their extreme rarity.  Below $\mbox{[Fe/H]} \lsim -4.5$, only seven of
them are known.  We present identification details for these in
Table~1, where the stars are arranged in order of increasing [Fe/H].
Detailed abundances based on model-atmosphere analysis of
high-resolution, high-$S/N$ spectra are available for six of the
stars, while analysis of the seventh is in progress. For future
reference, we draw the reader's attention to the very large
overabundances of carbon relative to iron in six of these seven stars
(see Column 6).  For the five C-rich stars in which iron is detected,
the relative carbon abundances lie in the range $\mbox{[C/Fe]} = +1.6$
to +4.3.  For {\smk}, in which iron is not detected, $\mbox{[C/Fe]} >
+4.9$. The seventh star is {\sdssc} (hereafter {\sdc};
\citealp{caffau11, caffau12}) for which $\mbox{[Fe/H]}= -4.8$ and
$\mbox{[C/Fe]} < +0.9$.  Its relatively low carbon
abundance provides a critical challenge for an understanding of the
diversity that existed at the earliest times.

\captionsetup[figure]{labelformat=empty}
\begin{figure}[!htbp]
\begin{center}
\includegraphics[width=1.00\textwidth,angle=0, clip=true]{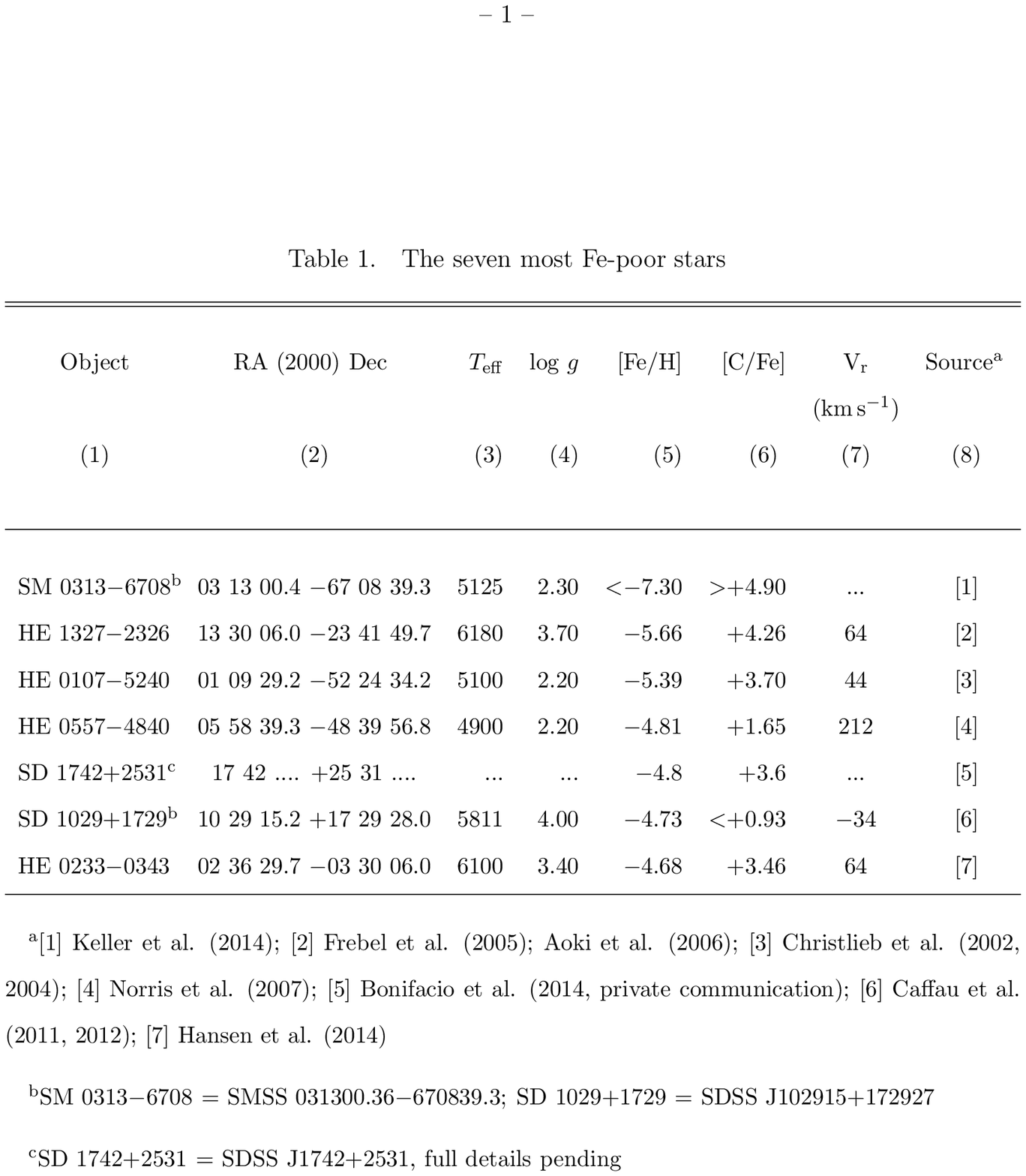}
  \caption{\scriptsize }
\end{center}
\end{figure}
\captionsetup[figure]{labelformat=default}

For comparison with the number of stars for which high-resolution,
high $S/N$ abundances are currently known at higher metallicity, we
estimate roughly that some 200 -- 300 stars have been discovered,
which have high-resolution estimates of [Fe/H] in the range
$-4.0<\mbox{[Fe/H]}<-3.0$.  Data are available for these, or soon will
be, suitable for analysis in a homogeneous and self-consistent manner.

To discover more stars with $\mbox{[Fe/H]} \lsim -4.5$ will prove a
challenge.  For the six stars in Table~1 with currently published
data, all but one are relatively bright, with $V < 15.5$. To find
more, such as the faintest of the six ({\sdc}) with $V = 16.7$
\citep{caffau11}, will require survey observations of larger volumes,
with follow-up high-resolution, high $S/N$ spectroscopy.

\begin{figure}[!htbp]
\begin{center}
\includegraphics[width=1.0\textwidth,angle=0]{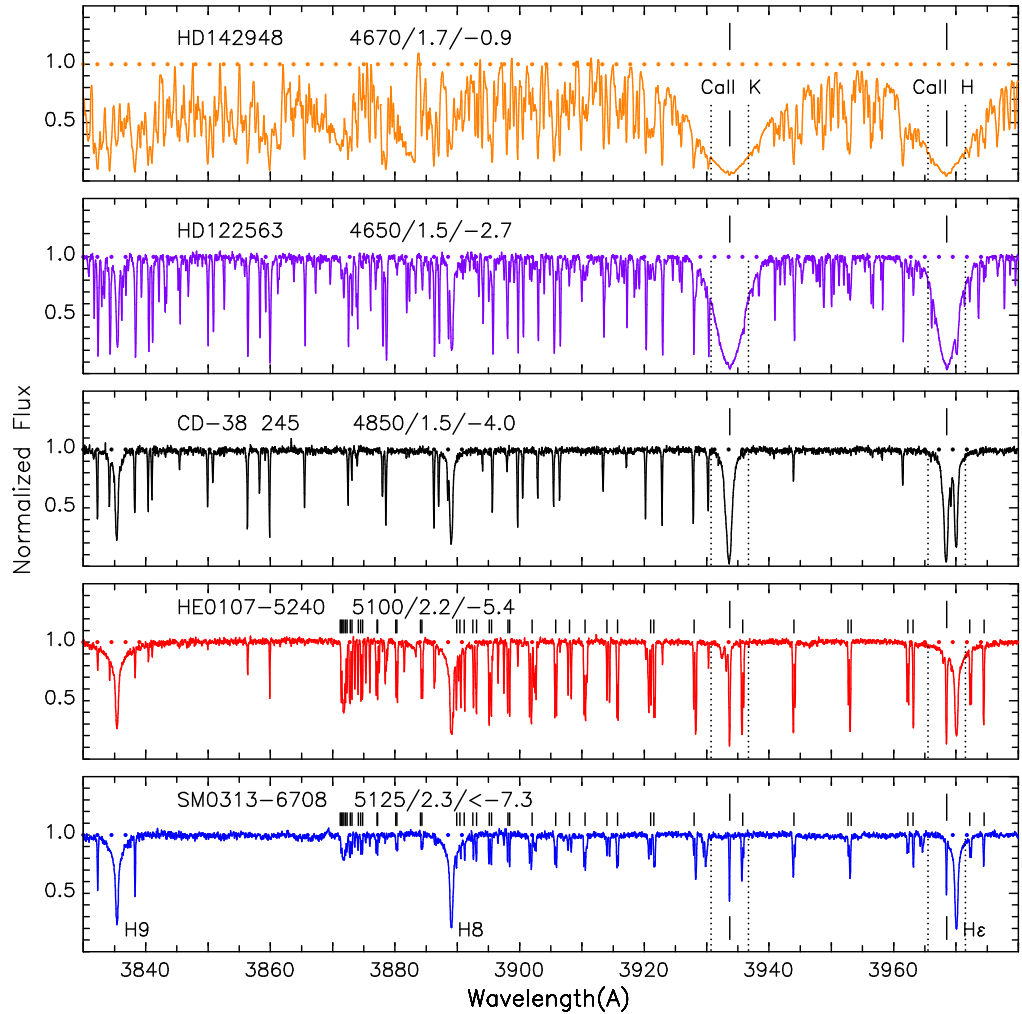}

  \caption{\label{fig:spectra_araa}\scriptsize High-resolution (R
    $\sim$ 40000), high $S/N$ spectra of metal-poor red giants having
    similar {\teff} and {\logg}, in the [Fe/H] range $<-7.3$ to $-0.9$
    and wavelength range 3830 -- 3980\,{\AA}.  The numbers in each
    panel to the right of the star's identification represent the
    atmospheric parameters {\teff}/{\logg}/[Fe/H], with metallicity
    decreasing from top to bottom.  Note that while the Ca\,II\,H and
    K lines become very weak in the two most iron-poor stars, {\hen}
    and {\smk}, many more lines have appeared.  These are features of
    CH (the positions of which are indicated immediately above the
    spectra) resulting from extremely large overabundances of carbon
    relative to iron in these objects.  For convenience, the vertical
    dotted lines are added to help delineate the positions of the H
    and K lines in the lower spectra.  Alternative representations of
    the spectra are provided in the Supplementary Material.}

\end{center}
\end{figure}

\subsubsection{Spectroscopic data} \label{sec:data} Figure~\ref{fig:spectra_araa} presents 
high-resolution, high-$S/N$ spectra of five metal-poor red giants, all
having approximately the same effective temperatures and surface
gravities, but with very different heavy element abundances,
decreasing as one moves from top to bottom in the figure.
(Atmospheric parameters {\teff}/{\logg}/[Fe/H] are also presented in
the figure.)  In the top panel, the spectrum of HD~142948, with
$\mbox{[Fe/H]} = -0.9$, is shown as an example of a relatively
metal-rich object, while each of the other four stars was the most
metal-poor object known at its time of discovery (see
Figure~\ref{fig:cah_epo}).  Alternative representations of these
spectra, together with their corresponding two-dimensional rainbow
colored spectra (for both metal-poor giants and near main-sequence
turnoff stars) are provided in the Supplementary Material.

There     are     two     important     points    to     take     from
Figure~\ref{fig:spectra_araa}.  First, as most clearly seen in the top
three  panels,  the  strength  of  the features  in  the  spectra,  in
particular the Ca\,II\,H  and K lines, decrease markedly  as one moves
from  top ($\mbox{[Fe/H]} =  -0.9$), to  middle ($\mbox{[Fe/H]}=-4.0$)
panel.   The  second point  is  somewhat  more  subtle, but  extremely
important for the discussion that  will follow.  As one moves from the
middle to  the bottom panel of  the figure, while the  Ca\,II\,H and K
lines clearly continue to weaken, the pattern of the weak lines in the
spectra changes in character.  This is  driven by the fact that in the
bottom two panels, lines of the CH molecule become stronger because of
enormous overabundances of carbon, not present in the stars in the top
three   panels.    Said   differently,   {\hen}   and   {\smk},   with
$\mbox{[Fe/H]} = -5.4$ and  $<-7.3$, have enormous relative abundances
of carbon,  with $\mbox{[C/Fe]} = 3.7$ and  $>+4.9$, respectively.  As
noted when introducing Table~1,  large overabundances of C relative to
Fe  are  a common  feature  of  the most  Fe-poor  stars.   This is  a
fundamental result, which we shall discuss at length in what follows.

\subsection{The Metallicity Distribution Function of the Halo}\label{sec:mdf}

Metallicity distribution functions (MDF) provide essential constraints
on galactic chemical enrichment models, and in the present context on
the nucleosynthetic yields of the first stars and the various putative
populations that existed at the earliest times.  Strictly speaking,
the term ``Metallicity'' Distribution Function refers to the
distribution of Z, the fraction of all elements heavier than lithium.
In practice, however, a more useful distribution function (both
observationally and theoretically) is Z$_{i}$, where ``i'' refers to
the element ``i''.  Observationally, in most cases i refers to Fe,
given that iron is one of the most abundant and readily measurable
elements in stars.  It is also observed to be closely related in its
abundance proportion to that of Ca, which is of critical importance as
one goes to lowest abundance, since the Ca\,II\,K line is
intrinsically much stronger than any of the lines of Fe I.  Accordingly, the
K line may be used more efficiently for discovery purposes, and at the
lowest metallicities, when Fe is no longer detectable in the spectra
of stars, for the determination of their chemical abundance.

In Section~\ref{sec:data} we discussed the increasing
disconnect between carbon and iron as one goes to lower and lower
abundances.  For this reason, the Carbon Distribution Function is also
of critical importance.  As we shall see in Section~\ref{sec:cno},
oxygen is important as well, but very difficult for practical reasons
to determine.  This point highlights the present impracticability of
the use of $Z$ as the independent variable for construction of an MDF,
given that not all of the major contributing elements are currently
measurable.  The following discussion thus addresses the distribution
functions of only Fe and C.

From a semantic point of view, we recall from Section~\ref{sec:defs}
that the term ``metal-poor''star is not necessarily synonymous with
``Fe-poor'' star.  It is also worth noting that when it comes to
relating chemical composition to age, it is not obvious which of two
stars having the same value of [Fe/H], but one carbon-rich (C-rich;
$\mbox{[C/Fe]}> +0.7$) and the other carbon-normal (C-normal;
$\mbox{[C/Fe]} {\le} +0.7$), might be older. We shall return to this
point in Section~\ref{sec:twopops}.

In an earlier review of metal-poor stars \citep{fn13} we discussed
(Fe)MDFs for the Galactic globular clusters, high-proper-motion
samples (\citealt{ryan&norris91a}, \citealt{carneyetal96}), and the
medium-resolution abundance samples of the Hamburg/ESO Survey (HES)
(\citealp{schoerck}, \citealp{lietal10}), to which we refer the reader
for background information.  There are two important points that
should be emphasized concerning the efficacy of these previous MDFs in
the context of low metallicities.  The first is that of these samples,
only the HES reaches well below $\mbox{[Fe/H]}<-3.0$.  The second is
that below $\mbox{[Fe/H]}\lsim -4.0$ at the low spectral resolution of
$R \lsim 2000$ adopted in these investigations it is difficult (if not
impossible) to determine reliable Fe abundances because of the
weakness of the lines, the contamination of CH lines, and an
occasional interstellar Ca\,II line in the spectra of these objects.
One needs high resolution to meet these challenges.  For these
reasons, we shall concentrate on recent high-resolution, high-$S/N$
analyses, and for which C and Fe abundances are now available.

\begin{figure}[!htbp]
\begin{center}
\includegraphics[width=0.8\textwidth,angle=0]{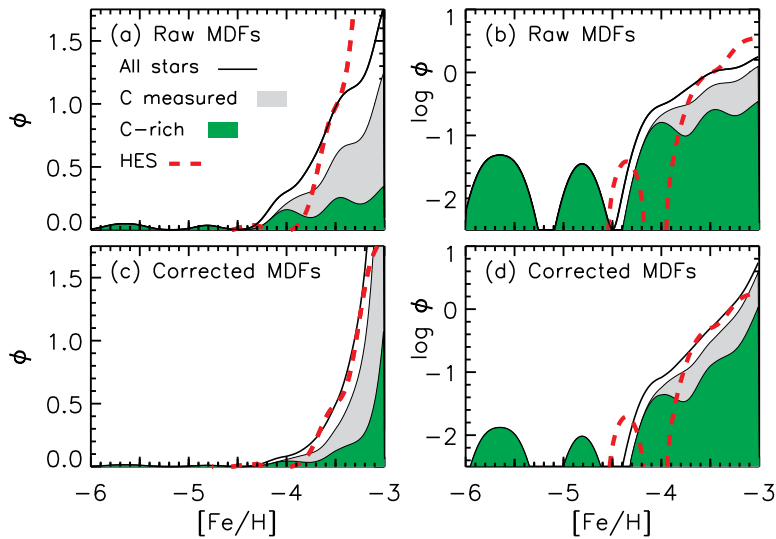}

  \caption{\label{fig:mdf} \scriptsize {The Fe metallicity
      distribution function, based on the high-resolution, high-$S/N$
      homogeneous abundance analysis of \citet{yong13a}.  The
      generalized histograms have been generated using a Gaussian
      kernel having $\sigma = 0.30$\,dex, and are presented here on
      linear (left) and logarithmic (right) scales.  Green and grey
      color-coding is used to present the contribution of C-rich and
      C-normal stars for which measurement was possible, respectively.
      The upper and lower panels refer to the raw data and those
      following completeness corrections on the range
      $-4.0<\mbox{[Fe/H]}< -3.0$, as described by \citet{yong13b}.
      The dashed line shows the HES MDF based on the data of
      \citet{schoerck}. Credit: D. Yong.}}

\end{center}
\end{figure}

Figure~\ref{fig:mdf} shows the (Fe)MDF from the work of
\citet{yong13b} for a sample of 86 extremely metal-poor stars
($\mbox{[Fe/H]} <-3.0$), based on their homogeneous chemical abundance
analysis of high-resolution, high-$S/N$ data, from both their own
observations and those available in the literature.  In this sample,
32 stars have $\mbox{[Fe/H]} \le -3.5$, and there are nine with
$\mbox{[Fe/H]} \le-4.0$.  (We note for completeness that while three
of their stars have $\mbox{[Fe/H]} <-4.5$, currently seven such stars
are known (see Table~1).)

Figure~\ref{fig:mdf} also shows the relative contributions of C-normal
and C-rich stars to the MDF for stars with $\mbox{[Fe/H]}<-3.0$, where
the increasing role of C-rich stars at lowest [Fe/H] is clearly
seen. (See Section~\ref{sec:xca} for a discussion of carbon richness.)
For three metallicity bins in the range $-4.5 <\mbox{[Fe/H]}<-3.0$
(with medians $\mbox{[Fe/H]}\sim-3.1$, $\mbox{[Fe/H]}\sim-3.4$ and
$\mbox{[Fe/H]}\sim-3.8$) containing roughly equal numbers of stars,
\citet[see their Figure 7]{yong13b} report increasing C-rich fractions
of 0.22, 0.32, and 0.33, as [Fe/H] decreases, and that for
$\mbox{[Fe/H]} <-4.5$ the fraction is 1.00.
More recently, \citet{placco14} have determined the C-rich fractions
using all available high-resolution abundances from the literature and
taking into account the decrease of the surface carbon abundance for
luminous stars on the upper RGB. Understandably, their
results yield higher fractions: 0.43, 0.60, 0.70, and 1.00,
respectively, for the median [Fe/H] values associated with the
  \citet{yong13b} fractions given above. For additional comparison,
  the uncorrected Placco et al. values are 0.32, 0.51, 0.65, and 1.00
  which are also higher than those of Yong et al. This clearly
  indicates that sample sizes and analysis methods still play a role in
  establishing these fractions. Homogeneously analyzed data of
  high-resolution samples, including abundance estimates of both C and
  Fe, and paired with correction for the surface carbon depletion, are
  needed to complete the picture. The up-coming and
on-going projects described in Section~\ref{sec:search} have the
potential to supply these data.  

We conclude by noting that the theoretical interpretation of the MDFs
will remain difficult.  Simple galactic chemical enrichment models are
of only limited usefulness.  We would argue that, at the very least,
theoretical models should address (i) the strong claims that the MDF
and C-rich fraction change with Galactocentric distance
(\citealp{frebel06}, Carollo et al. 2010, 2012, \citealp{an13}) and
(ii) the need for element yields which recognize that C and Fe are
decoupled for abundances below $\mbox{[Fe/H]} = -3.0$ (most likely
because of the existence of two different stellar progenitor
populations) before a meaningful comparison between theory and
observation could be achieved.

\subsection{Lithium and the Fragility of the Spite Plateau for Stars with [Fe/H] $<$ --3.0}\label{sec:lithium}

A fundamental prediction of the Big Bang paradigm is the result of
\citet{wagoneretal67} that ``$^{7}$Li [is] produced in a universal
fireball''.  SBBN, constrained by the results of WMAP \citep{WMAP},
predicts that the primordial lithium abundance is A(Li)$_{\rm P}$ =
2.72$_{-0.06}^{+0.05}$ \citep{cyburt08}. \citet{spite82} first
demonstrated that when the Li abundance of metal-poor, near
main-sequence-turnoff (MSTO) stars are plotted in the (A($^{7}$Li)
(hereafter A(Li)) vs. {\teff}) -- plane, the Li abundances appear
constant for those stars that lie in the effective temperature range
5500\,K $<$ {\teff} $< 6250$\,K.  They determined that the lithium
abundance of these stars is A(Li) $ = 2.05 \pm 0.16$.  

\begin{figure}[!t]
\begin{center}
\includegraphics[width=0.95\textwidth,angle=0]{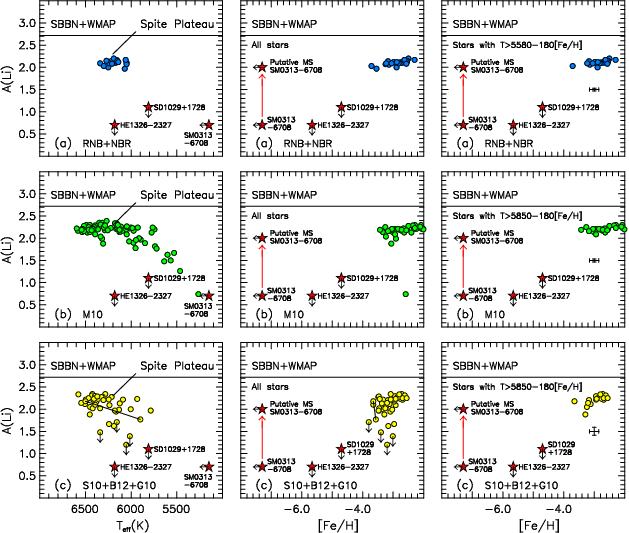}

  \caption{\scriptsize{\label{fig:li3}  A(Li) vs. {\teff} (left) and
    [Fe/H] (middle and right) for mear-main-sequence stars.  The
    circles represent data from: (a) \citet[RNB]{ryan99} and
    \citet[NBR]{norris2000} (1D/LTE), (b) \citet[M10]{melendez10}
    (1D/NLTE), and (c) \citet[S10]{sbordone10} (1D/LTE, \teff(IRFM)),
    \citet[B12]{bonifacio12a} (3D/NLTE), and \citet[G10]{gonzalez08}
    (1D/LTE).  ({\teff}-scale details are given when authors present
    multiple results.)  The Li abundances in the right panels are a
    subset of those presented in the middle panels, after exclusion of
    stars which may have experienced Li destruction, following
    \citet{melendez10}. (Here we use slightly different zeropoints for
    the {\mel} criterion to allow for differences in {\teff} scales.)
    The red star symbols refer to three stars with [Fe/H] $<$ --4.5,
    and are described in the text.  The most Fe-poor star, {\smk},
    with [Fe/H] $<$ --7.3 and A(Li) = +0.7, is connected by a long
    upward arrow to the Li abundance it may have had when on the main
    sequence.  See text for discussion.  The horizontal line in each
    panel represents the primordial lithium abundance. }}

\end{center}
\end{figure}

Since the discovery of the Plateau, there has been an enormous effort,
both observationally and theoretically, to refine and understand its
implications.  We refer the reader to \citet{thorburn94},
\citet{ryan99}, \citet{charbonnel05}, \citet{asplund06},
\citet{bonifacio07}, \citet{aoki09b}, \citet{melendez10},
\citet{sbordone10}, \citet{bonifacio12a}, and references therein, for
details of some of the observational contributions.  There are very
divergent conclusions among the interpretations of the derived lithium
abundances in these studies.  In Figure~\ref{fig:li3}(a) -- (c) we
present A(Li) for near-main-sequence stars (with one exception) as a
function of {\teff} (left column) and [Fe/H] (middle and right
columns) in order to illustrate some of the tensions among these
works.  The middle column presents all of the results from a given set
of authors, while on the right only those stars that satisfy the
\citet{melendez10} criterion {\teff} $> 5850$\,K $-
180\times\mbox{[Fe/H]}$, to take into account truncation of the cool
end of the plateau (by the convective destruction of Li in the stellar
outer layers), are plotted.  While the subject is extensive, we here
limit the discussion to three main points: (i) the most fundamental
result to be taken from Figure~\ref{fig:li3} is that the Spite Plateau
lies well below the primordial Li abundance predicted by SBBN/WMAP, by
a factor of $\sim3$ ($\sim0.5$\,dex).  Perhaps the most widely held
view is that, given the accuracy of the WMAP/SBBN primordial Li
abundance, the value obtained from analysis of observed Li line
strengths in near-MSTO metal-poor stars is not the primordial value,
and that an explanation of the difference will lead to a deeper
understanding of the astrophysics of stars.  See \citet{asplund06} and
\citet{fn13}, and references therein, for more detail; (ii) further
work is urgently needed to thoroughly explore and understand the
``meltdown'' of the Spite Plateau reported by Bonifacio, Sbordone, and
co-workers (Figure~\ref{fig:li3}c); and (iii) it is critical to limit
the putative {\teff} span of the Spite Plateau to ranges for which the
cool end does not includes stars that have experienced Li destruction
in their convective outer layers.

More relevant to the focus of the present review, we also plot in each
of the panels in Figure~\ref{fig:li3} the two most Fe-poor
near-main-sequence stars ({\hea} with $\mbox{[Fe/H]} = -5.7$ and
{\sdc} with $\mbox{[Fe/H]} = -4.7$; see Table~1).  Both stars are
extremely Li-poor, falling well below the Spite Plateau.  One should
thus be alive to the possibility that the Li was already depleted in
the material from which these stars formed.  That said, there are some
restrictions for a scenario in which such a star-forming cloud results
from the admixture of the ejecta of a supernova or a rotating massive star
into an existing interstellar medium having primordial Li. For
example, even if the resulting cloud comprised equal masses of these
two components, the first with no Li and the second with the
primordial value, the resulting star would have a lithium abundance
only 0.3~dex below the primordial value. A more far-reaching global
solution that may be relevant here is the suggestion of
\citet{piau06}, in the context of the primordial Li problem, that a
significant fraction of the earliest star forming clouds was processed
through Population\,III stars.

Also important in this context is the Li abundance of the red giant
{\smk}, also plotted in all panels of Figure~\ref{fig:li3}, which has
$\mbox{[Fe/H]} < -7.3$, {\teff} = 5125\,K, {\logg} = 2.3, and
$\mbox{A(Li)} = +0.7$ \citep{keller14}.  Given that evolution from
main sequence to red giant branch (RGB) destroys a considerable amount
of lithium in a star's convective envelope during its evolution on the
RGB, one might expect that the lithium abundance of {\smk} would have
been much higher when it was on the main sequence.  If, for example,
one uses the result of \citet{korn07} for the globular cluster
NGC~6397 ($\mbox{[Fe/H]} = -1.9$) as a guide, the Li depletion factor
between the MSTO and {\teff} = 5100\,K on the RGB is $\sim 1.3$\,dex,
which leads to a main sequence abundance for {\smk} of A(Li)$_{\rm
  MS}\sim 2.0$, not unlike the Spite Plateau values, as demonstrated
here schematically in Figure ~\ref{fig:li3}.  This is a rather
significant result, suggesting that the Spite Plateau may have existed
at the earliest times for low mass stars having [Fe/H] {\lsim}
--7.0. From stellar isochrones we expect that {\smk} was at the MSTO
relatively recently, some $\sim1$\,Gyr ago.  That is to say,
relatively little ``meltdown'' appears to have happened at the MSTO
and/or any putative truncation of the cool edge of the Spite Plateau
appears not to have reached the MSTO for stars having [Fe/H] $\sim$
--7.0 or less, at the earliest times.

A further basic result of Big Bang nucleosynthesis is that very little
$^{6}$Li accompanied the production of the initial $^{7}$Li (e.g.,
$^{6}$Li/$^{7}$Li \lsim 10$^{-4}$, \citealp{wagoner73}). Recent
predictions by \citet{coc12} yield $^{6}$Li/$^{7}$Li = 10$^{-4.6}$.
For a detailed discussion of the primordial $^{6}$Li/$^{7}$Li ratio
and corresponding observations, we refer the reader to the thorough
3D/NLTE analysis of four near-MSTO stars by \citet[see also references
  therein]{lind13}, who find that none of them has a significant
(2$\sigma$) detection of $^{6}$Li.

\subsection{An Overview of Relative Abundances -- [X/Ca] vs. [Ca/H]}\label{sec:xca}

In what follows we shall be primarily interested in stars having
$\mbox{[Fe/H]}<-3.0$.  To set the scene, we discuss relative
abundances, [X/Fe] and [X/Ca], obtained from high-resolution,
high-$S/N$, model atmosphere analysis, for a representative set of
elements in the metallicity range $\mbox{[Fe/H]}\lsim -2.0$.

As discussed in previous sections, one of the most intriguing, and
potentially most important, results concerning the abundance
signatures of metal-poor stars is the high incidence of stars with
abnormally high relative carbon abundances [C/Fe], a fact that was
first appreciated by \citet{bps92}.  \citet{beers&christlieb05}, to
whom we refer the reader for more details, classified C-rich objects
having $\mbox{[C/Fe]}> +1.0$ into a number of subclasses of {\bf
  C}arbon {\bf E}nhanced {\bf M}etal {\bf P}oor (CEMP) stars by using
the following taxonomy involving the relative abundances of the heavy
neutron-capture elements: CEMP-s (s-process element enhancement),
CEMP-r (r-process enhancement), CEMP-r/s (both r- and s-
enhancements), and CEMP-no (no enhancement of either s- or r-process
elements).  The CEMP-s subclass represents a large fraction of the
C-rich population, the chemical enrichment of which is almost
certainly driven by mass transfer across a binary system containing an
erstwhile asymptotic giant branch (AGB) star (e.g., \citealp[and
  references therein]{lucatello2005}).  Such stars are, however,
extremely rare below $\mbox{[Fe/H]}=-3.0$ (\citealp{aoki10},
\citealp{norris13}), and given that this is the abundance range under
discussion in the present review, we exclude the CEMP-s subclass by
rejecting C-rich stars that have $\mbox{[Ba/Fe]} > 0$.  Stars of the
CEMP-r and CEMP-r/s subclasses, which mainly have $\mbox{[Fe/H]} \gsim
-3.0$ were also excluded.  We thus expect the C-rich stars in the
following figures to belong to the CEMP-no subclass. In the range
$-4.0 <\mbox{[Fe/H]}<-2.0$, the majority of stars have relative
abundance ratios that do not differ wildly from the well-defined
trends of the bulk of halo stars, which may be regarded as ``normal''.
This is, however, not the case for the some 20 -- 30\% of stars that
are C-rich.  For $\mbox{[Fe/H]}<-4.0$, most of the stars are C-rich.

\begin{figure}[!t]
\begin{center}
\includegraphics[width=0.96\textwidth,angle=0]{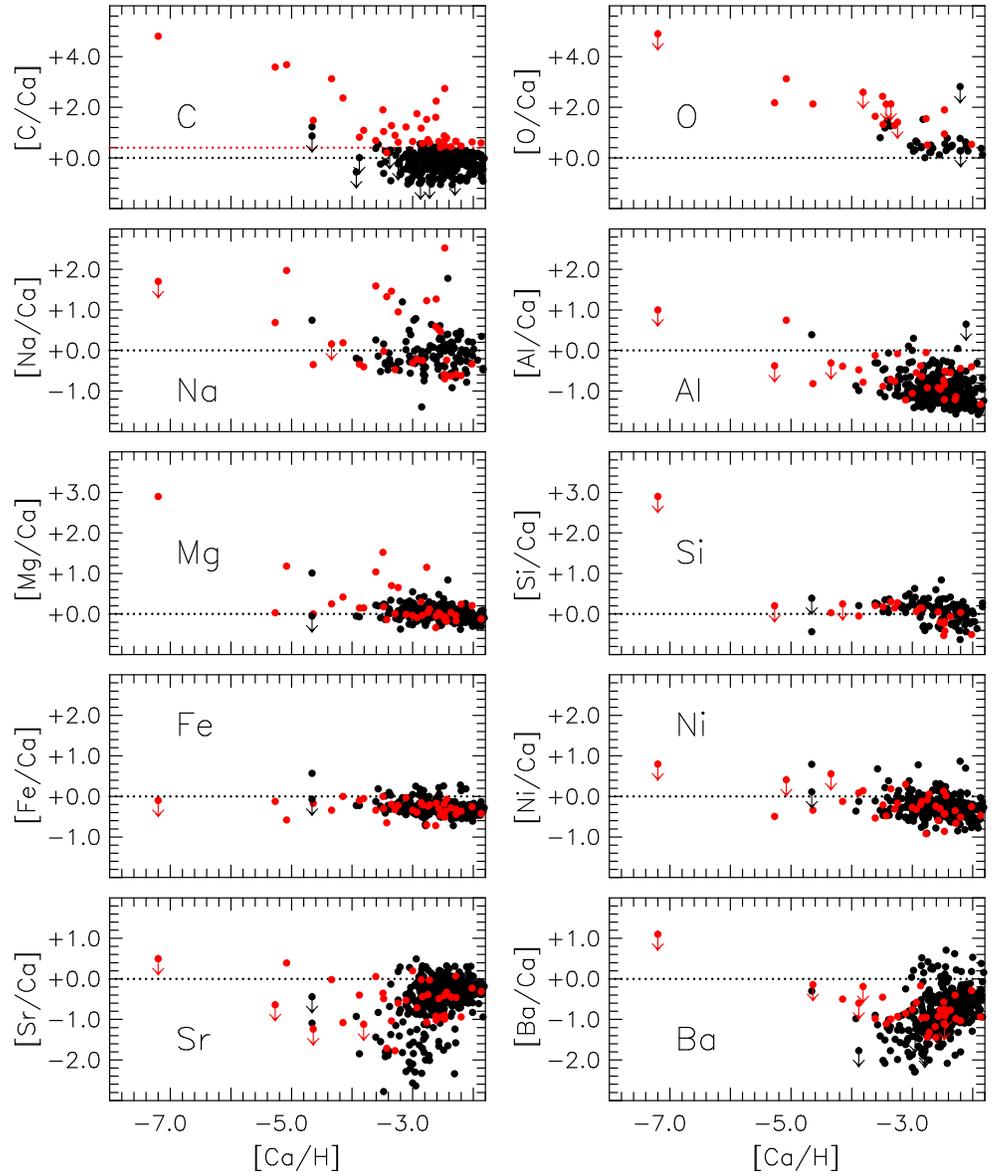}
\caption{\label{fig:xca}\scriptsize Relative abundances [X/Ca] as a
  function of [Ca/H], where red and black symbols refer to C-rich and
  C-normal stars, respectively.  From top to bottom the five pairs of
  panels are representative of the CNO elements, the light odd-elements,
  the $\alpha$-elements, the Fe-peak, and the heavy neutron-capture
  elements, respectively. In plotting the diagram the C-normal stars
  were plotted first, followed by the C-rich objects.  (The data have
  been taken from \citealt{frebel10}, \citealt{yong13a},
  \citealt{norris13}, \citealt{cohen13}, \citealt{hansen14}, and
  \citealt{keller14}.) }
\end{center}
\end{figure}

Figure~\ref{fig:xca} presents the dependence of [X/Ca] as a function
of [Ca/H] over the range $-8.0 < \mbox{[Ca/H]} < -2.0$. Normally, the
abscissa of choice in this sort of diagram is [Fe/H], and in the
sections that follow we shall revert to this convention.  That said,
we choose to use [Ca/H] as abscissa in Figure~\ref{fig:xca} in order
to highlight the newly discovered star {\smk} \citep{keller14}. With
$\mbox{[Ca/H]} = -7.2$, this star is so metal poor that only four
elements other than hydrogen (Li, C, Mg, and Ca) have currently been
observed in its spectrum. Red and black symbols in
Figure~\ref{fig:xca} are used to identify C-rich and C-normal stars,
here defined to have carbon abundances larger or smaller than
$\mbox{[C/Fe]} = +0.7$, respectively, following
\citet{aoki_cemp_2007}.  In plotting the diagram, the C-normal stars
were set down first, followed by the C-rich objects.  From top to
bottom, the five rows in the figure contain pairs of elements
representative of different nucleosynthesis processes: (i) C, O; (ii)
the light odd-elements Na, Al; (iii) the $\alpha$-elements Mg and Si;
(iv) the iron-peak Fe and Ni; and (v) the heavy neutron-capture
elements Sr and Ba.  Here are some highlights of these panels.

\begin{itemize}

\item

In the top-left panel, the upper dotted (red) line delineates the
boundary between C-rich and C-normal stars ($\mbox{[C/Fe]} = +0.7$,
and assuming $\mbox{[Ca/Fe]} = 0.4$). As one proceeds to lower values
of [Ca/H] (and [Fe/H]), the fraction of C-rich stars, as well as the
enhancement of C relative of the heavier elements Ca and Fe, increase
in comparison with the values found in C-normal stars.  Below
$\mbox{[Ca/H]} \sim -4.0$, almost all stars are C-rich, with enormous
relative overabundances $\Delta$[C/Ca] (and $\Delta$[C/Fe]), of order
1 -- 4\,dex, relative to the values found in C-normal stars.

\item

For the light elements O -- Mg, there are also very large relative
overabundances in the C-rich stars for [X/Ca] below $\mbox{[Ca/H]} =
-4.0$, with values up to 1 -- 2\,dex higher than those found in
C-normal stars.

\item

In contradistinction to what is seen for C -- Mg, in the range Si -- Ni
there appear to be no major differences between the relative
abundances [X/Ca] of C-rich and C-normal stars as a function of
[Ca/H].

\item

There is a large spread in [Sr/Fe] and [Ba/Fe] at all values of
[Ca/H], with no obvious dependence on carbon abundance. 

\end{itemize}

We shall discuss the role of these elements in more detail in
Sections~\ref{sec:co} and \ref{sec:dwarfs} in our
quest to understand what their patterns in the Milky Way's halo and
dwarf galaxy satellites have to tell us about conditions and
origins at the earliest times.

\subsection{The C-rich and C-normal Populations with $\mbox{[Fe/H]}<-3.0$}\label{sec:co}

\subsubsection{Carbon, Nitrogen, and Oxygen}\label{sec:cno}

\begin{figure}[!htbp]
\begin{center}
\includegraphics[width=1.0\textwidth,angle=-90]{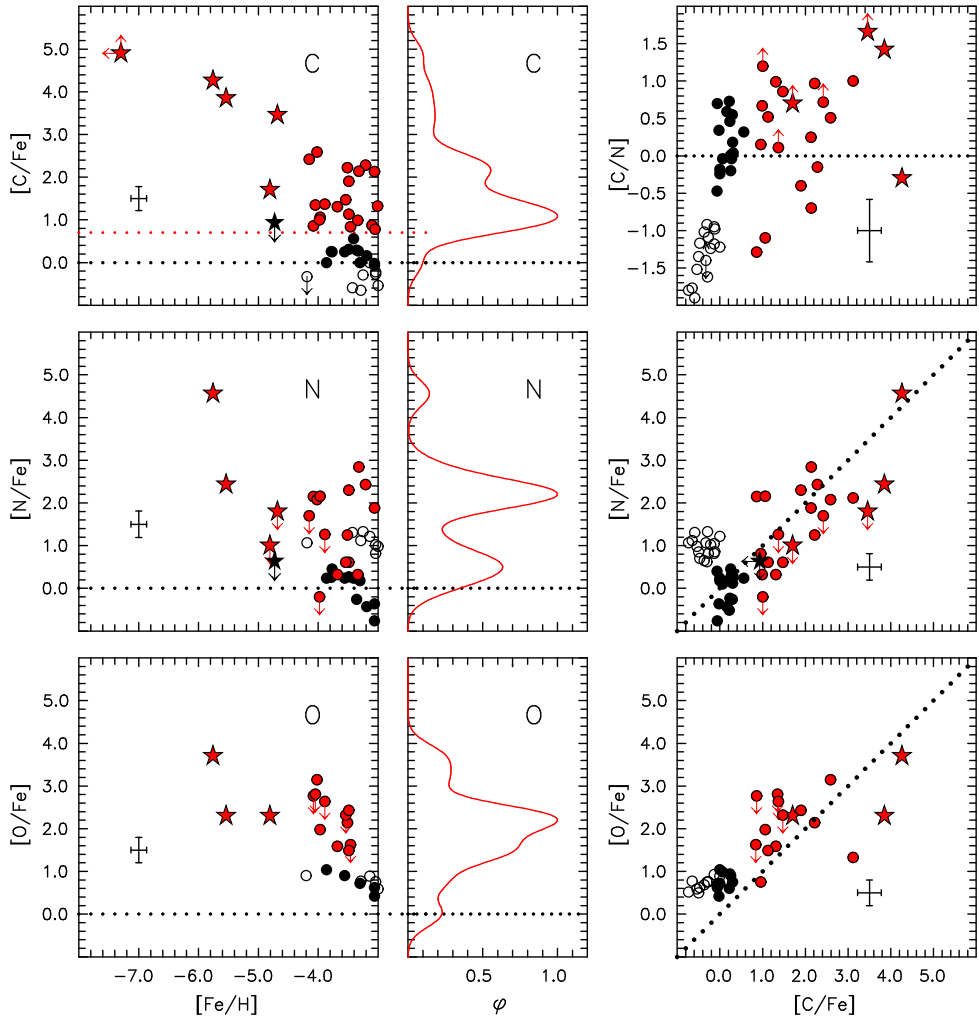}

\caption{\label{fig:cno}\scriptsize [C/Fe], [N/Fe], and [O/Fe] as a
  function of [Fe/H] (left) and [C/Fe] (right).  Red and black symbols
  refer to C-rich (excluding CEMP-s, CEMP-r, and CEMP-r/s sub-classes)
  and C-normal stars, respectively, while circles and star symbols
  stand for objects with [Fe/H] above and below $-4.5$. The middle
  column contains generalized histograms pertaining to the abundances
  to the left (Gaussian kernel, with $\sigma = 0.30$).  (Data from:
  \citealp{cayrel2004}, \citealp{spite05}, \citealp{siva06},
  \citealp{caffau12}, \citealp{yong13a}, \citealp{norris13},
  \citealp{cohen13}, \citealp{hansen14}, and \citealp{keller14}).  The red
  dotted line in the upper-left panel is the boundary between C-rich
  and C-normal stars adopted in the present work.  See text for
  discussion. }

\end{center}
\end{figure}

Figure~\ref{fig:cno} presents the behavior of C, N, and O, as
functions of [Fe/H] and [C/Fe], for stars in the range $-8.0
<\mbox{[Fe/H]}<-3.0$.  We are interested here, again, to highlight the
differences between the C-rich and C-normal stars.  The black symbols
represent C-normal stars, where open and filled circles are used to
denote the ``mixed'' and ``unmixed'' red giant stars of
\citet{spite05}.  The red symbols stand for C-rich stars (excluding
CEMP-s, CEMP-r, and CEMP-r/s sub-classes).

The leftmost panels show [C/Fe], [N/Fe], and [O/Fe] as a
function of [Fe/H], while the middle panels present the generalized
histograms of the abundances of these elements in the C-rich stars.
On the right, [C/N], [N/Fe], and [O/Fe] are plotted as a function of
[C/Fe].  

The important point to take from these data is that for the C-rich
stars in which oxygen has been measured this element also shows large
overabundances, commensurate with those determined for carbon.  (We
note for completeness that estimates of the oxygen abundance are not
available for several of the stars in Figure~\ref{fig:cno}.  While, in
part, this may be due to the greater difficulty of measuring the
abundance of O in comparison with that of C, it could in principle be
due to lower values of [O/Fe] than might be expected from the
correlation seen in the figure. Further investigation is necessary to
constrain this possibility.)  For nitrogen, on the other hand, there
exist both stars with large enhancements, and others with only small
excesses; its generalized histogram is suggestive of bimodality.

This behavior of the C-rich stars is in some contrast to the results
seen in Figure~\ref{fig:cno} for the C-normal red giants (filled and
open black circles), where one sees a bimodal behavior of the panels
involving C and N, particularly evident in the top-right panel.
\citet{spite05} have convincingly argued that this effect is driven by
mixing, on the RGB, of material that has been processed by the
CN-cycle, from interior regions into the outer layers of the observed
stars.  For the C-rich stars, on the other hand, which comprise not
only giants, but also dwarfs such as {\hea} (see Table~1), which has
$\mbox{[Fe/H]} = -5.7$, $\mbox{[C/Fe]} = 4.3$, $\mbox{[N/Fe]} = 4.6$,
and $\mbox{[O/Fe]} = 3.7$, the presence of such large overabundances
of all of C, N, and O is suggestive of the need of more than just the
CN-cycle.  These results have been discussed in more detail by
\citet{norris13}, to whom we refer the reader.  These authors also
investigate the behavior of $^{12}$C/$^{13}$C in the C-rich stars as a
function of [Fe/H], [C/Fe], and [C/N], and note ``one sees perhaps the
suggestion of a positive correlation between $^{12}$C/$^{13}$C and
[C/N], in the sense that would be expected from the processing of
hydrogen and carbon in the CN cycle. The large values of [C/Fe] seen
in [the ([C/Fe] vs. [Fe/H]) -- plane], however, suggest that, if this
were the case, one would require two processes, involving not only the
CN cycle, but also helium burning as well.''

\subsubsection{The Lighter Elements -- Sodium through Calcium}\label{sec:naca}

\begin{figure}[!htbp]
\begin{center}
\includegraphics[width=1.0\textwidth,angle=0]{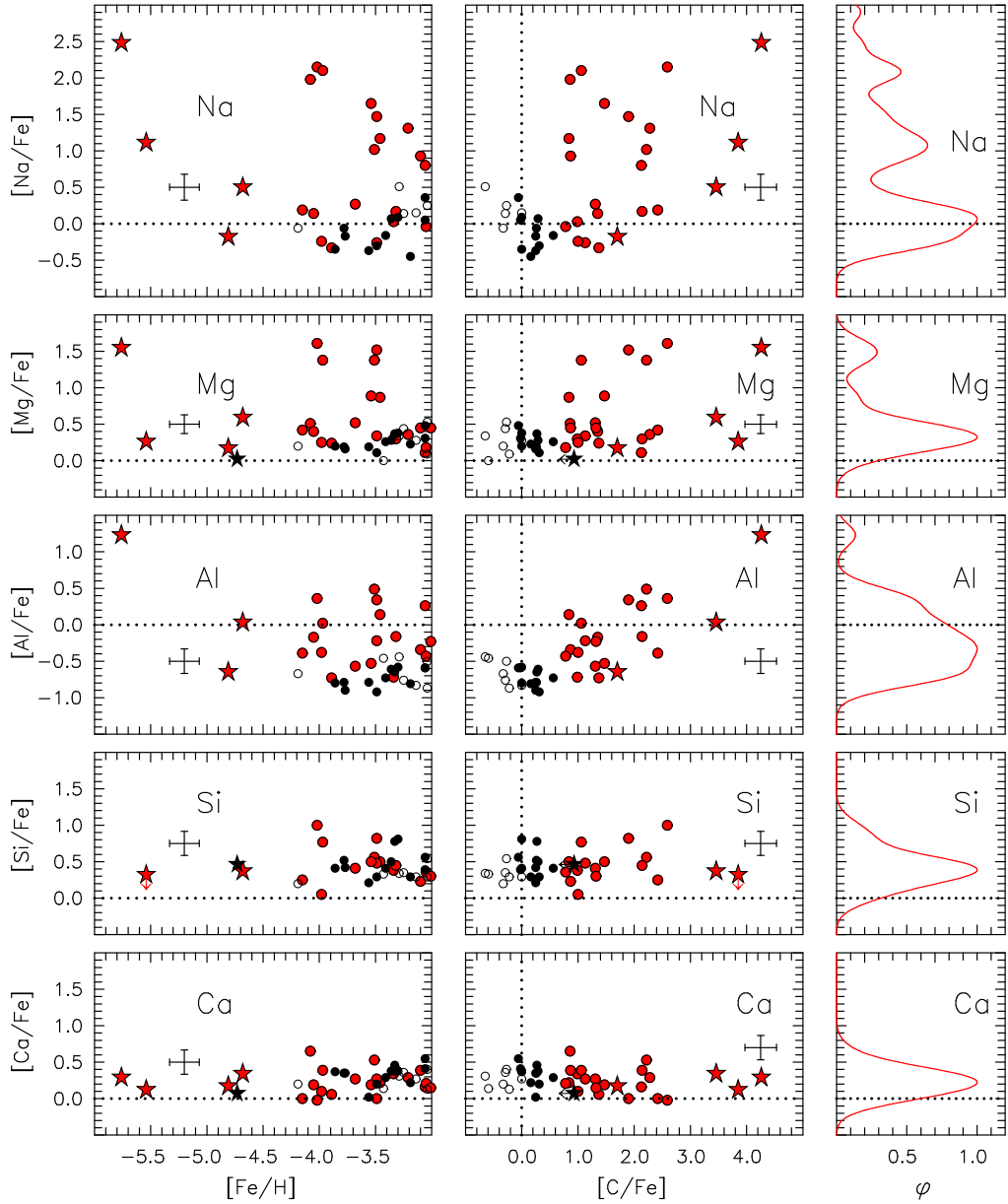}

  \caption{\label{fig:naca}\scriptsize The relative abundances of the
    light elements Na, Mg, Al, Si, and Ca versus [Fe/H] (left) and
    [C/Fe] (middle) for C-rich and C-normal Galactic halo stars.  The
    symbols and data sources are the same as in Figure~\ref{fig:cno}.
    The right column contains generalized histograms pertaining to the
    abundances of the C-rich stars in the left panels (Gaussian
    kernel, with $\sigma = 0.15$).  See text for discussion.}

\end{center}
\end{figure}

The abundances of Na -- Ca contain essential information for the
understanding of the C-rich stars with $\mbox{[Fe/H]}<-3.0$.
Figure~\ref{fig:naca} presents [Na/Fe], [Mg/Fe], [Al/Fe], [Si/Fe], and
[Ca/Fe] as a function of [Fe/H] and [C/Fe] for the C-rich and C-normal
stars of the Galactic halo, together with generalized histograms for
the former.  (With the exception of {\hen}, abundance limits are not
included.)  The data sources are the same as described for
Figure~\ref{fig:cno}.

There are a number of important points to be taken from
Figure~\ref{fig:naca}.  First, while the data for the C-normal stars
show low dispersions for all elements, there are large spreads in Na,
Mg, and Al for the C-rich stars, of order 1.5 -- 2.5\,dex.  Second,
there seem to be little, in any, spread for Si and Ca.  While the case
for this appears strong for Ca, the abundances for Si depend
essentially on only one line and the result may not be as strong.
Comparison of the present result with the apparent range and trend in
[Si/Fe] reported by \citet[their Figure~4]{norris13} suggests that the
larger data set and improved data quality now available demonstrate
that more work on this element is necessary and warranted. Our third
and final point is that not all C-rich stars exhibit large
overabundances of Na, Mg, and Al.  According to these authors only
about half show the effect.

We conclude the present section by noting that the abundance
signatures seen in Figures~\ref{fig:cno} and \ref{fig:naca} are
suggestive of the need of the admixing of material partially processed
by nucleosynthetic H-burning into regions experiencing He-burning.
Taken together, these data led \citet{norris13} to suggest that there
are two distinct stellar populations below $\mbox{[Fe/H]}=-3.0$: one
C-rich, the other C-normal.  We shall return to this topic in
Section~\ref{sec:twopops}.

\subsection{Unusual C-normal stars with $\mbox{[Fe/H]}\lsim -3.0$}\label{sec:diversity}

The diversity of chemical properties is not confined to C-rich stars.
Rare examples exist of individual (or small numbers of) extremely
metal-poor, non-CEMP, stars that have chemically unusual abundance
ratios.  An incomplete list of such stars, with $\mbox{[Fe/H]}
\lsim-3.0$, includes the following:

\subsubsection{Divergent $\alpha$-Element Abundances}\label{div}

\begin{itemize} 

\item
The Mg-enhanced BS~16934-002, with $\mbox{[Fe/H]} = -2.8$,
$\mbox{[Si/Fe]} = +0.44$, and $\mbox{[Ca/Fe]} = +0.35$, but
$\mbox{[Mg/Fe]} = +1.23$ \citep{aoki07b}.

\item
The $\alpha$-element challenged HE~1424$-$0241, with
$\mbox{[Fe/H]}=-4.0$ and $\mbox{[Mg/Fe]} = +0.44$, but $\mbox{[Si/Fe]}
= -1.01$ and $\mbox{[Ca/Fe]} = -0.44$ \citep{cohen07}.  Recently,
\citet[see their Figure~12]{cohen13} have identified a further nine
stars that share this characteristic, albeit at a lower significance
level than HE~1424$-$0241.

\item
The $\alpha$-element ambivalent SDSS J234723.64+010833.4, with
$\mbox{[Fe/H]} = -3.2$ and $\mbox{[Mg/Fe]} = -0.10$, but
$\mbox{[Ca/Fe]} = +1.11$ \citep{lai09}.

\item
A star, likely related to the above three (but which has only an upper
limit of $\mbox{[C/Fe]} <+1.7$) is HE~2136$-$6030: with $\mbox{[Fe/H]}
= -2.9$, it has $\mbox{[Mg/Fe]} = 0.08$, but $\mbox{[Si/Fe]} = +1.20$
  \citep{yong13a}.

\end{itemize}

\subsubsection{``Fe-rich'' stars}

Stars with a large number of elements having [X/Fe] lower by
$\sim0.3$\,dex compared with ``normal'' halo stars of the same [Fe/H],
such as CS\,22169-035, with $\mbox{[Fe/H]}=-3.0$ \citep{cayrel2004},
and HE~1207$-$3108, with $\mbox{[Fe/H]} = -2.7$ \citep[see their
  Figure 49]{yong13a}.

\subsubsection{Outlying abundances}

\citet{yong13a} and \citet{cohen13} have also examined their databases
for individual elements that appear significantly distinct from
``normal'' stars of the same [Fe/H].  In the range Na -- Ni,
\citet{yong13a} reported that $21 \pm 5$\% of stars were anomalous
with respect to one element, while $4 \pm 2$\% were anomalous with
respect to at least two.  \citet[see their Table 14]{cohen13} found a
similar result: ``Ignoring the C-stars, this leads to approximately
15\% of the sample being strong outliers in one or more elements
between Mg and Ni''.  

Further work is clearly necessary to understand the significance of
all of the above results.  One might expect that these
  outliers are most likely indicative of incomplete mixing of the
  ambient medium from which the stars formed, and contain clues
  concerning the nature of the ejecta, perhaps from only a limited part of the
  initial mass function of the previous enriching generation.

\subsubsection{Progeny of Pair-Instability Supernovae?}\label{sec:pisn}

One of the holy grails of metal-poor star research is the discovery of
stars showing the signatures of pair-instability supernovae (PISNe)
(see \citealp{heger2002}).  Very recently \citet{aoki14}
have reported the discovery of {\sdssa}, with $\mbox{[Fe/H]}=-2.5$,
which they suggest is such an object.  We shall defer consideration of
this object to Section~\ref{sec:firstnucleo}, following discussion of
theoretical modeling of the first stars.  

\newpage

\section{DWARF GALAXY ARCHAEOLOGY}\label{sec:dwarfgal}

\vspace{-0.3cm}
Over the last few decades, metal-poor halo stars have successfully
been used to study the early Universe. But without {\it a priori}
knowledge of where these stars actually formed and how they may have
entered the Galactic halo through their host system's accretion,
detailed information on their star forming environments remains
elusive. Studying the stellar content of the surviving dwarf galaxies
thus offers the opportunity to learn about ancient stellar systems.
Moreover, this offers the opportunity to investigate whether there
were particular conditions present in both these early systems and the
proto-Milky Way that led to the formation of extremely metal-poor
stars which are now found in the halo and the dwarf galaxies.  In
addition, reconstructing the early chemical evolution of these dwarf
galaxies sheds light on the star formation and supernova metal
enrichment processes within perhaps their first billion year. In this
manner, dwarf galaxy archaeology provides missing information
complementary to stellar archaeology. The challenge lies in obtaining
sufficient observational information of these faint and often sparsely
populated systems, which are at the technological limits of current
telescopes and instrumentation. In this section, we focus on the
discussion of the extent to which dwarf galaxy extremely metal-poor
stars are now being discovered and how the inferred chemical histories
compare with that of the halo.  Whether any of the surviving dwarfs
are analogs of the so-called ``Galactic building blocks'' will be
addressed in detail in Section~\ref{sec:ufd}.

\vspace{-0.5cm}

\subsection {The Discovery of Extremely Metal-poor Stars in the Dwarf Galaxy Satellites}

Given that the dwarf galaxies are at distances of $\sim20$ --
200\,kpc, their stars are faint (typically $V\gsim17$) and only the
red giants can currently be observed with high resolution
spectroscopy.  The fact that these systems are relatively compact,
however, has meant that large telescopes with multi-object
spectrographs have provided a very efficient way of obtaining larger
samples of stars at medium resolution to confirm membership and to
obtain [Fe/H] estimates, with follow-up at high resolution on somewhat
smaller samples of radial-velocity members.

Before the discovery of the ultra-faint galaxies (roughly $10^3 <
$L$_{\odot} < 10^5$), the medium-resolution efforts centered on the
relatively more luminous ``classical'' dSph systems (roughly $10^5 <
$L$_{\odot} < 10^8$). The Ca\,II infrared triplet was utilized to
obtain estimates of [Fe/H], employing a calibration based on globular
cluster metallicities.  These efforts led to the conclusion that such
systems contained no stars with $\mbox{[Fe/H]}<-3.0$ \citep{helmi06}.
However, after the first ultra-faint galaxies were discovered (see
\citealp[and references therein]{belokurov07}), \citet{kirby08}
developed a method involving the least squares fitting of
medium-resolution spectra (covering the range 6500 -- 9000\,{\AA})
against model atmosphere synthetic spectra to obtain abundances that
directly measured [Fe/H], and reported the existence of a total of 15
stars with $\mbox{[Fe/H]}<-3.0$ in seven ultra-faint systems. In
addition, \citet{norris08}, using blue medium-resolution spectra of
stars in the Bo\"otes\,I ultra-faint dwarf, reported a star with
$\mbox{[Fe/H]} = -3.4$ based on the Ca\,II\,K line.  These works
suggested that the calibration of the Ca\,II triplet in the
above-mentioned works on the dSph systems was unsafe for
$\mbox{[Fe/H]}< -2.5$.

Subsequently, the Ca\,II infrared triplet method has been revised
\citep{starkenburg10}. Stars with abundances as low as
$\mbox{[Fe/H]}\sim-4.0$ were soon discovered in the classical dSphs
Sculptor and Fornax (\citealp{scl}, \citealp{tafelmeyer10}), based on
both medium- and high-resolution techniques. Additional
discoveries are forthcoming. Extremely metal-poor stars are now
frequently compared with equivalent halo stars to investigate
differences and similarities  between the early
  nucleosynthesis histories of the dwarf galaxies and the halo.

\subsection{Early Chemical Evolution in the Dwarf Galaxies}\label{sec:dwarfs}

\begin{figure}[!t]
\begin{center}
\includegraphics[width=0.93\textwidth,angle=0]{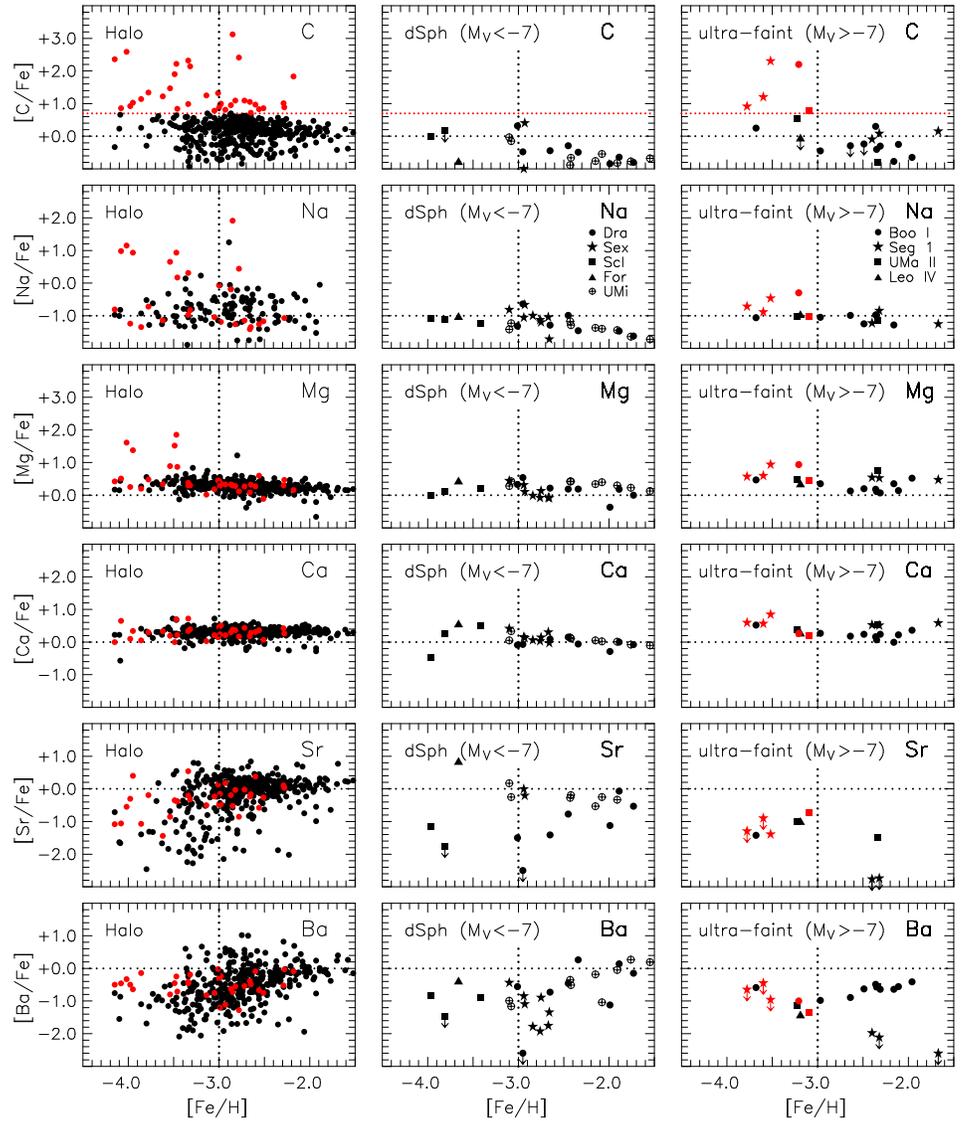}
  \caption{\label{fig:xfe}\scriptsize A comparison of the relative
    [X/Fe] vs. [Fe/H] between Galactic halo stars (left) and dSph
    (middle) and ultra-faint (right) dwarf satellites.  Red and black
    symbols refer to C-rich and C-normal stars, respectively.  In the
    left panels the data sources are the same as for
    Figure~\ref{fig:xca}, while in the middle and right panels the
    data have been taken from \citet{fulbright_rich}, \citet{norris08,
      norris10booseg, norris10seg}, \citet{aoki09}, \citet{cohen09,
      cohen10}, \citet{ufs, scl}, \citet{simon10},
    \citet{tafelmeyer10}, \citet{lai11}, \citet{gilmore13},
    \citet{frebel14}, and \citet{ishigaki14}.}
\end{center}
\end{figure}

Figure~\ref{fig:xfe} presents the relative abundances [C/Fe], [Na/Fe],
[Mg/Fe], [Ca/Fe], [Sr/Fe], and [Ba/Fe] as a function of [Fe/H].  To
keep the discussion more focussed, we restrict ourselves to dwarf
galaxies that contain extremely metal-poor stars
($\mbox{[Fe/H]}<-3.0$) for which carbon abundances are available.
Hence, only such stars are shown in the figure.  We refer the reader
to \citet{tolstoy_araa} and \citet{venn12} and their extensive
references on dwarf galaxy stars having high-resolution abundances for
elements other than carbon.  As in previous figures, red and black
symbols refer to C-rich and C-normal stars, respectively.  We also
include data for Galactic halo stars, and for ease of comparison, the
figure contains three columns which refer to the halo (left), dSph
systems (middle), and ultra-faint dwarfs (right).  While the dwarf
galaxy sample meeting the above criteria is still small relative to
that available for the halo, we make the following comparisons:

\begin{itemize}

\item

In large part, there appears to be a general similarity between the
lighter element abundance ratios. This implies that the chemical
enrichment history of the birth environments from which the extremely
metal-poor stars in both the Galactic halo and the various
  dwarf galaxies formed must have been driven in the main by massive
  stars (see \citealp{ufs} for details).

\item The metallicity range covered by the bulk of stars in the halo
  and in dSph is similar, reaching from $\mbox{[Fe/H]}\sim-4.0$ to
  $\mbox{[Fe/H]}\sim0.0$. The ultra-faint dwarfs, however, seem to
  lack stars with $\mbox{[Fe/H]}\gsim-1.5$ \citep{kirby08} which
  indicates some level of truncated chemical enrichment and star
  formation history. 

\item 

Considering the populations of both dSph and ultra-faint dwarfs, it
has been shown that as one goes to fainter luminosities in these
systems the mean value of [Fe/H] decreases while its dispersion
increases (\citealp{kirby08}, \citealp{norris10booseg},
\citealt{kirby11},  \citealt{leaman12}). For additional
details on the MDFs of these systems we refer the reader to the first three
references.

\item 

C-rich stars exist in both the halo and the ultra-faint dwarfs,
  but not (at least so far) in the dSph. It is important to note that
  there are commensurate numbers (albeit small) of extremely
  metal-poor stars in both the ultra-faint and dSph systems.  We shall
  discuss this result in Section~\ref{sec:ufd}.

\item 

For the $\alpha$-elements, a distinctly different behavior is known to
occur in the classical dSph systems, at higher values of [Fe/H].
There, solar values and even sub-solar values of [$\alpha$/Fe]
predominate (see \citealp[and references therein]{tolstoy_araa}). This
behavior is readily understood in terms of a slower chemical evolution
in the dSph galaxies compared with that found in the halo.  As the
result of this longer timescale, Type\,Ia supernovae have enriched the
gas with additional iron but at somewhat lower metallicity compared
with the halo, resulting in a decrease in the [$\alpha$/Fe] ratio. We
note in passing that low [$\alpha$/Fe] ratios are occasionally found
in the extremely metal-poor regime (see also Section~\ref{div}) which
can be explained with core-collapse supernova and/or hypernova yields
when preferentially sampling the lower mass end of the massive star
spectrum (\citealt{mcwilliam13}, \citealt{kobayashi14}). In contrast,
at least up to $\mbox{[Fe/H]}\sim-1.5$, the ultra-faint dwarf galaxy
stars show halo-like [$\alpha$/Fe] ratios.  More stars at higher
metallicities in the ultra-faint systems would be required to test if
they show the same behavior. Since, however, these systems seem to
contain hardly any such stars, this is a new kind of challenge beyond
that of a purely observational/technical one.

\item 

For the neutron-capture elements, generally lower abundance ratios are
observed in dwarf galaxy stars than in the halo. We shall discuss this
interesting behavior further in the following section.0*

\end{itemize}

\subsection{A Unique Dwarf Galaxy Signature: Low Neutron-Capture Element 
Abundances} \label{sec:ncaps}

\begin{figure}[!htbp]
\begin{center}
\includegraphics[width=0.95\textwidth,angle=0]{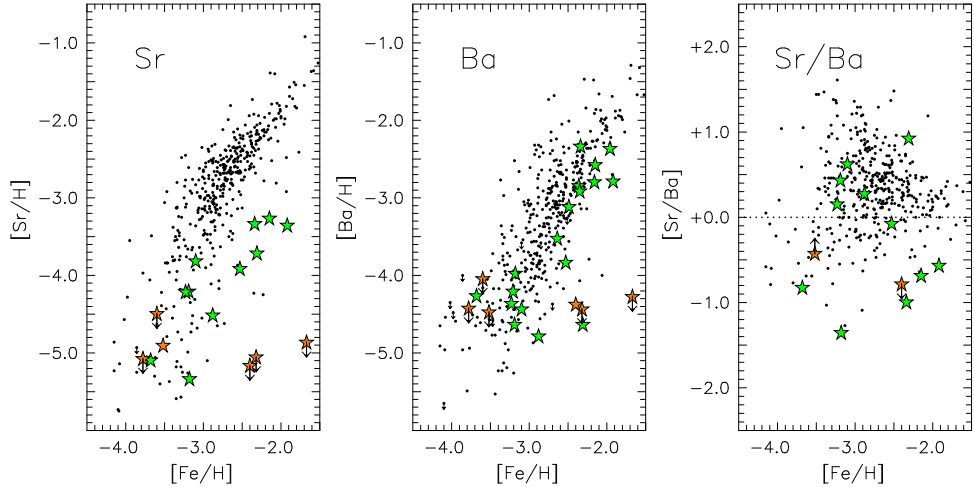}
  \caption{\label{fig:srba}\scriptsize [Sr/H] (left), [Ba/H] (middle)
    and [Sr/Ba] (right) as a function of [Fe/H] for the Galactic halo
    (small black circles) and its ultra-faint dwarf galaxies (star
    symbols: orange for Segue\,1, green for Bo\"otes\,I, Coma
    Berenices, Leo\,IV, and Ursa Major\,II).  The data have been taken
    from \citet{ufs}, \citet{simon10}, \citet{gilmore13},
    \citet{frebel14}, and \citet{ishigaki14}.  See text for
    discussion.}

\end{center}

\end{figure}

While there are numerous similarities between extremely metal-poor
stars in the halo and (in particular) the ultra-faint dwarf galaxies,
there also appears one curious difference which is worth examining
separately. It pertains to the behavior of heavy neutron-capture
element abundance ratios. Figure~\ref{fig:xfe} shows [Sr/Fe] and
[Ba/Fe] (which can be considered representative neutron-capture
elements) as a function of [Fe/H].  A potentially more insightful
comparison is that of [Sr/H] and [Ba/H] vs. [Fe/H] as shown in
Figure~\ref{fig:srba} for Galactic halo stars and the red giants of
the ultra-faint dwarf galaxies.  In this case, the axes in the panels
are nucleosynthetically decoupled. The production of neutron-capture
elements is, after all, independent of that of Fe and other lighter
elements \citep{sneden_araa}. Moreover, here one can more easily
follow the build up of neutron-capture element material as iron
increases.

As can be seen in Figure~\ref{fig:srba}, the halo exhibits
well-defined [Sr/H] and [Ba/H] trends for $\mbox{[Fe/H]}>-3.0$. Below
this value, the scatter increases and an overall trend is less well
defined and somewhat erratic, especially in the case of Sr. It is
important to note that in the range $-4.0<\mbox{[Fe/H]}<-3.0$, both
elements show variations of more than 3\,dex. That said, the
ultra-faint dwarf galaxy stars exhibit even lower neutron-capture
abundances.  That is to say, while they have values commensurate with
the lowest halo stars abundances, extreme values are found in these
systems at (nearly) all values of [Fe/H]. This behavior is more
extreme for Sr, where the ultra-faint dwarf galaxy stars appear to
form some kind of sub-population. Their [Sr/Ba] ratios (see
Figure~\ref{fig:srba}, right panel) also show a curiously distinct
pattern with values that are either somewhat offset from the main halo
trend (for stars with $\mbox{[Sr/Ba]} > 0$) or much lower than most
halo stars altogether (for $\mbox{[Sr/Ba]} < 0$).

Subsolar [Sr/Ba] ratios at low metallicity could be understood in
terms of a non-standard s-process operating in rotating massive stars
of a previous stellar generation. \citet{frischknecht12} found a
metallicity dependence of a (decreased) Sr and (increased) Ba net
production for models of such progenitors: [Sr/Ba] varied between
+2.05 and +0.42 for stellar models with $Z = 10^{-5}$, and between
+0.02 and $-0.54$ at $Z = 10^{-7}$.  The latter results are
qualitatively in agreement with the dwarf galaxy stars having
$\mbox{[Sr/Ba]} < 0$, although most of those observed [Sr/Ba] values
are even lower than $\sim-0.5$. Even the stars with
$\mbox{[Sr/Ba]} > 0$ may have resulted after $Z = 10^{-5}$ progenitors
enriched their birth gas clouds.

Interestingly, in Figure~\ref{fig:xfe}, the classical dwarfs paint a
somewhat intermediate picture. They appear to contain stars with
halo-like neutron-capture abundances, but also show examples of
extreme deficiency in these elements, especially in the case of Sr
(e.g., a star in Draco, see below). Perhaps this behavior illustrates
a transition from the earliest systems to the more evolved ones, such
as the most luminous dwarfs Sculptor and Fornax.

While additional element measurements are needed, the current body of
data already indicates that ultra-faint dwarf galaxy stars almost
exclusively have distinctly low neutron-capture element abundances,
and relatively low [Sr/Ba] values, whereas brighter systems and the
halo do not. In this context, it becomes interesting to speculate
about these observed differences and possible interpretations.
Perhaps the most obvious explanation would be that low neutron-capture
element abundances (i.e., lower than the typical halo trend) are a
signature of the earliest star forming clouds.

Since neutron-capture elements are produced independently of the
lighter elements (including Fe) and likely to occur in a different and
more restricted progenitor mass range (8 to 10\,M$_{\odot}$;
\citealt{wanajo_rs2006}), produced by the r-process or other
neutron-capture processes for stars with $\mbox{[Fe/H]}\lsim-2.8$.
(Below this value, general AGB stellar winds are not expected to
contribute to chemical evolution).  Hence, relatively little or even
no neutron-capture element material may have been created by the first
stellar generations, especially if most stars were more massive than
$\sim$10\,M$_{\odot}$. Observational examples might include the higher
metallicity stars in Segue\,1 with upper limits of $\mbox{[Sr,Ba/H]}$
of $\sim-5.0$ and $\sim-4.3$, respectively \citep{frebel14}, or the
Draco star D119 with $\mbox{[Fe/H]}\sim-3.0$ and upper limits of
$\sim-5.5$ for both elements \citep{fulbright_rich}. These upper
limits indicate no more than one production event (or even none) in
which these heavy elements may have formed
\citep{farouqi10}. Whether such an event originated in a
  supernova or a neutron-star merger is of ongoing discussion.

Applying this idea to halo star abundances leads to an hypothesis with
potentially far-reaching implications. Perhaps the halo stars (and
also classical dwarf galaxy stars) with the lowest neutron-capture
element abundances (and/or with the lowest [Sr/Ba] value) are those
that formed in the earliest gas clouds, similar to those that resulted
in extremely metal-poor stars present in the ultra-faint dwarfs. Later
gas clouds might have had additional neutron-capture element
enrichment but not necessarily increased Fe enrichment (or
vice-versa), due to different mass functions of the enriching
progenitors, or other factors. Such a scenario might be able to
explain the increased halo star scatter in Sr and Ba abundances at
$\mbox{[Fe/H]}<-3.0$. As a consequence, neutron-capture element
abundances might be suitable for discriminating halo stars that
originated in systems similar to the surviving faintest dwarf
galaxies, and thus even be able to separate out the ``oldest'' stars
formed in the earliest star forming regions for a given [Fe/H] value.
We note that in Section~\ref{sec:ufd}, we shall more broadly discuss
the role of neutron-capture elements as part of the question of
whether any of the surviving ultra-faint dwarfs could be a surviving
first galaxy.

It would be interesting to obtain abundances of additional
neutron-capture elements in any of the dwarf galaxy stars to further
test this idea. However, given their low abundances to begin with, and
often only moderate data quality due to the faintness of these stars,
this is a challenging task. Sr and Ba can be measured because they
exhibit relatively strong lines in the spectra while the other
neutron-capture elements exhibit much weaker features. Even in the
cases of Sr and Ba, limited spectral coverage has resulted in only a
few Sr measurements (the two lines are located in the harder to access
blue spectral regions at 4077\,{\AA} and 4215\,{\AA}). Fortunately,
several Ba lines further in the red region (reaching up to
$\sim6500$\,{\AA}) have led to Ba abundances in most dwarf galaxy
stars.

\newpage

\section{NEAR-FIELD COSMOLOGY}\label{sec:nearfield}

Observations of metal-poor stars in the Galactic halo and the dwarf
galaxies provide a wealth of information about the local chemical
abundance ratios at the time of star formation in the early
Universe. These local conditions are driven by cumulative effects,
such as those arising from the mass distribution of Population\,III
stars, the subsequent star formation rate, supernova yields, infall,
outflows, etc.  In this way, the assembly of the chemical elements can
be reconstructed for a given host system. Established chemical
evolution trends as well as any outliers, however, need to be
understood and interpreted based on physical principles of (supernova)
nucleosynthesis and the host galaxy's environment in which the
enrichment events occurred.  After all, the key to deciphering stellar
abundance signatures lies in unraveling how these nuclei were formed,
how they were expelled and mixed in the stellar birth gas cloud, how
they were incorporated into a star in its natal galaxy, and how that
star then ended up in the Galactic halo as we observe it today.

\subsection{Properties of the First Stars and the Nature of the First Chemical Enrichment Events}\label{sec:firststars}

Based on the current state of observational searches for the most
metal-poor stars, no truly metal-free first star has yet been
discovered.  While a few stars with exceptionally low Fe abundances
($\mbox{[Fe/H]}<-5.0$) have been found, their accompanying large
carbon abundances strongly suggest that they are some of the first
Population\,II stars to have formed in the Universe
(e.g., \citealp{UmedaNomotoNature}, \citealp{keller14}), rather than
being true Population\,III first stars.

These searches have shown that the most metal-poor stars are extremely
rare. Is the apparent absence of Population\,III stars due to
observational difficulties or is there a physical reason for their
absence? The nature and properties of the first stars have been much
debated over the past decade as simulations of these objects have
become more and more detailed (\citealp{abel_sci}, \citealp{bromm02},
\citealp{yoshida_bromm2004}, \citealp{yoshida08}, \citealp{turk09},
\citealp{stacy10}, \citealp{clark11}, \citealp{greif11},
\citealp{hosokawa11}, \citealp{turk12}, \citealp{susa13},
\citealp{stacy14}). The subject has been extensively reviewed by
\citet{brommARAA} and \citet{bromm13}, to whom we refer the reader for
a more in-depth discussion. Here we summarize the latest findings of
the field and highlight results relevant to an understanding of the
beginning of chemical enrichment and how metal-poor stars provide
constraints on the existence of Population\,III stars.

\subsubsection{The mass range of the first stars}   

Consensus exists that due to the lack of cooling agents in primordial
gas (i.e., metals or dust), significant fragmentation of the available
gas was largely suppressed so that these first objects must have been
very massive. Since the beginning of this field about 15 years ago,
however, there has been a slow but steady paradigm shift concerning
the typical mass scale of a Population\,III star. While early works
suggested single stars had masses of order 100\,M$_{\odot}$, the more
recent simulations that follow the formation of the protostar for a
much longer time to the important accretion phase, paint a more
complex and varied picture.

Recent detailed modeling of the formation process of Population\,III
stars (e.g., \citealt{hosokawa11}, who follow the formation of an
eventual 43\,M$_{\odot}$ star until the beginning of nuclear fusion)
has generally not yielded stars with masses above 100\,M$_{\odot}$.
Instead, following the accretion process which appears to be the main
determinant of the final mass of the Population\,III star, eventual
fragmentation of the accretion disk due to gravitational instabilities
can lead to the formation of small multiples (e.g., \citealp{turk09},
\citealp{stacy10}, \citealp{clark11}) in most simulated halos. As a
result, secondary stars with lower masses, of order 1\,M$_{\odot}$,
can form. The primary star, however, remains massive (at least a few
tens of M$_{\odot}$).  For example, \citet{susa13} obtained masses of
4.4\,M$_{\odot}$ and 60\,M$_{\odot}$ in their
simulation. \citet{stacy14} even found a system with just two
Population\,III stars of mass $<1$\,M$_{\odot}$ and
$\sim5$\,M$_{\odot}$, although in a very unusual environment with very
high angular-momentum content. The protostellar disk in this system is
thus unusually extended, with reduced fragmentation and small
protostellar accretion rates.

While the early results for exclusively very massive Population\,III
stars implied extremely short lifetimes (a few Myr), and hence, no
chance of their ever being observed, the prospect of Population\,III
stars with $M<1$\,M$_{\odot}$ and correspondingly long lifetimes
($>13$\,Gyr) has re-opened this possibility. It thus remains to be
seen whether a true low-mass Population\,III star will be
discovered. Efficient large-scale surveys such as that carried out
with SkyMapper will play a vital role in answering this important
question from an observational point of view, while simulations will
deliver more refined estimates concerning the possible formation
channel of these stars.

Knowledge about the lower mass end of Population\,III 
  stellar masses is going to be as important as establishing what the
overall mass function of the first stars may have been. This initial
mass function (IMF) -- the number of stars per mass bin -- is the most
important characteristic of this population and a vital ingredient for
essentially any interpretation or simulation associated with
Population\,III stars, early star and galaxy formation, chemical
enrichment and evolution. Knowledge of the IMF is particularly
important for stellar archaeology because stellar mass determines a
star's fate. This immediately implies that the IMF determines the
overall chemical output of this population and thus how the chemical
enrichment of the Universe began.

In summary, contrary to the present-day mass function, which is
dominated by low-mass stars, the Population\,III IMF was likely
top-heavy and dominated by massive stars. Due, however, to various
modeling difficulties and global challenges such as cosmic variance,
it remains poorly determined. That said, first attempts are now being
made to quantify the IMF by using large suites of
simulations. \citet{susa14} find a top-heavy IMF that peaks at several
tens of \,M$_{\odot}$, and with stars in the range of 10\,M$_{\odot}$
to 100\,M$_{\odot}$.  Similarly, \citet{hirano14} find a reasonably
flat distribution over the range of 10\,M$_{\odot}$ to
1000\,M$_{\odot}$, based on 100 first stars simulations. Only more
detailed and dedicated modeling will result in a reliable IMF
determination; but in the meantime, explorations on the low and high
mass ends will advance our understanding of the full range of
Population\,III stellar masses and what the corresponding implications
are.

For completeness, it should be noted that stars having masses of
thousands of M$_{\odot}$ are being predicted within the
cosmological framework. This follows earlier work on stars with
  $M\sim$ hundreds of M$_{\odot}$ (e.g.,
  \citealt{fryer01}). \citet{hosokawa13} simulated Population\,III
  stars with masses of $10^{4-5}$\,M$_{\odot}$ which are expected to
  collapse directly into a supermassive black hole. While not
  contributing to chemical enrichment, these monstrous objects might
  be luminous enough to be observable with JWST, and moreover, be the
  seeds of future galaxies' supermassive black holes.  In the same
  context, calculations have been made of the luminosity of the first
  massive Population\,III star cluster and its observability with JWST
  \citep{johnson10}. These massive objects might offer the only chance
  of ever directly observing Population\,III stars in the
  high-redshift Universe -- if such a suitable cluster, or massive
  star, ever existed.  \citet{johnson13b} then modeled a
  55000\,M$_{\odot}$ star that completely disrupts in a gigantic
  supernova with an explosion energy of up to $10^{55}$\,erg instead
  of collapsing to a black hole. This would presumably leave behind
  enormous instantaneous metal enrichment that would trigger the
  formation of metal-rich low-mass second-generation stars. These,
  however, would likely be too metal-rich to be found in surveys for
  metal-poor stars. Uncovering observational evidence of these
  behemoths will thus be rather difficult.

\subsubsection{First Star Nucleosynthesis}\label{sec:firstnucleo}

Stars with masses of 8\,M$_{\odot}$ or more will, in the main, end
their lives as supernovae.  Depending on the stellar mass and rotation
state, however, the stellar remnants may be vastly different. Some
explosions will lead to chemical enrichment while others will not
contribute elements at all.  We refer the reader to \citet{heger2002}
for an overview of the likely outcomes for non-rotating stars that
initially had heavy element abundance $Z = 0$.  As with the analysis
of stellar atmospheres, the large majority of modeling of the
evolution of stars is undertaken within the framework of one
dimensional analyses, based in this case on the assumption of
spherical symmetry.  We caution also that the modeling of final
explosive events, of necessity, makes fundamental and {\textit{ad
    hoc}} assumptions about, in particular, the energy of the
explosion and the mass-cut above which material is potentially ejected
(e.g., \citealp{nomoto13}).  In the present context, stars that die as
element-contributing pair-instability (PISNe) and core-collapse
supernovae are of prime interest for stellar archaeology and dwarf
galaxy archaeology.  Hence, we shall not discuss the neutron stars and
black holes that are also end products of stellar evolution.

Detailed calculations of the yields of $Z = 0$, non-rotating,
core-collapse supernovae, in the mass range 10 -- 40\,M$_{\odot}$ have
been presented, for example, by \citet{woosley_weaver_1995} and
\citet{kobayashietal06}. Galactic chemical enrichment models using the
above yields, together with additional ones covering $0 < Z <
Z_{\odot}$, have been used by these authors and others for comparison
with observations of Galactic metal-poor stars.

While the role of core-collapse supernovae in producing the observed
chemical abundance signatures in metal-poor stars is clear, the same
may not be said concerning pair-instability supernovae (PISNe),
predicted to be the end product of $Z = 0$, non-rotating, high-mass
Population\,III stars with $M\sim140$ -- 260\,M$_{\odot}$.
Calculations of the chemical yields of these objects predict a very
strong ``odd-even'' effect in their abundance patterns (large
differences between elements with neighboring atomic
numbers), together with enhancements of calcium (and Mg and Si)
relative to iron (see \citealp[their Figure~5]{heger2002}). Certainly,
until very recently no stars have been discovered with PISNe abundance
signatures.  \citet{karlsson08} have argued that observational
selection effects inherent in the use of the Ca\,II\,K line in the
discovery of metal-poor stars, have militated against detection of
metal-poor stars (with $\mbox{[Fe/H]}\lsim-2.5$) and PISN abundance
characteristics.  We note that systematic efforts are now underway by
\citet{ren12} to search for this type of signature.  This fundamental
question has remained unanswered for some time.

As we noted in Section~\ref{sec:diversity}, very recently
\citet{aoki14} have reported the discovery of an object, {\sdssa},
with $\mbox{[Fe/H]}=-2.5$ and unusually low, subsolar C, Mg and Co,
together with some  of the PISN odd-even pattern.  The critic would
note that while the  odd-even effect is clear for their values of Sc/Ti
and Co/Ni in this star, in reasonable agreement with their PISN model
calculations, this is not the case for Na/Mg, Al/Si, V/Cr, and Mn/Fe.
Indeed, an alternative model presented by them (in their Figure~2)
which includes chemical enhancement by both Type\,II (core-collapse)
and Type\,Ia supernovae ejecta, seems to us to provide a better fit to
the observations. (We would also note, for completeness, that their
dismissal of their model including Type\,Ia enrichment seems open to
question in light of the Type\,Ia-like signatures that have been noted
by \citealp{yong13a} for stars with $\mbox{[Fe/H]}\sim-2.5$
(CS\,22169-035 and HE~1207$-$3108).)

Finally, in the past 10 -- 15 years, driven in large part by observed
chemical abundance signatures of the kind discussed above in
Section~\ref{sec:archaeology}, theoretical attention has increasingly
focused at $Z = 0$, and in the mass range $25 - 60$\,M$_{\odot}$, on
the effects of (i) stellar rotation on nucleosynthesis
(e.g., \citealt{fryer01}, \citealt{meynet06}, and \citealt{meynet10}),
and (ii) ``mixing and fallback'' of the material above the mass-cut of
the supernova explosion and the resulting (non-canonical) ejecta (see
\citealt{nomoto13}).  That said, many details regarding, for example,
the masses and explosion energies involved are still being debated.
We shall return to the interpretation of the observed abundances of
the most metal-poor stars in Section~\ref{sec:abund_int}.

\subsection{The First Galaxies and the Formation of the First Low-mass Stars}\label{sec:firstgalaxies}

Early structure formation led to the existence of so-called
``minihalos''  with masses of $\sim 10^6$\,M$_{\odot}$ of
dark matter that collapsed at high redshifts of $z \sim 20 - 30$
\citep{tegmark97} entraining baryonic matter with it, about 100 to 200
million years after the Big Bang (e.g., \citealp{greif11}).  Each
minihalo hosted one first star (or a small number of them) which
evolved and provided chemical enrichment to the minihalo (depending on
the star's mass-dependent element yields, and other relevant
parameters, as outlined above).  

Throughout their brief life, massive stars provided vast amounts of
ionizing radiation that changed the conditions of the surrounding
primordial gas, including that in neighboring minihalos, and thus the
details of subsequent star formation. In particular, whereas cooling
in the first phase of star formation was provided by H${_{2}}$, in
this second phase cooling would be driven by HD leading to lower gas
temperatures and smaller stellar masses (e.g., \citealp{johnson06} and
\citealp{greif06}).  In this context, it is interesting to note that
in minihalos which had experienced no metal enrichment there could
have formed a generation of metal-free stars, some
having typical masses of order $\sim10$\,M$_{\odot}$.

As structure formation and Population\,III stellar evolution
proceeded, of order 10 such minihalos merged to form so-called
``atomic cooling halos'' (\citealt{wise08}, \citealt{greif08}) with
dark matter masses of $\sim 10^8$\,M$_{\odot}$ at redshifts of $z\sim
10 -15$. While molecular hydrogen was the primary coolant in
minihalos, now atomic hydrogen took over due to a larger virial
temperature of the gas ($\sim10^4$\,K, as opposed to $\sim$1000\,K in
minihalos). Metals were immediately present in these systems, brought
in from the merging minihalos that hosted core-collapse
Population\,III supernovae or stars that were about to explode.
Atomic cooling halos can be considered as first galaxies since they
were massive and metal-rich enough to host Population\,II star
formation -- including the formation of the first low-mass metal-poor
stars -- and able to withstand supernova explosions of Population\,III
and massive Population\,II stars and other feedback effects without
immediately disrupting. We refer the interested reader to the review
of \citet{bromm_araa11} for additional details on the first
galaxies. Alternatively, it may have been possible that at least some
of the first low-mass stars already formed in minihalos. Depending on
the strength of the explosion and mass ($<40$\,M$_{\odot}$) of the
Population\,III star in such a minihalo, the recovery time of the
system may have been rather short ($\sim10$\,Myr), leading to prompt
Population\,II star formation (\citealt{cooke14}, \citealt{jeon14}).

Simulations of first galaxies that model metal enrichment by
Population\,III stars and take into account subsequent metal mixing
have found these systems to become significantly enriched to average
metallicities of $Z > 10^{-3}\,$Z$_{\odot}$ (\citealt{greif10},
\citealt{safranekshrader14a}). Metal mixing, however, occurs in a
rather inhomogeneous fashion resulting in large abundance spreads of
several dex across the system (\citealt{greif10},\citealt{wise12}).
In this context, it is not unlikely that some of the first galaxies
received ``external'' metal enrichment from neighboring halos. In the
process of undergoing massive supernova explosions that evacuate their
own minihalos, metal enrichment may occur over large length scales
that affect neighboring minihalos regardless of whether or not they
host their own Population\,III stars.

These recent works have thus all shown that the first galaxies were
enriched in metals, possibly provided via different avenues, and most
certainly in an inhomogeneous fashion. This naturally set the scene
for the formation of Population\,II stars, since the first galaxies
provided the first star-forming environments which were ready for the
formation of metal-poor, low-mass stars with $M <
1$\,M$_{\odot}$. Metal cooling became possible due to the metallicity
floor set by the Population\,III stars during or shortly after the
formation of the atomic cooling halo (or even in a
  minihalo). With the availability of metals, dust formation (and in
turn, subsequent dust cooling) became possible (e.g., in supernova
ejecta; \citealp{todini01}, \citealp{nozawa03}, \citealp{schneider04},
\citealp{omukai05}, \citealp{cherchneff10}, \citealp{schneider12a},
\citealp{klessen12}).

While the formation mode of the first low-mass stars remains an
ongoing debate, the details largely revolve around the concept of a
``critical metallicity''.  Above the critical metallicity, vigorous
fragmentation of the gas occurs. In such environments it could be
possible to form long-lived low-mass stars as a result of the N-body
dynamical ejection of accreting protostars before these objects had
the chance to accumulate their thermal Jeans mass. If indeed formed,
they would be able to survive to the present day and be observable in
the Milky Way or dwarf galaxies.  In turn, below such a critical
metallicity, only massive, short-lived stars would be able to form.
Two different pathways for low-mass star formation have thus been
suggested: the first is based on atomic fine structure line cooling,
the second on dust thermal cooling which results in a drop in the
Jeans mass to $<1$\,$M_{\odot}$. The critical metallicities are
$Z_{crit} \sim 10^{-3.5}$\,$Z_{\odot}$ for the line cooling
\citep{brommnature} and $Z_{crit} \sim 10^{-5}$\,$Z_{\odot}$ for dust
cooling (\citealp{schneider_nature03,schneider12a},
\citealp{omukai05}). In both cases sufficient fragmentation is assumed
to follow, and with it, low-mass star formation.

Given the already complex nature and interplay of the metal content of
supposedly simple systems such as the first galaxies, many details of
the cooling induced fragmentation process(es) remain unsettled. One
way forward, however, to test whether the mechanisms might be feasible
in facilitating Population\,II star formation, is to compare relevant
predictions with observational chemical abundance results of the most
iron-poor stars. After all, these are the stars that can be
  assumed to have formed in the first, or at least
earliest, galaxies, out of sparsely enriched gas and
could thus provide information about their formation mechanism.

\citet{dtrans} developed an ``observer-friendly'' formulation of early
C\,I and O\,II metal line cooling \citep{brommnature}. They predicted
that all metal-poor stars that formed through this mechanism should
have carbon and/or oxygen abundances \textit{in excess} of the line
cooling critical metallicity. To date, all stars with
$\mbox{[Fe/H]}<-3.0$ indeed show sufficiently enhanced levels of
carbon and oxygen, with only one exception, {\sdc} (see Table~1 and
Section~\ref{sec:mmps}). The existence of this star, however, has been
successfully explained in terms of a dust cooling mechanism
instead (\citealp{schneider12b}, \citealp{klessen12},
  \citealp{ji14}). A more recent study has suggested that while both
mechanisms may be important for forming the first low-mass stars, they
may occur as part of separate formation pathways and/or environments
\citep{ji14}. Detailed numerical simulations of star formation in
atomic cooling halos have already suggested where these pathways may
occur, although details so far remain unclear (\citealp{wise07},
\citealp{greif08}, \citealp{ritter12}).

With additional metal-poor star observations becoming available, the
details of the cooling and star formation processes can be tested
further. It will be particularly interesting to investigate the
specific physical reasons as to why the two cooling mechanisms would
be mutually exclusive \citep{ji14}. Understanding the underlying
physics of early star formation will then provide a much improved
framework for understanding the nature and origin of the most
iron-poor stars as well as for future modeling of star and galaxy
formation within large-scale simulations of galaxy growth and
evolution.

In the meantime, it is important to further explore the implications
of the fact that all but one of the seven most iron-poor stars (with
$\mbox{[Fe/H]}<-4.5$) are C-rich (see Table~1), and that some 20 --
30\% (or 40 -- 70\%, following \citealp{placco14}) of the metal-poor
stars in the range of $-4.5<\mbox{[Fe/H]}<-3.0$ are also C-rich. Is
this really a signature of a specific gas cooling process or could
there be another explanation?  Said differently, were there specific
supernova yields or other processes that led to the population of C-
and O-rich metal-poor stars?

\subsection{Origins of the Chemical Composition of C-rich Stars with $\mbox{[Fe/H]}<-3.0$}\label{sec:origins} 

No universally accepted hypothesis currently exists to explain the
origins of the C-rich stars with $\mbox{[Fe/H]} < -3.0$
(which are almost exclusively CEMP-no stars).  Given
that all but one of the seven most iron-poor stars ($\mbox{[Fe/H]} <
-4.5$) are C-rich, any acceptable scenario should be able to explain
fundamental aspects of the first chemical enrichment events in the
Universe.  This topic has been considered by many authors in the
recent literature, and we refer the reader to \citet{fujimoto00},
\citet{suda04}, \citet{beers&christlieb05}, \citet{meynet06},
\citet{nomoto06}, \citet{he0557}, \citet{frebel07}, \citet{lai08},
\citet{cohen08}, \citet{meynet10}, \citet{heger10},
\citet{joggerst10}, \citet{masseron10}, \citet{nomoto13},
\citet{fn13}, \citet{norris13}, and \citet{tominaga14} for previous
discussions. Here (following \citealp{norris13}) are scenarios that
have been suggested to have played a role in the production of stars
having abnormally large carbon enhancements from material initially
having zero or very low heavy-element content.

\subsubsection{Fine-structure line transitions of C\,II and O\,I as a
  major cooling agent in the early Universe {\citep{brommnature}}} C-
and/or O-rich gas forms stars, through fragmentation, on
shorter timescales than in regions where these elements
  are of lower abundances, leading to the formation of long-lived,
low mass C- and/or O-rich stars still observable today
\citep{dtrans}.

\subsubsection {Supermassive ($100 \lsim M \lsim300$\,M$_\odot$), rotating stars
  \citep{fryer01}} In some mass ranges, rotation leads to mixing, by
meridional circulation, of C and O from the He-burning core into the
H-burning shell. This results in extra N production, and then enhanced
N as well as C and O surface abundances and subsequent yields.

\subsubsection {Multiple  Type\,II supernova involving
  ``fallback'' ($M \sim10-40$\,M$_\odot$) (\citealp{limongi03})}

The ejecta from a ``normal'' supernova is combined with that from one
of low energy in which the outer layers (rich in light elements) are
expelled, while much of the inner layers (rich in the heavier
elements) ``fall back'' onto the central remnant.

\subsubsection {Mixing and fallback Type\,II supernovae ($M\sim10-40$\,M$_\odot$)
  (\citealt{UmedaNomotoNature}, \citealt{iwamoto05})}

Low energy supernovae eject material preferentially from their outer
regions, which are enhanced in light elements, with the expulsion of
only relatively small amounts of the heavier elements formed deeper in
the star.  During the explosion, internal mixing occurs in an annulus
outside the mass cut at which the expansion in initiated.  A small
amount of mixed material is eventually expelled from the star, with
most of it falling back into the central regions.  See
\citet{tominaga14} for a recent ``profiling'' of extremely metal-poor
stars.

\subsubsection {Type\,II supernovae with relativistic jets (\citealp{tominaga07b})}

A relativistic jet-induced black-hole-forming explosion of a
40\,M$_\odot$ supernova leads to infall of inner material that
``decreases the [ejected] amount of inner core material (Fe) relative
to that of outer material (C)''.

\subsubsection {Zero-metallicity, rotating, massive ($\sim60$\,M$_\odot$)
  and intermediate mass ($\sim7$\,M$_\odot$) stars
  (\citealp{meynet06}, \citealp{meynet10}, \citealp{hirschi07})}

Rotationally-driven meridional circulation leads to CNO enhancements
and large excesses of $^{13}$C (and hence low $^{12}$C/$^{13}$C
values), Na, Mg, and Al, in material expelled in stellar winds.  The
essential role of rotation is to admix and further process the
products of H and He burning. 

\subsubsection {A combination of rotation, mixing, and fallback}

For investigations of the combined effects of mixing, fallback, and
rotation in massive stars over wide parameter ranges, we refer the
reader to \citet{heger10} and \citet{joggerst10}.  

\subsubsection {Nucleosynthesis and mixing within low-mass, low-metallicity, stars (\citealp{fujimoto00}, \citealp{campbell10})}

Driven by a helium flash, carbon is mixed to the outer layers of
low-mass, extremely metal-poor giant stars, while protons are
transported into the hot convective core. Enhancements of Na, Mg, Al,
and heavy neutron-capture elements are also predicted.

\subsubsection {Population III binary evolution with mass transfer, and
  subsequent accretion from the interstellar medium (\citealp{suda04},
    \citealp{campbell10})}

The primary of a zero-heavy-element binary system is postulated to
transfer C- and N-rich material, during its AGB phase, onto the
currently-observed secondary, which later accretes Fe from the
interstellar medium to become a CEMP star.  Currently, radial velocity
monitoring of C-rich (CEMP-no) stars with $\mbox{[Fe/H]} \lsim -3.0$
does not support universal binarity among this group of stars
(\citealp{hansen13}, \citealp{norris13}, \citealp{starkenburg14}).
Rather, the reported binary fraction is roughly similar to the value
found for C-normal Galactic halo stars.

\subsubsection {Separation of gas and dust beyond the stellar surface
  during stellar evolution, followed by the accretion of the resulting
  dust-depleted gas (\citealp{venn08}, \citealt{venn14})}

Peculiar abundance patterns result from fractionation of the elements
onto grains, as determined by their condensation temperatures, during
stellar evolution, rather than being due to ``natal variations''.
Subsequent examination of the critical elements sulfur and zinc in the
Fe-poor, C-rich stars CS~22949-037 ($\mbox{[Fe/H]} = -4.0$) and {\hea}
($\mbox{[Fe/H]} = -5.8$) by \citet{spite11} and \citet{bonifacio12},
respectively, shows that these elements are detected in neither
object, in contrast to what might be expected in this model. A search
for mid-infrared excesses in these objects has been relatively
unsuccessful, with five out of six C-rich stars showing no mid-IR
excesses, while a small mid-IR excess near 10\,$\mu$m was detected at
the 2$\sigma$ level in the seventh (\hen) \citep{venn14}.

\vspace{0.2cm}
In summary, detailed comparisons of the stellar abundances with these
diverse approaches to the calculations of yields will help to
constrain Population\,III star properties. Investigating the results
within different environments will hopefully reveal the underlying
processes that drove star formation in the earliest galaxies.

\subsection{Interpreting the Abundance Signatures of stars 
with $\mbox{[Fe/H]}<-3.0$}{\label{sec:abund_int}}

\subsubsection{Rationale and limitations}
The most metal-poor stars very likely reflect the earliest enrichment
events by just one or a few prior Population\,III supernovae. This
concept can be understood by considering the following simple
example. Assume an average iron yield of a core-collapse supernova to
be 0.1\,M$_{\odot}$ (e.g., \citealt{heger10}) that is
homogeneously and instantaneously mixed into a typical pristine star
forming cloud of mass of $10^5$\,M$_{\odot}$. The resulting
metallicity of the gas, and hence the star formed in this cloud, will
be $\mbox{[Fe/H]}=-3.2$. This implies that stars with
$\mbox{[Fe/H]}\lsim-3.0$ can, in principle, be second generation stars
whose heavy elements were provided by just one supernova.  Various
effects such as mixing mass and mixing efficiency of the supernovae
ejecta into the ambient gaseous medium, as well as different supernova
yields, would of course alter this threshold value.

The most metal-poor stars are thus the most important and readily
available diagnostic tool for learning about Population\,III
supernovae. In turn, this provides information about the nature and
properties of the progenitor first stars. The critical
  challenge lies in correctly and sufficiently interpreting the
stellar chemical abundance patterns with the help of Population\,III
nucleosynthesis model calculations. Unfortunately, there are many
problems that arise on several fronts:

i) Observational: a good match between observed abundances and model
predictions depends crucially on having available as many elements
measured in the star as possible.  With the most iron-poor stars this
can become the limiting factor, as the recent example of {\smk}, with
$\mbox{[Fe/H]} < -7.3$, shows.  For this star, abundances could be
determined for only four elements (together with several upper limits)
due to its overall low abundance, causing the comparison between
observation and theory to be poorly constrained (\citealp{keller14},
\citealp{ishigaki14}). Complicating things further are systematic
abundance uncertainties arising from the use of 1D/LTE model
atmospheres to obtain the abundances rather than 3D/NLTE modeling. As
highlighted in Section~\ref{sec:assumptions}, for example, corrections
for C, N, and O can exceed 0.5\,dex which can change the results of the
abundance, yields matching significantly.  All of these
issues can thus lead to different conclusions about the progenitor
properties: for example, \citet{ishigaki14} infer solutions for
25\,M$_{\odot}$ and 40\,M$_{\odot}$ progenitors, whereas
\citet{keller14} find a best fit using a 60\,M$_{\odot}$ model.

ii) Theoretical: current nucleosynthesis model calculations are
plagued by many free parameters and gross uncertainties
(\citealt{UmedaNomoto:2005}, \citealt{heger10}). Essential knowledge
such as the explosion mechanism (e.g., 1D vs. 3D models, energy
injection through a piston approach), explosion energy, mass cut,
stellar rotation, existence/non-existence/extent of nucleosynthesis
reactions and processes, etc. are in many cases poorly
constrained. Moreover, treatment of the explosion has been largely
limited to spherical symmetry which simplifies current procedures but
also introduces further systematic uncertainties. All these caveats
prevent actual {\it ab initio} nucleosynthesis
\textit{predictions}. Rather, to this point, in most cases only
\textit{post-dictions} have been feasible, by searching for the best
match of theory with observations.

iii) Others: heterogeneous gas mixing processes in the star forming
cloud could lead to different amounts of local dilution and hence
different abundance patterns across one stellar generation
(\citealp{greif10}, \citealp{ritter14}). Accretion of interstellar
matter during the lifetime of a star could change its surface
abundance pattern (\citealp{suda04}, \citealp{hattori14}), as would
any surface pollution by a binary star companion.

\subsubsection{Sensitive and robust abundance probes}

In the absence of more refined knowledge or more exact solutions to
many or even all of these issues, the focus has been on reproducing
certain abundance ratios that appear to be particularly sensitive to
overall progenitor properties and nucleosynthesis calculations (but
which are robust to different overall modeling
assumptions).  Some of these are listed in Table~2, where we present
abundance ratios, how large (or small) the observed abundance ratio in
question needs to be in order to effectively constrain the scenario,
what property they constrain, and corresponding references to which we
refer the reader for additional details on the role of these abundance
ratios. Regarding their interpretation, the individual abundance
patterns of the most iron-poor stars with $\mbox{[Fe/H]}<-4.5$ have
already been described extensively in Sections~\ref{sec:mmps} and
\ref{sec:co}. What remains to be discussed is the extent to which
their elemental patterns have been reproduced by supernova
nucleosynthesis calculations and what has been concluded thus far
about the properties of the first star progenitors. We summarize the
most important findings below.

\captionsetup[figure]{labelformat=empty}
\begin{figure}[!htbp]
\begin{center}
\includegraphics[width=1.00\textwidth,angle=0]{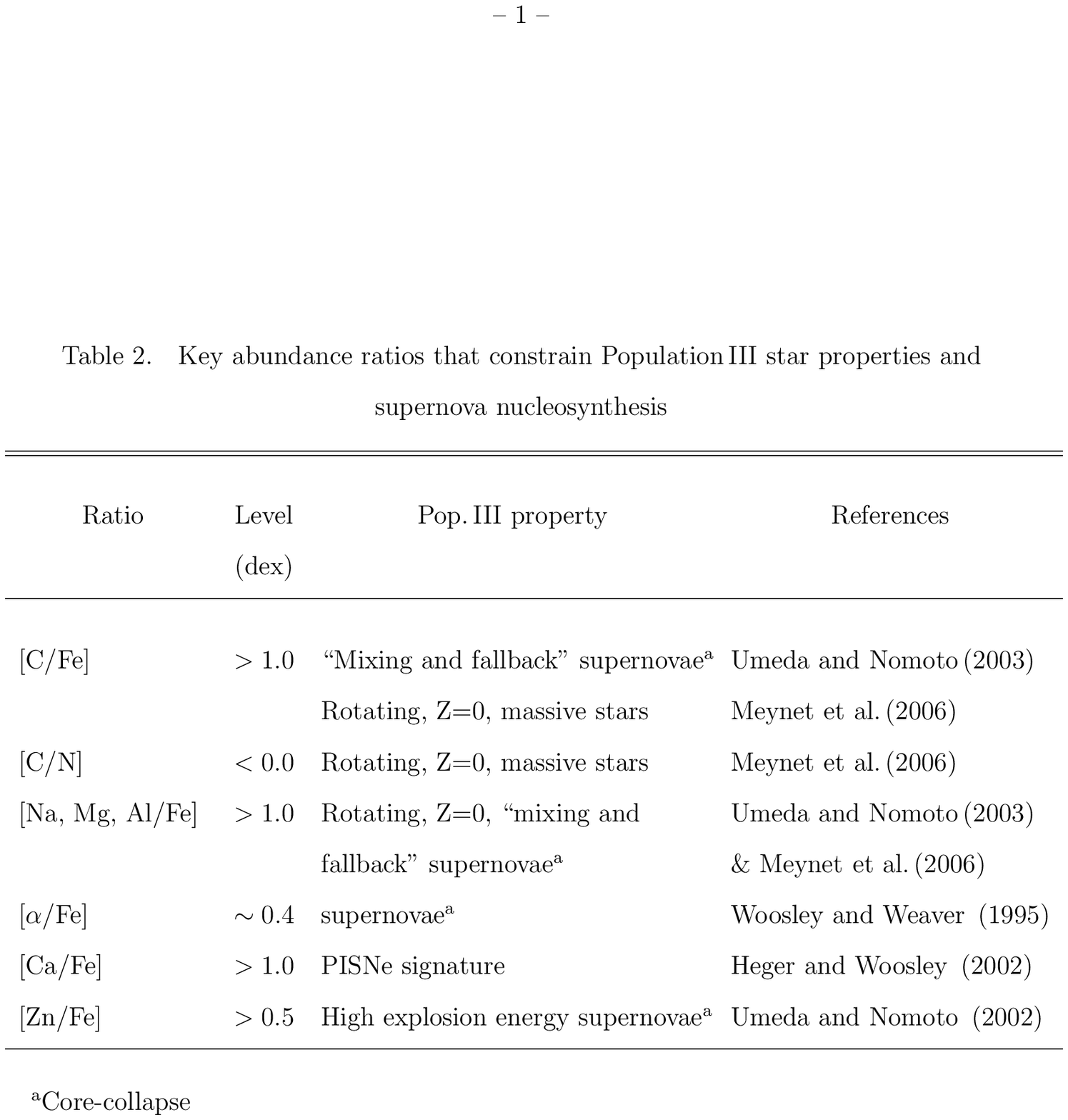}
  \caption{\scriptsize }
\end{center}
\end{figure}
\captionsetup[figure]{labelformat=default}

\subsubsection{Fitting with Mixing and Fallback models}
Low Fe abundances coupled with a high [C/Fe] ratio are a striking
feature of six of the seven most iron-poor stars, which
we have discussed in great detail in Section~\ref{sec:origins}. Many
interpretations are listed which are all potentially applicable for
the $\mbox{[Fe/H]}\lsim-4.5$ stars.
Perhaps the most promising, or at least the most extensively studied
scenario, is the mixing and fallback model. There, due to a
relatively low explosion energy ($\lsim10^{51}$\,erg;
\citealt{UmedaNomotoNature}), only the outer layers of the exploding
star containing principally lighter elements, made in the earlier
phases of stellar evolution, are ultimately ejected. The innermost
layers containing iron-peak elements, and especially iron from the
last burning stage, remain close to the core and fall back onto the
newly created black hole. Only an (arbitrarily chosen) small fraction
is then ejected, resulting in little or even no enrichment in these
elements.  The origin of the elements in the six C-rich stars in
Table~1 (all with $\mbox{[Fe/H]}<-4.5$) for which adequate abundance
data are available may all be explained in this manner.

Comparison of the abundance patterns of stars with $\mbox{[Fe/H]} <
-3.0$ with theoretical predictions by \citet{tominaga07_b},
\citet{tominaga14}, and \citet{ishigaki14} demonstrate that generally
good fits can be obtained with the yields of mixing and fallback
core-collapse supernovae.  The modeling involves some five free
parameters: (i) the explosion energy, E$_{51}$ (in units of 10$^{51}$
ergs); (ii) the mass of the inner boundary of the mixing region M$_{\rm
  cut}$; (iii) the mass of the outer boundary of the mixing region
M$_{\rm mix}$; (iv) the ejection factor f$_{\rm ej}$ (the fraction of
the mass M$_{\rm mix}$ -- M$_{\rm cut}$ that is ejected from the star;
and (v) a ``low-density'' factor f$_{\rho}$ defined by
\citet{tominaga14}.  We refer the reader to these works for details,
and in particular to \citet[their Figure~3]{tominaga14} for the range
of fitting parameters adopted in their investigation of 48 stars with
$\mbox{[Fe/H]} <-3.5$.

\begin{figure}[!t]
\begin{center}
\includegraphics[width=0.96\textwidth,angle=0]{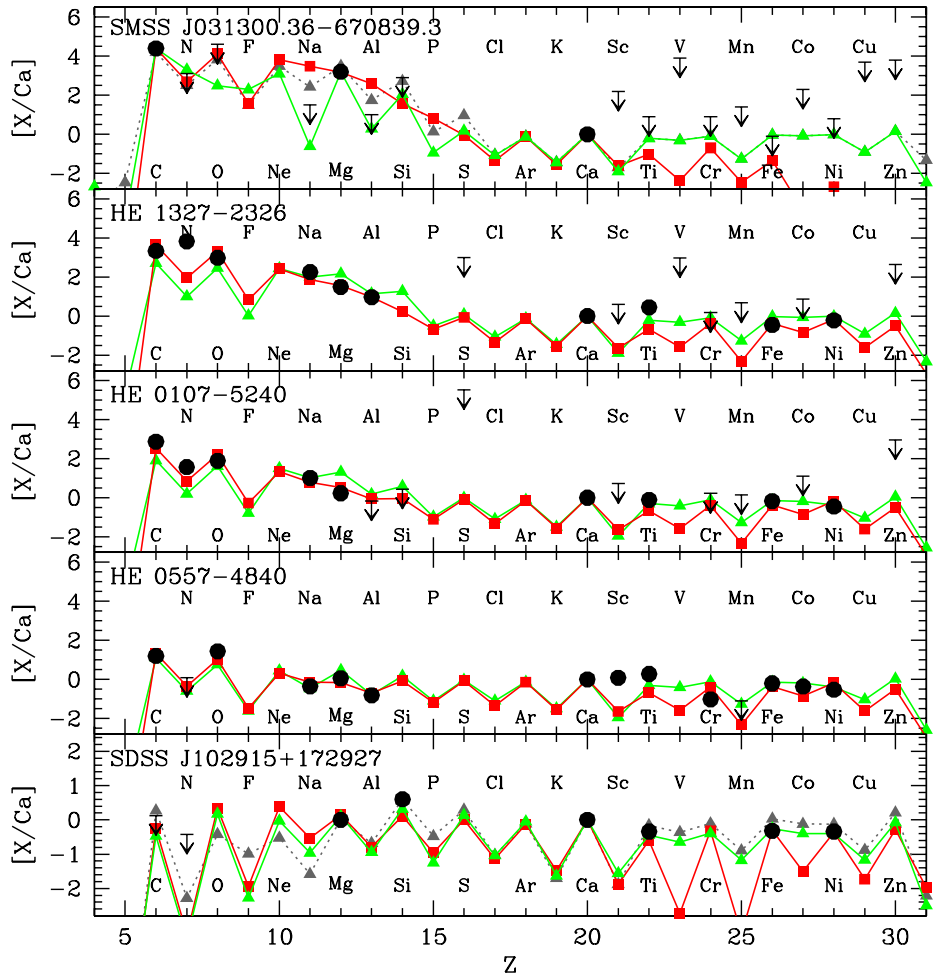}

  \caption{\scriptsize{\label{fig:ishigaki} Examples of the abundance
    fitting procedure for five of the most iron-poor stars
     with $\mbox{[Fe/H]}<-4.5$ (black circles) to infer properties of the
    Population\,III stars that enriched the respective birth gas
    clouds (adapted from \citealp{ishigaki14}). (From top to bottom,
    the observed stellar iron abundances are $\mbox{[Fe/H]} <-7.3$,
    $-5.7$, $-5.4$, $-4.8$, and $-4.7$.)  Red squares and green
    triangles show 25\,M$_{\odot}$ model results for supernovae with
    E$_{51} = 1$ (in units of 10$^{51}$\,erg) and E$_{51} =
    10$, respectively, for the carbon-enhanced stars in
    the top four panels. The bottom panel shows a non-carbon enhanced
    star fit with a 40\,M$_{\odot}$ model having energies of E$_{51} =
    1$ and 30 (red squares and green triangles, respectively). Note
    the factor of two change on the [X/Ca] axis for this panel.  (The
    gray triangles with the dotted line indicate a higher energy model
    having an alternative set of parameters.) Credit: M. Ishigaki.}
}
\end{center}
\end{figure}

Figure~\ref{fig:ishigaki}, adapted from the work of
\citet{ishigaki14}, presents the abundances, [X/Ca], (filled black
circles) as a function of atomic number for five of the most iron-poor
stars currently known (four C-rich, and one with $\mbox{[C/Fe]} <
+0.9$).  Comparison among the observed abundances shows the
interesting and obvious stark contrast between the overall abundance
patterns of C -- Al in the C-rich stars in the top four panels and the
C-normal star, SD 1029+1729, in the bottom panel. While the C-rich
stars have large overabundances of C -- Al (ranging over several dex),
such a spread does not exist in SD 1029+1729. (Note also that the
y-axis abundance scale in the bottom panel is only half that of the
top four panels).

Each observed abundance pattern is overplotted with the mixing and
fallback nucleosynthesis calculations of \citet{ishigaki14} for two
different explosion energies (red squares and green triangles). For
the C-rich stars, in the top four panels, 25\,M$_{\odot}$ models with
E$_{51} = 1$ and 10 are used.  For the C-normal star, {\sdc}, in the
bottom panel, 40\,M$_{\odot}$ models with E$_{51} = 1$ and 30 are
presented.  While the authors were unable to determine one best
fitting energy for {\smk} due to the small number of elements
observed, they do report best fits of E$_{51}$ = 1 for the three other
C-rich stars, and E$_{51} = 30$ for the C-normal star.

\subsubsection{The case for rotation}

An important exception to the good fits in Figure~\ref{fig:ishigaki}
is nitrogen in the warm C-rich subgiant HE~1327$-$2326 ({\teff} =
6190\,K, {\logg} = 3.7, $\mbox{[Fe/H]} = -5.7$, $\mbox{[C/Fe]} =
+4.3$, $\mbox{[N/Fe]} = +4.6$, and $\mbox{[O/Fe]} = +3.7$,
\citealt{aokihe1327}, \citealt{o_he1327}), which is severely
underproduced in the model. Nitrogen thus defies explanation in terms
of mixing-and-fallback. Its overabundance, however, is readily
explained in terms of the zero-metallicity, rapidly-rotating, massive
($\sim60$\,M$_\odot$) stars of \citet{meynet06} and \citet{meynet10},
in which rotationally-driven meridional circulation leads to large
amounts of surface enhancements in C, N, O, Na, Mg, and Al in material
which will subsequently be expelled in stellar winds prior to the star
exploding.

The essential role of rotation is to spatially admix and further
process the products of H and He burning. In Section~\ref{sec:naca},
we reported large variations of Na, Mg, and Al that correlate with
carbon in some 50\% of the C-rich stars.  As also demonstrated by
Meynet and coworkers, one naturally expects enhancements of Na, Mg,
and Al as the result of the further nuclear processing of the
admixture of the products of H and He burning, via (p,$\gamma$)
reactions.  In mixing and fallback models, on the other hand, this
could result from the admixing of different radial zones, their
subsequent nuclear burning, and the expulsion of material that
contains different relative amounts of synthesized Na, Mg and Al. The
prospect of observing nucleosynthetic signatures of rotation-related
element patterns is furthermore supported by recent simulations of
Population\,III stars showing that they may have been rapidly rotating
(e.g., \citealt{stacy13}).

\subsubsection{A test between mixing-and-fallback and rotation}

There is a basic difference between the rotating star models, on the
one hand, and those that experience mixing and fallback on the other.
In principle at least, the two cases sample different regions of the
progenitor stars that produce the enrichment.  In the rotating models,
the regions providing the enrichment are the outer layers that mix via
meridional circulation, and much of the ejecta are expelled in stellar
winds, before exhaustion of the nuclear fuel in the central regions
leads to a potential explosion.  In the other class of model, the
entire enrichment patterns are determined in the supernova phase,
during which there is mixing and expulsion, potentially at least, of
material from all parts of the star outside the core.  Insofar as Si
and Ca are produced deeper in a star than are the lighter elements,
they present the opportunity to test the
predictions of the different models more closely.  In particular, more
accurate abundances for Si in a larger sample of C-rich stars are
needed for comparison with more detailed predictions of the two
classes of models.  In Figure~\ref{fig:naca}, we see no evidence for
enhancement of Si and Ca, as might be expected, in the rapidly
rotating scenario.  It would, however, be interesting to have the
detailed abundance predictions of the mixing and fallback models.
This could be a very useful future avenue of
investigation.

\subsubsection{Open questions}

We comment, finally, on two signatures that are absent from the
abundance patterns of the six most iron-poor stars for which detailed
abundances exist.  First, no star exhibits a pronounced ``odd-even''
effect or a significantly high [Ca/Fe] ratio, as has been predicted to
be produced by PISNe enrichment (see Section~\ref{sec:firstnucleo}).
As we noted in our earlier discussion of PISNe, the most iron-poor
stars in Table~1 appear not to have been enriched by
ejecta from these extremely massive explosions.

Second, no zinc measurement is available in any of the stars with
$\mbox{[Fe/H]}<-4.5$ because the only two available Zn lines are too
weak at these metallicities. Zn, however, is an indicator of the
explosion energy \citep{UmedaNomoto:2002}. Higher [Zn/Fe] (and also
[Co/Fe]) abundances can be explained with explosions of higher energy,
(E$_{51}>10$). Perhaps future Zn measurements in these stars could
provide an independent estimate of the explosion energy, in addition
to the more indirect method of inferring the energy based on the
combination of low [Fe/H] and high [C/Fe] ratios discussed here.

\subsubsection{Summary}

Significant progress in this area will likely depend on a more refined
understanding of supernova nucleosynthesis, and the availability of
additional metal-poor stars for a broad mapping of the full observed
abundance parameter space. One promising possibility for progress in
this respect would be a more comprehensive treatment of multiple
mechanisms involving mixing and fallback of rapidly rotating stars.

\subsection{The Relationship between the C-rich and C-normal Populations}\label{sec:twopops}

What is the basic difference between the origins of the C-rich and
C-normal populations with $\mbox{[Fe/H]}<-3.0$?  As discussed in
Section~\ref{sec:firstgalaxies}, an attractive possibility is that
there were two distinct cooling processes -- C\,I and O\,II metal line
cooling, on the one hand, and dust thermal cooling, on the other --
that operated at the earliest times, after the first Population\,III
stars had exploded and enriched the material from which Population\,II
formed.  What was not clear is why the two processes should operate to
different extents, with the C and O cooling apparently being
predominant.  

\citet{norris13} advocated a simple scenario, in which the first
Population\,II star forming clouds which fragmented to produce
low-mass stars that still exist today contained large amounts of
carbon and oxygen, relative to their heavier element content.  The
enriching Population\,III stars may have been some or all described by
the stellar evolutionary models in Section~\ref{sec:origins} -- the
rotating 250 -- 300\,M$_{\odot}$ models of \citet{fryer01}; the mixing
and fallback models of \citet{UmedaNomotoNature} and
\citet{iwamoto05}; the relativistic jet-induced explosions of
\citet{tominaga07b}; and the rapidly-rotating stars of
\citet{meynet06} and \citet{meynet10}.  Following \citet{dtrans}, it
was also suggested that during the subsequent star formation within
the second generation (the first Population\,II generation), the
material with large enhancements of carbon and oxygen fragmented to
form low-mass, long-lived stars that are still observable today.  They
identified the C-rich population as having been formed from
carbon-enriched material. In contrast, to explain the C-normal star
{\sdc}, with $\mbox{[Fe/H]} = -4.7$ and $\mbox{[C/Fe]} < +0.9$
\citep{caffau11, caffau12}, it was suggested that some fraction of the
Population\,III stars did not produce large amounts of carbon (as the
result perhaps of canonical supernovae explosions without fallback, or
slower rotation), but instead produced ejecta with chemical abundance
patterns that were rather more solar-like in nature.  The gas enriched
by this material then experienced dust-induced star formation and
fragmentation once the dust critical metallicity is exceeded (e.g.,
\citealt{debennassuti14}).

\citet{cooke14} considered the formation of C-rich stars using
cosmological simulations of chemical enrichment within minihalos.
Adopting the supernova model yields of \citet{heger10}, they found
that low-energy, mixing and fallback supernovae, which produced
material with a high value of [C/Fe], were not powerful enough to
evacuate the gas from their host minihalos.  In their simulations, the
authors assumed ``that each minihalo that is able to retain its gas
will form a second generation of stars''.  This subsequent
Population\,II star formation then resulted in a population of C-rich
stars.  At the other extreme of the energy scale, the authors reported
that high-energy PISNe would have sufficient energy to remove all gas
from the minihalo (see also \citealt{greif07}). This would preclude
such subsequent star formation that would carry the characteristic
PISN signature of the odd-even effect and large [Ca/Fe] values
discussed above in Section~\ref{sec:firstnucleo}.  In the general
case, the assumed supernova energy and Initial Mass Function, together
with their yields, will thus determine the enrichment patterns in the
minihalos.  Within this framework the simulations were able to explain
(i) the essential features of the distribution of CEMP-no stars with
$\mbox{[Fe/H]}<-3.0$ in the ([C/Fe], [Fe/H]) -- plane, and (ii) the
increasing fraction of C-rich stars with decreasing iron abundance.
We refer the reader to \citet{cooke14} for more details of this work.

A semantic question that remains unanswered in the discussions of both
\citet{norris13} and \citet{cooke14} is which of the C-rich and the
C-normal stars formed first.  Hopefully, insight into this question
will be provided by future simulations that model the timescales
associated with C\,I and O\,II metal line cooling, on the one hand,
and with dust cooling, on the other.

\subsection{The Ultra-faint Dwarfs: Survivors of the First Galaxies?}\label{sec:ufd}

The metallicity-luminosity relation for dwarf galaxies \citep{kirby08}
clearly shows that the faintest galaxies have average metallicities of
$\mbox{[Fe/H]}\sim-2.5$ with underlying [Fe/H] spreads of
up to $\sim2.5$\,dex. In addition, these systems
are lacking stars with $\mbox{[Fe/H]}>-1.0$ which is perhaps the main
reason why they contain, relatively speaking, such a large fraction of
metal-poor stars. Having found galaxies with average metallicities
close to the extremely metal-poor regime is an exciting new prospect
to study early star forming environments and associated conditions. In
fact, instead of just using individual metal-poor stars, as done in
stellar archaeology, now entire galaxies have become the fossil record
that can be studied together with its stellar content. Thus, dwarf
galaxy archaeology has become the latest tool in the field of
near-field cosmology. Topically, it elegantly bridges the gap between
observational stellar archaeology and the theoretical simulation
studies of the first stars and first galaxies. After all, the
immediate question that has arisen with the discovery of ultra-faint
galaxies with as few as $\sim$1000 stars is whether any of these
faintest systems could be related to Galactic building blocks and/or
the earliest galaxies to have formed in the Universe. The answer is of
great importance to observers and theorists alike.

All of the Milky Way's dSph galaxies, and now also the ultra-faint
dwarfs, have been subject to intensive studies, with a common goal:
what can these systems tell us about galaxy formation and chemical
evolution on small scales, and what is their relationship with the
building blocks of the Milky Way? With their large relative fraction
of extremely metal-poor stars, the metal-deficient ultra-faint dwarfs
likely formed at the earliest times and are thus ideal test objects
for answering these questions  \citep{bovill09}.

Let us first briefly recall that the first galaxies must have harbored
the first long-lived low-mass metal-deficient Population\,II
stars. Learning about the environment in which these stars formed
would help to understand the origin of the most metal-poor stars found
in the halo. Based on first galaxy simulation results,
\citet{frebel12} explored what the chemical tell-tale signatures of a
first galaxy might be, assuming it survived until the present day. The
basic idea rests upon the assumption that a first galaxy would have
only experienced a Population\,III star generation plus one additional
first generation of Population\,II stars (formed from somewhat
metal-enriched gas) before losing its gas and the possibility of
subsequent star formation (through a possible blow-out of gas by the
more massive Population\,II stars, or reionization). The
Population\,II star generation would contain the first long-lived
low-mass stars. The corresponding chemical make-up of such a galaxy
with this heavily truncated star formation history and which underwent
only chemical \textit{enrichment} and no chemical \textit{evolution}
is straightforward to predict. In brief, we list the criteria that
have to be fulfilled in order for a system observed today to qualify
as a candidate surviving first galaxy.

\begin{itemize}

\item Large [Fe/H] spread with low average metallicity and existence
  of stars with $\mbox{[Fe/H]} < -3.0$.  This can be explained with
  inhomogeneous mixing in the early gas cloud.

\item Generally, halo-style chemical abundance pattern that signal
  core-collapse supernova enrichment.

\item No signs of AGB star driven enrichment of heavy
  neutron-capture elements, such as Sr and Ba.  (No long-lived stars
  would have formed from material enriched by the ejecta of the AGB
  stars of the first Population\,II stars.)

\item No downturn in [$\alpha$/Fe] at any (higher, e.g., $\mbox{[Fe/H]}
  > -2.0$) metallicity due to the onset of iron-producing supernovae
  Type\,Ia. (No long-lived stars would have formed after the Type Ia
  explosions).

\end{itemize}

We note that as research continues into the nature and evolution of
the first galaxies, these criteria may need to be refined or even
extended in the future, but overall these basics should remain valid.
Comparison of these criteria with abundances of stars in the
ultra-faint dwarf galaxies should thus already shed light on the
question of whether any of today's least luminous dwarfs are perhaps
surviving first galaxies. In turn, this would provide hints as to
whether it might be possible for any of the first galaxies to have
survived until the present day -- an exciting prospect for near-field
cosmology.

The issue can be approached by using the Segue\,1 system as the most
relevant example. With $L\sim1000$L$_{\odot}$, Segue\,1 is the
faintest galaxy yet discovered \citep{belokurov07}. It was found
because it is only 23\,kpc away in the halo. Its average metallicity
is $\mbox{[Fe/H]}\sim-2.5$ which is, however, difficult to accurately
determine given its fairly flat and sparsely populated distribution of
[Fe/H] (e.g., \citealt{simon11}). Some controversy about the nature of
Segue\,1 occurred soon after its discovery. While \citet{niederste09}
advocated its being a star cluster, additional studies have shown that
it is a highly dark matter dominated galaxy (\citealt{geha09},
\citealt{simon11}).  This result is supported by the (i) large [Fe/H]
and [C/Fe] spreads ($\Delta\mbox{[Fe/H]} = 2.4$ and
$\Delta\mbox{[C/Fe]} = 2.4$), (ii) the existence of stars with
$\mbox{[Fe/H]}<-3.0$, (iii) halo-like chemical abundances of light
elements, and (iv) consistently low neutron-capture element abundances
of all its stars studied with high-resolution spectroscopy
\citep{frebel14}. Without doubt, these are all signatures associated
with galactic chemical enrichment rather than with any star cluster
that does not show signs of chemical evolution. It was thus concluded
that Segue\,1 is most likely the most primitive galaxy known.

The detailed chemical abundances of seven stars in Segue\,1 were thus
compared to the above criteria \citep{frebel14}. All of them were met,
making Segue\,1 the first viable surviving first galaxy candidate.  In
contrast, inspection of the limited abundance data available for stars
in the somewhat more luminous ultra-faint dwarfs indicates that
systems with $L\gsim10^{4.5}$L$_{\odot}$ already show signs of
chemical evolution (in particular, some lower, solar-like
$\alpha$-element abundances) and thus do not fulfill all of the
criteria.  This suggests that these systems already had multiple
generations of stars and that some chemical evolution had occurred,
albeit still very little in comparison with the luminous dSphs which
clearly show extended star formation and chemical evolution. This
luminosity limit is not unlike that suggested by \citet{bovill11} of
$L<10^6$\,L$_{\odot}$, (based on simulation work), as a limit for the
fossil nature of the ultra-faint systems.

Surviving dwarf galaxies like Segue\,1 thus give unique clues as to
what the conditions of early galaxies may have been. At the same time,
it has to be considered that Segue\,1 is a building block-type galaxy
that survived the Milky Way's assembly process. Whether it survived
for a particular reason will remain unanswered for now (although
structure formation simulations may shed light on this more general
issue of the survival rate and preference of building blocks during
the evolution of large galaxies). 

That said, accepting Segue\,1 as an analog of the accreted building
blocks offers tantalizing insights.  Considering its stellar content,
it appears plausible that Segue\,1-like objects may have populated the
low-metallicity tail of the halo metallicity distribution
function. Given its low mass and the current incompleteness of the
metal-poor end of the halo metallicity distribution function, it
remains unclear how many such systems would be required to populate
the entire tail.  It thus seems likely that the bulk of the halo stars
must have come from more massive systems. This is supported by the
excellent agreement between the abundance signatures of the most
metal-poor Segue\,1 stars and those of the metal-poor halo stars that
have similar values of [Fe/H]. This means that we can use
Segue\,1-type systems to learn about the origin of the most metal-poor
halo stars since they likely formed in such systems. It closes an
important loop since we can only speculate otherwise that halo stars
must have formed in some kind of early systems.

This example illustrates that studying the surviving dwarf galaxies is
helping to fill in the missing information that the most metal-poor
halo stars by themselves cannot offer us: where they formed and how
they made their way into the halo of the Milky Way. See
  also \citet{revaz12}.  Moreover, as the more sophisticated
simulations become available dwarf archaeology can be established as a
major empirical constraint on the formation process of the first
galaxies and the first long-lived low-mass stars.

Finally, we note that after having found one candidate first galaxy in
the northern hemisphere, further progress on the observational side
will depend on finding more of these faintest Segue\,1-type dwarf
galaxies in the south, for example, with the SkyMapper Telescope and
the Large Synoptic Survey Telescope (see Section~\ref{nearterm}), and
also with search techniques involving stellar proper motion
measurements \citep{fabrizio14}. The availability of new systems is
important because all dwarf galaxy stars are extremely faint. Often
only 1 -- 3 stars per galaxy can be observed with high-resolution
spectroscopy, and by now essentially all of the accessible stars in
the currently-known ultra-faint dwarfs have been observed. More new
dwarfs are needed to provide a fresh supply of sufficiently bright
stars for high-resolution spectroscopy with current facilities for the
detailed abundance work that may lead to the discovery of additional
first-galaxy candidates.

\newpage
\section{NEAR-FIELD MEETS FAR-FIELD COSMOLOGY}\label{sec:nearfar}

The study of the most metal-poor stars provides information about the
early Universe and the conditions at the time and place of their
birth.  Near-field results, however, rest on the implicit assumption
that the most metal-poor stars did indeed form from low-metallicity
gas within the first billion years or so after the Big Bang.  While
this conclusion is clearly supported by the age dating of the rare
group of r-process enhanced stars which have ages of 13 -- 14\,Gyr and
$\mbox{[Fe/H]} \sim -3.0$ (e.g., \citealp{Hilletal:2002},
\citealp{he1523}, \citealp{sneden_araa}), it is of critical importance
to verify this, by investigating high-redshift early gaseous systems
and their metal content via direct observations.

Metallicity measurements have been obtained for some 200 high redshift
Damped Lyman-$\alpha$ (DLA) systems with $z\sim2 - 5$ (e.g.,
\citealt{jorgenson13}, \citealt{rafelski14}, and references therein),
observed along the sightline to a given quasar (where it remains
unknown whether the nucleus or the diffuse outer halo is being
probed).  The data of Rakelsky et al., on average, reveal a slow but
steady decrease in metallicity with increasing redshift, and by
$z\sim$ 4.7 the mean metallicity is $\mbox{[M/H]}\sim -2.2$.  While
the distribution of abundances has a floor at $\mbox{[M/H]} \sim-2.5$,
there are important exceptions. \citet{cooke11b} have reported a small
number of objects in the extremely metal-poor regime, with
$\mbox{[Fe/H]}<-3$, at $z = 2.3$ -- 3.7.  This is indicative that at
least some of these early gas clouds could have hosted the formation
of the first/early low-mass Population\,II stars which we observe as
the most metal-poor halo stars today.

We have, moreover, reached an interesting point in the field of metal
abundance determinations pertaining to both stellar archaeology and
high-redshift DLA systems and other gas clouds of the Lyman-{$\alpha$}
forest.  Objects in both disciplines are now being discovered that
have such low abundances that only upper limits can be determined. The
stellar archaeology example is {\smk}, with $\mbox{[Fe/H]}<-7.3$,
which has already been discussed in Sections~\ref{sec:archaeology} and
\ref{sec:nearfield} (see also Table~1, and
Figure~\ref{fig:spectra_araa}).  We now address the far-field
low-abundance limits.

\subsection{The Most Metal-poor High-redshift Gaseous Systems}

\citet{fumagalli11} have reported Lyman limit systems (LLSs) along two
quasar sightlines having redshifts $z\sim3.4$ and $z\sim3.1$ with
metallicity upper limits of $Z < 10^{-4.2}$\,Z$_{\odot}$ and $Z <
10^{-3.8}$\,Z$_{\odot}$ (i.e., $\mbox{[M/H]} = -4.2$ and $-3.8$)
respectively.  These redshifts imply an epoch about 2\,Gyr after the
Big Bang.  It seems reasonable, therefore, to suggest that systems
like these could be representative of the initial clouds in the early
Universe which hosted some of the first Population\,II stars.
However, this is not {\it that} early in the evolution of the Universe
(compared with the epochs earlier than 500 Myr or interest in the
present discussion) and possibly implies that low-metallicity gas, or
even metal-free clouds, may have survived for this long.  It is
unclear, of course, whether these particular LLSs have actually formed
metal-poor stars.

An even more extreme case, reported by \citet{simcoe12}, is a gaseous
cloud with $z\sim7$, observed along the sightline to the quasar ULAS
J1120+0641.  This redshift corresponds to an age of the Universe of
only $\sim$\,800\,Myr.  No metal absorption is found in the spectrum
of the cloud, leading to metallicity limits of $Z <
10^{-3.0}$\,Z$_{\odot}$ if the gas is simply diffuse and unbound, or
$Z < 10^{-4.0}$\,Z$_{\odot}$ if it is a gravitationally bound
protogalaxy (i.e., $\mbox{[M/H]} = -3.0$ and $-4.0$, respectively).

\subsection{Abundances ([X/H] and [X/Fe]) in the High Redshift Gas Clouds}\label{sec:dlas}

A critical point of comparison between near-field and far-field
endeavors lies in their detailed relative abundances. In what follows
we shall restrict our attention to quasar absorption line systems
having redshifts $z>$ 2, in particular the metal-poor DLA systems at
lower redshifts 2 $<z$ {\lsim} 4 (e.g., \citealp[and references
  therein]{cooke11b}) and the so-called sub-DLAs over the range 4
{\lsim} $z <$ 6 (e.g., \citealp[and references
  therein]{becker12}).  In the DLA systems, below redshifts of 
$z\sim$ 4 the amounts of hydrogen and several elements, X, can be
determined along the sightline to the background quasar, and both [X/H]
and [X/Fe] can be obtained.  For the sub-DLAs, however, in the range
4 {\lsim} $z<$ 6, the amount of hydrogen is not well measured and
relative abundances of only the other elements (e.g., [X/Fe]) are
available.

In the lower redshift regime, \citet{cooke11b} report column densities
for some 11 elements.  In a sample of 21 objects with $\mbox{[Fe/H]}
\lsim -2.0$, the three most metal-poor systems have [Fe/H] in the
range $-3.5$ to $-3.0$.  They also report that the ratios of C/O and
O/Fe are consistent with values determined for stars in the Galactic
halo.  Of particular interest, in the present context, is the result
that one (J0035--0918) of the 10 systems having both C and Fe
abundances has the composition of a CEMP-no star ($\mbox{[Fe/H]}=-3.0$
and $\mbox{[C/Fe]} = +1.5$) (see also \citealp{cooke11}). (We note
that this result has been challenged by \citealp[their
  Section~4]{becker12}.  See also \citealp{carswell12}.)  In a later
paper \citet{cooke12} report a second carbon-enhanced object
(J1358+6522), with $\mbox{[Fe/H]} = -2.8$ and $\mbox{[C/Fe]} = +0.6$.

The results of \citet{becker12} for the sub-DLAs extend the dataset to
redshift $z = 6.3$, and provide abundance information for C, O, Si,
and Fe.  These authors supplement their results with those of others
at lower redshift (but excluding the C-enhanced system of
\citealp{cooke11}) to provide a sample of 20 objects over the redshift
range $z = 2$ -- 6.  In their Figure 11 \citet{becker12} plot [C/O],
[Si/O], [C/Si], [C/Fe], [O/Fe], and [Si/Fe] as functions of redshift.
In this diagram there is no evidence for a large variation in any of the
relative abundances.  In particular, for their four systems having C
and O abundances in the redshift range $4.7 < z < 6.3$, they report
mean values $\langle\mbox{[C/Fe]}\rangle = +0.17 \pm 0.07$ and
$\langle\mbox{[O/Fe]}\rangle =+0.50\pm 0.05$, respectively.  That is
to say, the C and O abundances of sub-DLA systems at the highest
redshifts currently observed are the same as those of ``normal''
non-carbon-enhanced Galactic halo stars.  Concerning the comparison
with the abundances of carbon in the most Fe-poor stars in the Milky
Way, \citet{becker12} suggest: ``If carbon-enhanced stars fairly
reflect their native ISM abundances, then these abundances are no
longer common by $z\sim 6$.  This raises the intriguing possibility
that most carbon-enhanced stars were formed at even earlier times.''

We have come to the crux of the matter.  If one includes the two
carbon-enhanced extremely metal-poor systems of \citet{cooke11} and
\citet{cooke12}) with those of \citet[their Figure 11]{becker12} one
finds (i) that in the range $2 < z < 6.3$ the fraction of carbon-rich
DLAs is $\sim$0.15 compared with the value of 0.20 -- 0.30 (or higher,
according to \citealp{placco14}, see Section~\ref{sec:mdf}) for C-rich
metal-poor stars at $\mbox{[Fe/H]}<-3.0$, and (ii) the carbon
abundances of the two DLA systems ($\mbox{[C/Fe]} = +0.6$ and +1.5)
are smaller in size and range in comparison with the values one finds
for the Galactic C-rich stars discussed in
Sections~\ref{sec:archaeology}, which span the range in [C/Fe] from
$\sim+0.7$ to $>4.9$.  The numbers of stars and DLA systems involved
here are clearly too small to permit any meaningful conclusion,
especially when keeping in mind that it is unknown which region of the
DLA is being studied with the observed sightline.  That said, the
tension will need to be resolved in the future.  More DLA abundance
measurements of iron and carbon should be obtained to ensure the
current result is not due to small-number statistics.  Given that the
redshifts of the DLAs discussed here are associated with epochs more
than 800 Myr after the Big Bang, an intriguing alternative is that the
DLA and sub-DLA abundances reflect the average abundances of the
medium at that time, depending on the level and length scales of
mixing in these early clouds.  Potentially, the stellar abundances
could reflect more detailed local conditions at earlier times.  In
this case, the abundance patterns of both the most metal-poor halo
stars and the high-redshift DLAs could be understood in terms of
chemical enrichment by the ejecta of massive stars exploding as
core-collapse supernovae, albeit with the stars and DLA systems being
at different phases in their evolution.

\newpage
\section{THE BRIGHT FUTURE OF NEAR-FIELD COSMOLOGY}\label{sec:future}

The last decade has seen a steep increase in activities related to
stellar archaeology and near-field cosmology. A number of exciting
discoveries have been made such as the stars with $\mbox{[Fe/H]}<-5.0$
and the ultra-faint dwarf galaxy population. This opened up entire new
lines of research as well as drawing attention to near-field cosmology
as an important area of study that connects stellar astrophysics with
galaxy formation and evolution.

Several observational large scale efforts to characterize the Milky
Way, its stellar content, satellite population and formation history
have contributed enormously to near-field cosmology, ushering in a new
era of exploration. Most notably among them is SDSS (primarily its
extensions focused on Galactic science), but there have also been the
numerous individual studies that have collectively led to the detailed
analysis of more than 1000 stars with $\mbox{[Fe/H]}<-2.0$. This
number is now expected to rapidly increase; we list major near-term
surveys for more metal-poor stars in Section~\ref{nearterm}.

Besides these and other observational advances that have brought much
attention to the topics of galactic astrophysics, stellar populations,
chemical evolution and galaxy formation, crucial progress has been
made in the area of first star and first galaxy simulations. These
works directly relate to metal-poor stars and their formation as well
as the nature and history of dwarf galaxies. We are thus at an
interesting crossover point right now where simulations of early star
forming environments can finally be compared, rudimentarily, with
observational near-field cosmology results and where stellar
archaeology, and in particular dwarf galaxy archaeology, inspire and
constrain these simulations. This close connection promises to be
vital for future progress in both areas, especially once the detailed
evolution of metals can be traced more generically in
simulations. Efforts towards this goal are well underway in this
respect (e.g., \citealt{greif10}, \citealt{wise12},
\citealt{safranekshrader14b}, \citealt{safranekshrader14a},
\citealt{jeon14}). In this manner, interpretion of the stellar
abundance patterns will soon be undertaken in the context of detailed
simulation results.

\subsection{Open Questions in Near-field Cosmology}

Despite the enormous progress both observationally and theoretically,
important questions remain to be answered. Many of these have only
arisen as the result of recent advances, which illustrate that
near-field cosmology is a vibrant field that is rapidly moving
forward. Below we list a number of open key questions that describe
the state of the field while simultaneously providing hints
as to what may be coming in the next decade.

What are the lowest stellar iron and carbon abundances, respectively?
Could there be stars with arbitrarily low Fe abundances?   What is the
fraction of carbon-rich stars at the lowest metallicities?  What is
the spatial distribution of the most metal-poor stars and do they
reflect an accreted halo component as opposed to an {\it in situ}
component?

Are there any surviving low-mass Population\,III stars? Can
Population\,II stars form in minihalos?  To what extent are
  stellar abundance patterns altered by accretion of ISM material or
  nearby supernova explosions? What are the nucleosynthesis yields of
the earliest supernova? How does metal dispersal operate in early star
forming environments? What are the details of the manner
  in which the Milky Way's halo was assembled?

How many ultra-faint dwarf galaxies are out there? Faint
  systems such as Segue\,1 can currently only be identified out to
$\sim30$\,kpc. How many are yet to be found in the unexplored southern
hemisphere? Are systems like Segue\,1 really undisturbed first/early
galaxies that have survived until today? How many of them could be
orbiting the Milky Way? Have the classical dSph galaxies already been
assembled from smaller (gaseous) fragments? 

Answers to some questions can already be estimated, at least when they
are pushing towards the technical limit of observations or
simulations.  One example is that of the lowest iron and carbon
abundances in halo stars. The current technical limit for an abundance
measurement based on just detecting the strongest iron line is
$\mbox{[Fe/H]}<-7.2$ \citep{frebel09}. Using {\smk} as an example, a
carbon abundance of down to $\mbox{[C/H]} \sim-4.0$ would be
measurable, based on the detection of the G-band at 4313\,{\AA} and
assuming the availability of a high $S/N$ ($\sim100$) spectrum. If
{\smk} was cooler (it has {\teff} = 5125\,K), for example, with
{\teff} = 4500\,K, a carbon abundance of $\mbox{[C/H]} \sim-4.2$ would
still be measurable.

The hypothetical spectrum of a star with no metal absorption lines in
its spectrum (just hydrogen lines), and assuming it to be a cool giant
would thus yield abundance limits of $\mbox{[Fe/H]}<-7.2$,
$\mbox{[Ca/H]} <-9.4$, $\mbox{[Mg/H]} <-6.0$,and $\mbox{[C/H]} <-4.0$
{\citealt{frebel09}, \citealt{fn13}}. This highlights an important
point regarding the nature and origin of carbon-rich metal-poor
stars. The current detection limit essentially prevents us from
measuring accurate carbon abundances that are very low, with
underabundances similar to those of iron. However, it might just be
enough to at least discriminate formation scenarios for these low-mass
stars. The $D_{trans}$ criterion of \citet{dtrans} is
$D_{\rm{trans}} = \log(10^{\rm{[C/H]}}
+0.3\times10^{\rm{[O/H]}})>-3.5$. It is based on a combined minimum
abundance of carbon and oxygen. A value of $\mbox{[C/H]} \sim-4.0$,
especially if additionally downcorrected by applying 3D abundance
corrections might already indicate that dust cooling and not
fine-structure line cooling might have led to the formation of the
object. However, since much lower carbon values cannot be determined,
a direct confirmation of a dust cooling scenario may not be possible
for these stars.

\subsection{Near-term Searches for Metal-poor Stars and Dwarf Galaxies} \label{nearterm}

Looking back some 40 years, one sees that the search for metal-poor
stars has transitioned from relatively small-scale projects into an
enormous enterprise.  Several surveys operating around the globe have
dedicated programs for metal-poor stars and galactic exploration while
others will be more generally usable for the characterization of the
Milky Way's structure, stellar populations and history. We briefly
list the main near-term projects below.

{\bf The Apache Point Observatory Galactic Evolution Experiment}
(APOGEE; \citealt{majewski10}) is part of the third extension of the
Sloan Digital Sky Survey (SDSS;
  http://www.sdss.org). Since 2011 it has been mapping the Galaxy
(bulge, disk, and even parts of the halo) with high-resolution,
near-infrared spectroscopy to establish the Milky Way's chemical and
kinematical evolution.

{\bf The Large Sky Area Multi-Object Fiber Spectroscopic Telescope}
(LAMOST; \citet{deng12};
http://www.lamost.org/public/dr1?locale=en) has a dedicated galactic
program called LEGUE which provides low
($R\sim1500$) resolution {spectroscopy of up to 4000 stars per
  pointing (facilitated by 16 linked spectrographs, each with 250
  fibers, in the northern hemisphere. Metal-poor candidates are
  already being selected from these data.

{\bf The SkyMapper Telescope} \citep{kelleretal07} is now efficiently
selecting metal-poor candidates based on its combination of
broad and narrow band filters designed to straightforwardly
characterize stellar properties. Its photometric Southern Sky Survey
will also provide deep images in the future to search for southern
ultra-faint dwarf galaxies and stellar streams.

{\bf The Anglo-Australian Telescope's high-resolution multi-object
  spectrograph HERMES} \citep{hermes12} is being used to observe
$\sim$1 million stars (400 at a time) to characterize the history of
star formation in the Galaxy through detailed chemical abundance
measurements as part of the Galah survey.

{\bf The satellite all-sky mission Gaia} (e.g., \citealp{cacciari09}) will
produce astrometry, and photometry for some 1 billion stars down to
magnitude $V = 20$ (about $3.5\times10^{5}$ sources to $V=10$,
$2.6\times10^{7}$ to $V=15$, and $2.5\times10^{8}$ to $V=18$).
Additional spectroscopy ($R \sim 10000$; over 8450 -- 8720\,{\AA}, the
region of the Ca\,II triplet) will enable stellar parameter
determinations of perhaps half a dozen million objects from which
metal-poor candidates can be selected. We refer the reader to
\citet{gilmore_gaia12} for a description of the synergy between Gaia
and ground-based facilities that will together provide basic detailed
chemical abundances for an unprecedented sample of the Galaxy's
metal-poor stars.

 {\bf The Large Synoptic Survey Telescope} (LSST; \citealt{ivezic08},
version 4) will carry out a deep southern sky photometric survey using
broad band filters. These data will allow a detailed assessment of
Galactic structure and enable the search for southern ultra-faint
dwarf galaxies and stellar streams.

\subsection{Next-generation Telescopes}
\vspace{-0.3cm}
Near-field cosmology will greatly benefit from the existence of the
three next-generation extremely large telescopes currently planned to
be operational from $\sim$2020.  In particular, the Giant Magellan
Telescope (GMT) promises great advances in stellar and dwarf galaxy
archaeology because it selected a high-resolution optical spectrograph
called G-CLEF \citep{szent12} as one of its first light
instruments. This fiber-fed spectrograph will enable detailed studies
of metal-poor stars at resolving powers of $R\sim20000$, 40000, and
100000 in the outskirts of the Milky Way and in dwarf galaxies. The
GMT will also have a low-resolution optical spectrograph, GMACS,
suitable for observation of the fainter stars in the dwarf galaxies
and deep low-resolution stellar spectroscopy within the Galaxy.

The European Extremely Large Telescope (E-ELT) and the Thirty Meter
Telescope (TMT) will not initially be equipped with an optical
high-resolution spectrograph but instead with optical and
near-infrared imagers and low-resolution spectrographs.  These
instruments will enable unprecedented observations of high-redshift
galaxies which will deliver complementary information to that provided
by near-field cosmology concerning the earliest epochs of star and
galaxy formation. In the same vein, the James Webb Space Telescope
(JWST) will allow the highest-redshift observations of
the earliest galaxies, perhaps even the first massive star clusters to
have formed in the Universe.

\vspace{-0.5cm}
\subsection{Theoretical Insight: The Journey from First Stars to the Milky Way}
\vspace{-0.2cm}

Observations of metal-poor stars located in the Milky Way permit
powerful insight into the earliest epochs of star and galaxy
formation. However, they cannot provide direct information on where
exactly these stars formed and how their respective host systems were
accreted by the Galaxy. Cosmological simulations of early star forming
processes and structure formation in the Universe are necessary to
reveal this type of global information. With the emergence of powerful
supercomputers tremendous progress has been made in this area. Within
the next decade near-field cosmology, paired with far-field cosmology
and supported by large-scale simulations of galaxy assembly, will
provide a comprehensive picture of how the Milky Way assembled and how
to interpret the nature and history of its stellar content.

\section*{Acknowledgements}

We have pleasure in thanking colleagues W. Aoki, T.C. Beers,
M.S. Bessell, V. Bromm, N. Christlieb, G. Gilmore, H.L. Morrison,
S.G. Ryan, J.D. Simon, R.F.G. Wyse, and D. Yong for collaborations in
these areas over many years.  We also gratefully acknowledge
T.C. Beers, V. Bromm, N. Christlieb, R. Simcoe, K.A. Venn, and D. Yong for
commenting on the manuscript, and M. Ishigaki, A.D. Mackey, and
D. Yong, for providing figures that have been included in the
presentation.  We thank P. Bonifacio and P. Fran{\c c}ois for
providing information concerning SDSS J1742+2531 in advance of
publication.  A.F. is supported by an NSF CAREER grant AST-1255160.
J.E.N.  acknowledges support from the Australian Research Council
(grants DP03042613, DP0663562, DP0984924, and DP120100475) for studies
of the Galaxy's most metal-poor stars and ultra-faint satellite
systems.

\nocite{christliebetal02, christliebetal04}

\end{document}